\documentclass[11pt,letterpaper]{article}
\usepackage{fullpage}
\usepackage{times}
\usepackage{enumitem}
\usepackage[dvipsnames]{xcolor}

\usepackage{authblk}

\usepackage[margin=.85in,footskip=0.32in]{geometry}
\usepackage{setspace}
\usepackage{pdflscape}
\usepackage{parskip}
\usepackage{fancyhdr}
\usepackage{lastpage}
\fancypagestyle{plain}{ % for page 1
    \fancyhf{}
    \fancyfoot[R]{\footnotesize\hfill \textit{Page \thepage\ of \pageref{LastPage}}}
}
\usepackage{amsfonts,latexsym,amsthm,amssymb,amsmath,amscd,euscript}
\usepackage{mathtools}
\usepackage{nicefrac, xfrac}
\usepackage{cancel}
\usepackage{tcolorbox}
\tcbuselibrary{theorems}
\tcbset{
  noborderbox/.style={
    colframe=white,     % border color
    boxrule=0pt,        % border thickness
    colback=Periwinkle!8, % background color
    sharp corners,      
    width=0.9\linewidth
  }
}

\usepackage[]{hyperref}
\hypersetup{
    pdftitle = {},
    pdfauthor = {}
    hidelinks,
    colorlinks = true,
    bookmarksnumbered,
    citecolor = black,
    linkcolor = black,
    urlcolor  = black,
    }

\usepackage{multirow}
\usepackage{longtable}
\usepackage{booktabs}
\usepackage{colortbl}
\usepackage{blkarray}
\usepackage[bf,margin=0.4cm,font=footnotesize]{caption} 
\usepackage[font=small]{subcaption}
\usepackage{tabularx}

\usepackage{float}

\usepackage[numbers,sort&compress,round]{natbib}
\renewcommand{\cite}{\citep}

\usepackage[capitalise,nameinlink]{cleveref}
\usepackage{nameref}

\usepackage{titlesec}
\newcommand{\Addperiod}[1]{#1.}

\titleformat{\subsubsection}[runin]
    {\normalfont\bfseries}{\thesubsection}{0.5em}{\Addperiod}
\titleformat{\subsection}
    {\normalfont\bfseries}{\thesubsection}{0.5em}{}
\titleformat{\section}
    {\large\bfseries}{\thesection.}{0.5em}{}[]

\titlespacing*{\section}{0pt}{2.5ex}{1.5ex}
\titlespacing*{\subsection}{0pt}{2ex}{1ex}
\titlespacing*{\subsubsection}{0pt}{1.5ex}{1ex}

\newcommand{\exec}[0]{\varepsilon}
\newcommand{\assess}[0]{\alpha}

\newcommand{\DR}{\textrm{local}}
\newcommand{\IR}{\textrm{global}}

\newcommand{\svDR}{\mathbf{s}^{\DR}}
\newcommand{\svDRX}[1]{\mathbf{s}^{\DR,\,#1}}
\newcommand{\svIR}{\mathbf{s}^{\IR}}
\newcommand{\svIRX}[1]{\mathbf{s}^{\IR,\,#1}}
\newcommand{\av}{\mathbf{a}}
\newcommand{\sDR}[2]{s^{\DR}_{#1#2}}

\newcommand{\sIR}[1]{s^{\IR}_{#1}}

\newcommand{\ant}[2]{a_{#1#2}}
\newcommand{\svDRt}{\mathbf{\tilde s}^{\DR}}
\newcommand{\svDRtX}[1]{\mathbf{\tilde s}^{\DR,\, #1}}
\newcommand{\svIRt}{\mathbf{\tilde s}^{\IR}}
\newcommand{\svIRtX}[1]{\mathbf{\tilde s}^{\IR,\, #1}}
\newcommand{\avt}{\mathbf{\tilde a}}
\newcommand{\sDRt}[2]{\tilde s^{\DR}_{#1#2}}

\newcommand{\sIRt}[1]{\tilde s^{\IR}_{#1}}

\newcommand{\at}[2]{\tilde a_{#1#2}}

\newcommand{\vh}{\mathbf{h}}
\newcommand{\vP}{\mathbf{P}}

\DeclareMathOperator{\sgn}{sgn}

\newcommand{\res}[1]{{\color[HTML]{6B8E23}#1}}
\newcommand{\resnn}[1]{{\color[HTML]{3d5114}#1}} 
\newcommand{\mut}[1]{{\color[HTML]{9B4E9B}#1}} 

\newcommand{\R}[0]{{\textrm{R-N}}}
\newcommand{\RNN}[0]{{\textrm{R-NN}}}
\newcommand{\M}[0]{{\textrm{M}}}

\newcommand{\ALLC}[0]{\textrm{ALLC}}
\newcommand{\ALLD}[0]{\textrm{ALLD}}
\newcommand{\SI}[0]{{\color{black}\textit{Supplementary Information}}}

\title{A model of local and global reciprocity}
\author[1,2,*]{Mari Kawakatsu}
\author[3,4,5]{Yohsuke Murase}
\author[1]{Taylor A.\ Kessinger}
\author[1,2]{Joshua B.\ Plotkin}
\affil[1]{Department of Biology, University of Pennsylvania, Philadelphia, PA, USA}
\affil[2]{Center for Mathematical Biology, University of Pennsylvania, Philadelphia, PA, USA}
\affil[3]{RIKEN Center for Interdisciplinary Theoretical and Mathematical Science (iTHEMS), Wako, Japan}
\affil[4]{RIKEN Center for Computational Science, Kobe, Japan}
\affil[5]{Graduate School of Science and Engineering, Saitama University, Saitama, Japan}
\affil[*]{Corresponding author: marikawa@sas.upenn.edu}

\date{\vspace{-2ex}}

\begin{document}
\addtocontents{toc}{\protect\setcounter{tocdepth}{-10}}

\maketitle

\begin{abstract}

\noindent
We often decide how to treat friends based on observations of their past behavior, whereas actions toward strangers are typically guided by their public reputations. These two kinds of information underlie two classical mechanisms for the evolution of cooperation---direct and indirect reciprocity---which have largely been studied in isolation. They are not interchangeable: we can recall the past actions of only a small circle of close contacts, whereas for the far larger pool of strangers we must rely on public reputations. Here we develop a mathematical framework built on this distinction. Each individual engages in direct reciprocity in local games within a finite neighborhood of friends, whose actions they observe directly, and in indirect reciprocity in global games with a large population of strangers, known only by reputation. Separating local and global interactions allows us to address two questions. First, can cooperation persist under a cognitively simple norm of judgment? We show that combining direct and indirect reciprocity resolves the scoring dilemma: conditional cooperators resist invasion by both unconditional cooperators and unconditional defectors, where indirect reciprocity alone would fail. Second, how should one treat a friend whose past behavior conflicts with their public reputation? We find that the strategies that maximize cooperation are forgiving---overlooking whichever piece of information is unfavorable---and that these forgiving strategies can often remain robust to invasion. By distinguishing between local and global scales of interaction and integrating information across them, our framework offers a more cognitively realistic account of how reciprocity sustains cooperation. 
\\
\begin{center}\textbf{Keywords}\end{center}
\begin{center}
social evolution; evolutionary game theory; reciprocity; image scoring; local and global interactions
\end{center}

\end{abstract}

\setcounter{secnumdepth}{3}
\normalsize

%%%%%%%%%%%%%%%%%%%%%%%%%%%
% MAIN TEXT 
%%%%%%%%%%%%%%%%%%%%%%%%%%%
\newpage
\section{Introduction}

How should we treat people we know well, in light of how they are viewed by a broader public? The question arises constantly. A scientist celebrated in their field may treat the trainees in their own group poorly; a neighbor with a reputation for being unfriendly may prove kind in private. In each case, our direct experience of a person conflicts with their public reputation, and we must decide which to act on. How to weigh personal experience against social reputation---and especially how to behave when the two disagree---is all the more pressing in a modern world where reputations are widely and easily accessible through ratings, reviews, and social media. Underlying this question is a related problem of enduring interest to social scientists and evolutionary biologists alike: how cooperation is sustained among self-interested individuals.

A leading explanation for cooperation among unrelated individuals is reciprocity---the tendency to condition one's behavior toward others on how they have behaved in the past. Reciprocity takes two classical forms, distinguished by the kind of information they use. Under direct reciprocity, we condition our behavior on direct experience: the past actions of partners we have observed firsthand \cite{axelrod_evolution_1981,axelrod_evolution_1984}. Under indirect reciprocity, we instead condition our behavior on reputations: social assessments that summarize how others have behaved toward third parties \cite{alexander_biology_1987,fehr_nature_2003,boyd_evolution_1989}. Direct reciprocity thus draws on private, firsthand knowledge, whereas indirect reciprocity draws on public, secondhand information---the two kinds of information that may conflict when we know someone both personally and by reputation. Yet theoretical accounts of cooperation have largely studied the two mechanisms in isolation.

In everyday life, direct and indirect reciprocity are intercalated as we interact with both friends and strangers. But the two mechanisms tend to operate at different social scales because they make very different cognitive demands. Models of indirect reciprocity based on private reputations, where every individual forms and stores their own opinion of everyone else \cite{uchida_effect_2010,uchida_effect_2013,sasaki_evolution_2017,radzvilavicius_evolution_2019,radzvilavicius_adherence_2021,michel-mata_evolution_2024,kawakatsu_mechanistic_2024,schmid_evolution_2021,schmid_quantitative_2023,schmid_unified_2021,perret_evolution_2021,okada_tolerant_2017,okada_solution_2018,fujimoto_evolutionary_2023,fujimoto_reputation_2022}, would require an individual to monitor, judge, and remember the standing of every other member of a large population; and they demand a memory more powerful than direct reciprocity itself requires. Public reputations relieve this burden: when assessments are shared across a population, an individual can know a stranger's reputation without ever having observed them \cite{leimar_evolution_2001,ohtsuki_how_2004,ohtsuki_leading_2006,pacheco_sternjudging_2006,santos_social_2016,sasaki_evolution_2017,santos_social_2018,murase_social_2022,murase_indirect_2023,sasaki_evolution_2024,glynatsi_exact_2025}. But public reputation alone is too coarse a description of human social life. For the small circle of people we interact with repeatedly---close friends, family, near neighbors, or colleagues---we possess something richer than a public reputation: direct, firsthand experience of how they have actually behaved toward us and toward one another.

Here we propose that direct and indirect reciprocity are best understood not as competing accounts of cooperation, but as complementary processes that naturally occur at distinct but coupled scales. We develop a model in which each individual belongs to a small, finite \textit{neighborhood} of local partners whose actions they can directly observe and recall, while also interacting with a much larger population of \textit{non-neighbors} known only through public reputation. Individuals engage in direct reciprocity in \textit{local} games with their neighbors and in indirect reciprocity in \textit{global} games with non-neighbors. The finiteness of the neighborhood is essential: it is small enough that remembering recent actions is cognitively realistic, whereas the global pool is far too large for anything but shared, public reputations. This separation of local and global interactions, each governed by the form of reciprocity that the available information supports, is the central idea of our framework. Recent efforts to integrate direct and indirect reciprocity have assumed private reputations and well-mixed populations \cite{schmid_unified_2021,pal_coevolution_2024,hubner_stable_2025a,yamamoto_tolerant_2025}; by separating a finite local neighborhood from the vast global population, our framework offers what we argue is a more faithful description of human social life.

This framework allows us to pose two key questions about reciprocity and its consequences. The first concerns the stability of cooperation under cognitively simple rules of judgment. In indirect reciprocity, reputations are assigned according to a social norm. The simplest norm, \textit{image scoring}, judges cooperation as good and defection as bad \cite{wedekind_cooperation_2000,leimar_evolution_2001,brandt_indirect_2005,sigmund_calculus_2010}. Intuitive as it is, image scoring famously fails to stabilize cooperation: discriminators who withhold help from bad-reputation partners are themselves judged badly, and cooperation collapses---a failure known as the \textit{scoring dilemma} \cite{okada_review_2020,okada_two_2020}. Can the combination of local and global interactions stabilize cooperation under image scoring, without recourse to more elaborate norms? And even if this combination works to stabilize cooperation, we are left with a second question: how should we treat a neighbor when direct experience and public reputation conflict, such as a well-regarded friend who has behaved badly toward us?

We find that the interplay of local and global reciprocity answers both questions. First, combining direct and indirect reciprocity resolves the scoring dilemma: a population of conditional cooperators who play tit-for-tat locally and discriminate by reputation globally can resist local invasion by both unconditional cooperators and unconditional defectors, so long as local and global interactions are suitably balanced and neighborhoods remain small. Cooperation is thereby stabilized under the simple image-scoring norm, without higher-order norms or other cognitively demanding machinery \cite{leimar_evolution_2001,ohtsuki_how_2004,ohtsuki_leading_2006,pacheco_sternjudging_2006,santos_social_2016,sasaki_evolution_2017,santos_social_2018,murase_social_2022,murase_indirect_2023,berger_learning_2011,berger_stability_2016,michel-mata_evolution_2024}. Second, when direct experience and public reputation conflict, the strategies that maximize cooperation are \textit{forgiving}: they overlook whichever piece of information is unfavorable. Forgiving a neighbor's poor private behavior tends to matter more than forgiving a poor public reputation. In some regimes, appropriately forgiving strategies can simultaneously maximize cooperation and remain robust against unconditional invaders. Together, these findings suggest that cooperation depends not on firsthand experience or public reputations alone, but on how individuals use and combine the two, an issue that is increasingly pressing as public reputations become ever easier to access.

\section{A model of local direct reciprocity and global indirect reciprocity}
\label{sec:model}

\subsubsection*{Neighborhood structure} 
We consider a large population of players partitioned into neighborhoods, each containing $n \geq 2$ players. For a given focal player, the $n-1$ other players in their neighborhood are called \textit{neighbors}, while all others in the population are called \textit{non-neighbors}. Each finite neighborhood represents a small collection of players who can observe and recall recent actions of one another within the neighborhood. By contrast, players cannot directly observe actions among the large pool of non-neighbors and know only their public reputations. Accordingly, players engage in direct reciprocity with their neighbors and in indirect reciprocity with their non-neighbors.

\subsubsection*{Local and global games} 
\begin{figure}[t!]
    \centering
    \includegraphics[width=0.95\linewidth,trim={0.25in 2.5in 0.25in 2.5in},clip]{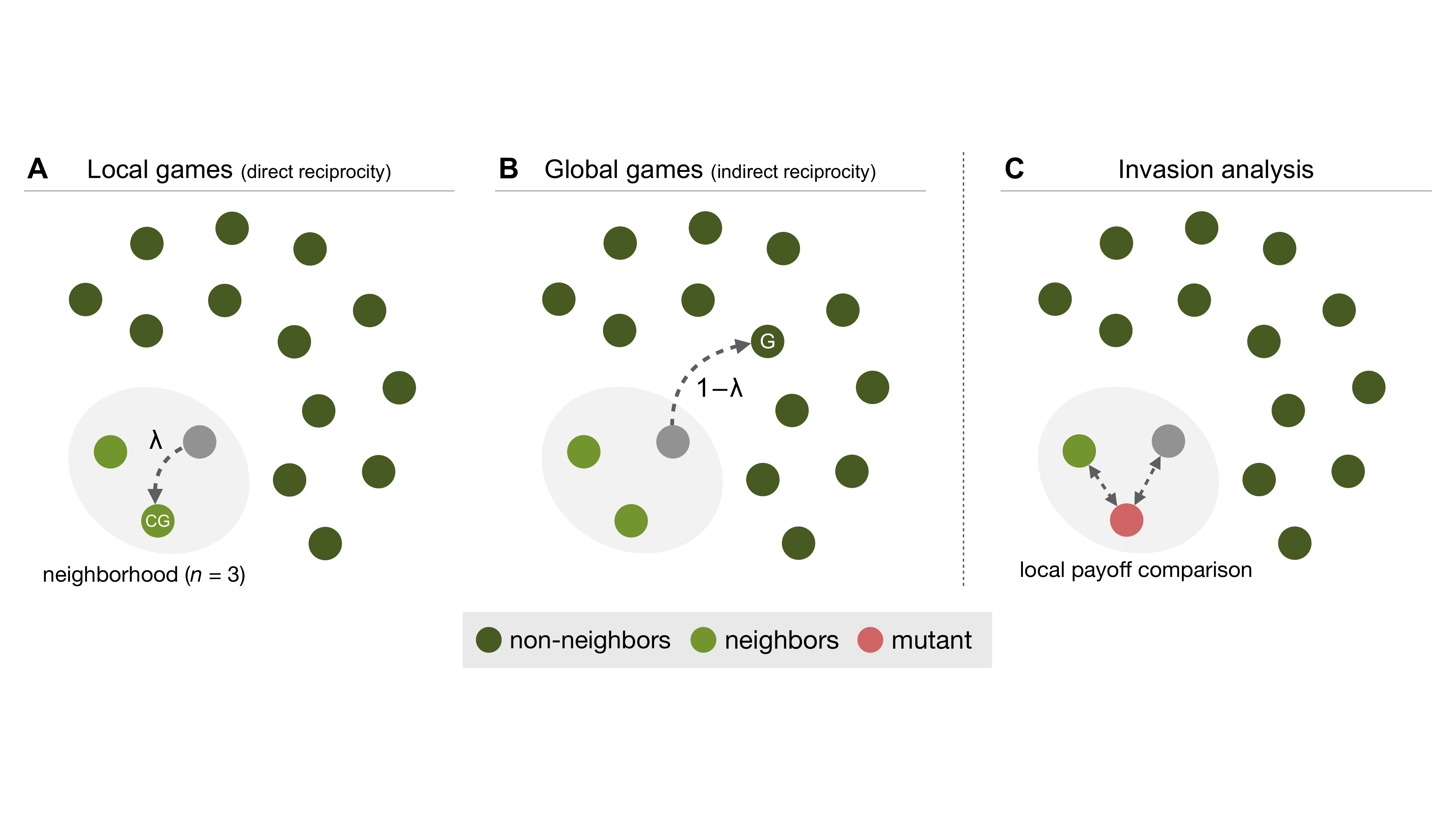}
    \caption{
        \textbf{A model of direct and indirect reciprocity.}
        We consider a large population of players partitioned into neighborhoods of $n$ players each. 
        \textbf{A, B}: In each round, a focal player (gray circle) is chosen uniformly at random from the population. From the perspective of the focal player, the $n-1$ other players in the focal player's neighborhood (light gray oval) are \textit{neighbors} (light green circles), while all others in the population are \textit{non-neighbors} (dark green circles).
        \textbf{A}: With probability $\lambda$, the focal player engages in a local game and is paired with a randomly selected neighbor. In a local game, the action of the focal player may depend on the co-player's action ($C$ or $D$) in the most recent local game or on the co-player's public reputation in the current round ($G$ or $B$).
        \textbf{B}: With probability $1-\lambda$, the focal player engages in a global game and is paired with a randomly selected non-neighbor.
        In a global game, the action of the focal player may depend on the co-player's public reputation ($G$ or $B$).
        \textbf{C}: A rare mutant (red circle) is introduced into a neighborhood. Resident-type players inside the mutant's neighborhood are called \textit{resident neighbors} (light green circles), while those outside are termed \textit{resident non-neighbors} (dark green circles). Because each neighborhood is finite but small relative to the population, the mutant affects the fitness of resident neighbors but not that of resident non-neighbors. 
        The resident type can resist local invasion by the mutant if and only if the fitness of resident neighbors exceeds that of the mutant.
        }
    \label{fig:schematic}
\end{figure}

More specifically, players engage in infinitely many rounds of pairwise interactions. In each round, a focal player is chosen uniformly at random from the population. An interaction type is then selected: with probability $\lambda$, the focal player engages in a \textit{local game} and is paired with a randomly selected neighbor (\cref{fig:schematic}A); with probability $1-\lambda$, the focal player engages in a \textit{global game} and is paired with a randomly selected non-neighbor (\cref{fig:schematic}B). 

Whether local or global, each interaction takes the form of a donation game. An interacting pair plays the game twice, with each player acting once as a \textit{donor} and once as a \textit{recipient}. In each game, the donor chooses one of two actions: cooperate ($C$), i.e., pay a cost $c>0$ to provide a benefit $b>c$ to the recipient, or defect ($D$), i.e., incur no cost and provide no benefit. 

This model reduces to classical indirect reciprocity with public reputations \cite{leimar_evolution_2001,ohtsuki_how_2004} when interactions are always global ($\lambda=0$), and it reduces to classical direct reciprocity \cite{axelrod_evolution_1981,sigmund_calculus_2010} when interactions are always local in a neighborhood of size two ($\lambda=1$ and $n=2$).

\subsubsection*{Strategies}

In a local game, the action of a focal player toward a neighbor may depend on two pieces of information about that neighbor: 
their past behavior within the neighborhood and their reputation at large (\cref{fig:schematic}A). We capture this dependence using reactive strategies that condition a player's behavior on the neighbor's \textit{most recent local action} ($C$ or $D$)---that is, their action in their most recent local game, regardless of whether that action was directed toward the focal player or another neighbor---as well as the neighbor's current \textit{public reputation} (good ($G$) or bad ($B$)). A strategy in local games is given by a vector $\svDR = \left(\sDR{C}{G}, \sDR{C}{B}, \sDR{D}{G}, \sDR{D}{B} \right)$, where $\sDR{A}{R} \in \left[0,1\right]$ denotes the probability that a focal player will cooperate with a co-player whose most recent local action was $A \in \{C,D\}$ and whose current public reputation is $R \in \{G,B\}$. For example, $\svDR = \left(1,1,1,1\right)$ and $\left(0,0,0,0\right)$ correspond to unconditional cooperation and unconditional defection, respectively. By contrast, the strategy $\svDR = \left(1,0,0,0\right)$ cooperates with a neighbor only if they \textit{both} cooperated in their most recent local game \textit{and} have a good public reputation, corresponding to a strict AND policy for combining local and global information, whereas strategy $\svDR = \left(1,1,1,0\right)$ cooperates with a neighbor if they \textit{either} cooperated in their most recent local game \textit{or} have a good public reputation, corresponding to a more forgiving policy for combining information.

For a neighborhood of size $n=2$, this formulation of strategies for local games includes memory-1 reactive strategies in classical models of direct reciprocity, such as tit-for-tat (TFT) and generous TFT (GTFT) \cite{sigmund_calculus_2010}. In our analysis, we will use TFT and GTFT to refer to strategies $\svDR = \left(1,1,0,0\right)$ and $\left(1,1,Q,Q\right)$ ($Q>0$), respectively, which are natural extensions of the two-player TFT and GTFT to local games with $n\geq2$.

In a global game, the action of a focal player toward a non-neighbor may depend on the non-neighbor's public reputation (\cref{fig:schematic}B).
A strategy in global games is thus given by a vector $\svIR = (\sIR{G}, \sIR{B})$, where $\sIR{R}\in \left[0,1\right]$ denotes the probability that a focal player will cooperate with a co-player whose reputation is $R\in\{G, B\}$. For example, $\svIR = \left(1,1\right)$ corresponds to unconditional cooperation, $\left(0,0\right)$ to unconditional defection, and $\left(1,0\right)$ to the discriminator strategy (DISC, cooperate only if the co-player has a good reputation) in models of indirect reciprocity \cite{sigmund_calculus_2010}.

\subsubsection*{Reputation assessment}

We assume that reputations are public knowledge, so that all players agree on each other's reputation at any given time. Public reputations can emerge through a variety of mechanisms that synchronize people's opinions of one another \cite{murase_indirect_2024}, including centralized monitoring and broadcast \cite{radzvilavicius_adherence_2021}, rapid gossip \cite{kawakatsu_mechanistic_2024}, and simultaneous observations of players' actions \cite{fujimoto_evolutionary_2023}. In addition, in real-world settings, an individual's reputation at large is often shaped not only by how they behave in public, but also by how they treat those close to them within relatively private contexts. For instance, reports of a scientist mistreating trainees or a manager exploiting employees can spread quickly through social networks, affecting the scientist's reputation within their field or the manager's standing within their workplace. We therefore assume that reputations are updated based on actions taken in both local and global games. 

After a round of games, the action of each donor is assessed according to an \textit{assessment rule}, and the donor is assigned a new reputation based on that assessment. An assessment rule (also called a \textit{social norm}) specifies how a donor is judged based on their action toward a recipient \cite{brandt_logic_2004}. We consider the simplest class of rules called \textit{first-order norms}, which assigns a donor a reputation based only on their action. A first-order norm is denoted by $\av = \left( a_C, a_D \right)$, where $a_{A}$ is the probability that a donor who takes action $A\in\{C,D\}$ is assessed as good. Our analysis focuses on the simple rule called \textit{image scoring} ($\av = \left(1,0\right)$), which judges a donor who cooperates as good and who defects as bad, regardless of the recipient's reputation \cite{sigmund_calculus_2010}.

\subsubsection*{Errors}
We allow for errors in reputation assessment and, independently, errors in strategy execution \cite{leimar_evolution_2001,panchanathan_tale_2003,ohtsuki_how_2004}. With probability $\assess \in (0,1/2)$ (\textit{assessment error rate}), a good reputation is accidentally assigned as bad, or vice versa. With probability $\exec \in (0,1/2)$ (\textit{execution error rate}), a player who intends to cooperate accidentally defects; however, as is standard in models of indirect reciprocity, a player intending to defect can never accidentally cooperate. 

The introduction of these errors effectively rescales strategies and  assessment rules.
For convenience, we denote \textit{effective strategies} that account for these execution errors by \smash{$\svDRt = \left(1-\exec\right)\svDR  \coloneqq (\sDRt{C}{G}, \sDRt{C}{B}, \sDRt{D}{G}, \sDRt{D}{B})$} and \smash{$\svIRt = \left(1-\exec\right)\svIR \coloneqq (\sIRt{G}, \sIRt{B})$}. Similarly, we denote \textit{effective assessment rules} that account for assessment errors by $\avt = \left(1-\exec\right)\av + \exec \left(\mathbf{1}-\av\right) \coloneqq (\at{C}{}, \at{D}{})$, where $\mathbf{1}$ is a vector of ones.

\subsubsection*{Analysis in a monomorphic population}
We first consider a monomorphic population in which all players adopt the same strategy pair $(\svDR,\svIR)$. In a given round, each player can be in one of four statuses: $CG$, $CB$, $DG$, or $DB$. The first letter denotes a player's most recent local action ($C$ or $D$); the second denotes a player's current public reputation ($G$ or $B$). 

Our analysis adopts a mean-field approach. We assume that the statuses of any two players in the population are independent, so that each player has status $AR \in \{CG, CB, DG, DB\}$ with probability equal to its frequency in the population as a whole, which we denote $h_{AR}$ with $\sum_{AR} h_{AR} = 1$.
We analyze the dynamics of status distribution $\vh = (h_{CG}, h_{CB}$, $h_{DG}, h_{DB})$ by a system of ordinary differential equations, as we describe below. In the \SI, we show that this system can also be derived as the continuous-time limit of a Markov chain describing the individual-level game play and reputation updates (\SI\ \cref{sec:markov,sec:markovtoode}). We also perform Monte Carlo simulations written in Julia \cite{bezanson_julia_2017} to verify that our mean-field approach provides a good approximation of finite populations (\textit{\nameref{sec:matmethods}}; \cref{fig:SI-comparison}).

The status of a player may change after each game they play.
After a local game, the new status of a donor depends on both their realized action in that game (i.e., most recent local action) and the assessment of that action (i.e., public reputation). By contrast, after a global game, only the donor's public reputation is updated, and their most recent local action remains unchanged.
Weighting these updates by the probability of local versus global games yields a system of ODEs describing the dynamics of $h_{AR}$:
\begin{equation}
    \dot h_{AR} = - h_{AR} 
    + \lambda \underbrace{
        \sum_{A'R'} h_{A'R'} \, P^{\DR}_{AR \mid A'R'}
    }_{\textrm{changes due to local games}}
    + \left(1-\lambda\right)\, \underbrace{
        h_{A\bullet} \sum_{R'} h_{\bullet R'} \, P^{\IR}_{R \mid R'} 
    }_{\textrm{changes due to global games}}
    \;.
    \label{eq:hsystem}
\end{equation}
Here, $\smash{P_{AR\mid A'R'}^{\DR}}$ denotes the probability that a donor is assigned status $AR$ after a local game with a recipient with status $A'R'$, and $\smash{P_{R\mid R'}^{\IR}}$ denotes the probability a donor is assigned reputation $R$ after a global game against a recipient with reputation $R'$ (see definitions in \cref{eq:pDRgeneral,eq:pIRgeneral} in \textit{\nameref{sec:matmethods}}); these are governed by the player's strategy as well as the scoring assessment rule.
The quantity $h_{A\bullet} \coloneqq h_{AG} + h_{AB}$ denotes the probability that the focal player's most recent local action was $A$, while $h_{\bullet R} \coloneqq h_{CR} + h_{DR}$ denotes the probability that the co-player's reputation is $R$.

\newpage
Analysis of long-term payoffs and rate of cooperation reduces to solving for the equilibrium distribution $\vh^* = (h_{CG}^*,$ $h_{CB}^*, h_{DG}^*, h_{DB}^*)$ from \cref{eq:hsystem}. The long-term rate of cooperation in a monomorphic population is then
\begin{equation}
    \label{eq:self_coop_monomorphic}
    \gamma = \sum_{AR} h^{\ast}_{AR} \left[ \lambda\, \sDRt{A}{R} + \left( 1-\lambda \right) \sIRt{R} \right] \;,
\end{equation}
and the corresponding average payoff per round is $\pi = (b-c) \gamma$.

\subsubsection*{Invasion analysis}

We also analyze whether a resident population with strategy pair $(\svDR ,\svIR)$ can be invaded by a mutant with a different strategy pair $({\svDR}' ,{\svIR}' )$, by considering a rare mutant introduced in a neighborhood within an otherwise monomorphic resident population.

The fate of a mutant invader hinges on the fact that a neighborhood is finite in size but vanishingly small relative to the total population. The fitness of the mutant's neighbors will assuredly be influenced by the mutant, because the neighborhood is finite; but the status and fitness of non-neighbors remain unaffected by a mutant, because there is a vanishingly small chance of interaction per capita. We therefore distinguish between two types of residents: the resident-type players inside the mutant's neighborhood are called \textit{resident neighbors}, and those outside are called \textit{resident non-neighbors} (\cref{fig:schematic}C).

Because the dynamics of resident non-neighbors decouple from those of the mutant and resident neighbors, we can solve separately for the equilibrium status distributions for the three player classes---the mutant, resident neighbors, and resident non-neighbors---and compute their long-term average payoffs (\textit{\nameref{sec:matmethods}}). The long-term average payoffs of the resident neighbors, $\pi_{\R}$, and the mutant, $\pi_{\M}$, are given by
\begin{equation}
    \begin{split}
        \pi_{\R} 
        & = \lambda 
        \left(
        \dfrac{1}{n-1} \left(b\, \gamma_{\M\,\to\,\R}^{\DR}
        -c \, \gamma_{\R\,\to\,\M}^{\DR} \right)
        + \dfrac{n-2}{n-1}\, \left(b-c\right) \, \gamma_{\R\,\to\,\R}^{\DR}
        \right) 
        \\
        & \qquad\qquad
        + \left(1-\lambda\right) \left(b\,\gamma_{\RNN\,\to\,\R}^{\IR} -c\,\gamma_{\R\,\to\,\RNN}^{\IR} \right)  \;,
        \\
        \pi_{\M} & = \lambda \left(b\, \gamma_{\R\,\to\,\M}^{\DR} -c\, \gamma_{\M\,\to\,\R}^{\DR} \right)  + \left(1-\lambda\right) \left(b\, \gamma_{\RNN\,\to\,\M}^{\IR} - c\,\gamma_{\M\,\to\,\RNN}^{\IR} \right)  \;,
    \end{split}
    \label{eq:fitnesseq}
\end{equation}
where $\gamma_{X \to Y}^{\DR}$ and $\gamma_{X \to Y}^{\IR}$ denote the long-term average cooperation rate of class $X$ with class $Y$ in local and global games, respectively (\textit{\nameref{sec:matmethods}}). The long-term average of resident non-neighbors, $\pi_{\RNN}$, is identical to that of a monomorphic population of the resident type (i.e., $\pi_{\RNN}=(b-c)\gamma$, where $\gamma$ is as defined in \cref{eq:self_coop_monomorphic}), because resident non-neighbors are unaffected by the rare mutant.

We say that the resident type resists invasion by a mutant if and only if $\pi_{\R} > \pi_{\M}$. This condition compares the mutant with the resident players whose payoffs are directly affected by the mutant; if resident neighbors earn more than the mutant, the mutant cannot grow in number within its neighborhood. 

\section{Results}

\subsection*{Combining direct and indirect reciprocity solves the scoring dilemma}

\begin{figure}[b!]
    \centering
    \includegraphics[width=0.95\linewidth,trim={0.9in 0 0.9in 0in},clip]{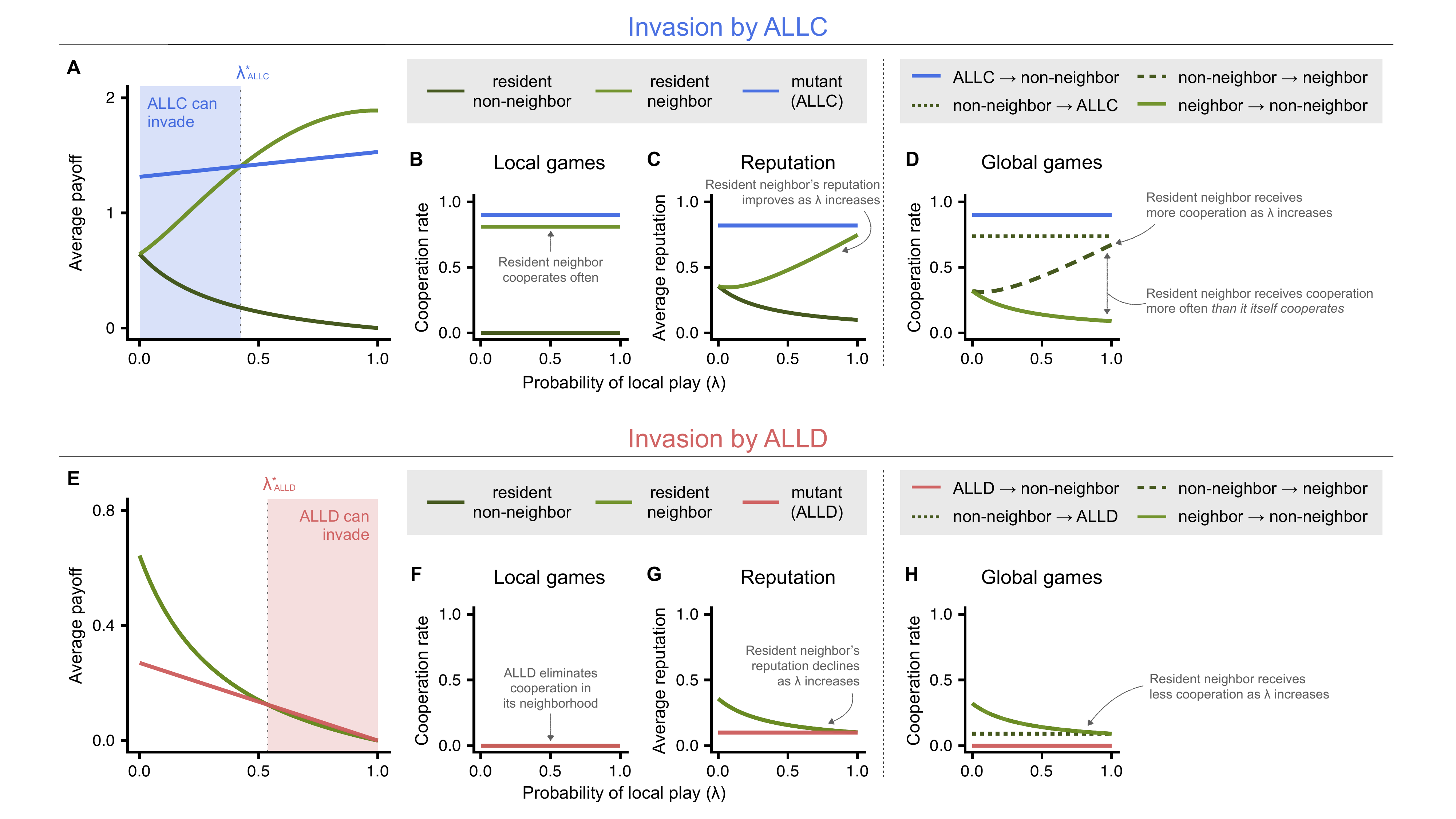}
    \caption{
        \textbf{Invasibility of tit-for-tat discriminators under direct and indirect reciprocity.}
        A rare mutant (\textbf{A--D}: ALLC; \textbf{E--H}: ALLD) is introduced to a resident population of TFT--DISC.
        The resident TFT--DISC players (both non-neighbors and neighbors) act as tit-for-tat (TFT) players in local games and as discriminators (DISC) in global games. 
        Colors indicate player classes (dark green for resident non-neighbors, light green for resident neighbors, and red or blue for the mutant); in \textbf{D} and \textbf{H}, line types distinguish between cooperation with different classes. 
        Panels show average payoff (\textbf{A}, \textbf{E}), cooperation rate in local games (\textbf{B}, \textbf{F}), average reputation (\textbf{C}, \textbf{G}), and cooperation rate in global games (\textbf{D}, \textbf{H}), each as a function of the probability of local play ($\lambda$).
        In particular, when all games are global ($\lambda = 0$), interactions reduce to pure indirect reciprocity, and the scoring dilemma is observed: discriminators can resist ALLD (\textbf{E}) but are vulnerable to ALLC (\textbf{A}) under the scoring assessment rule. 
        (Curves overlap in \textbf{E--H}:
        In \textbf{E}, non-neighbors and neighbors have identical payoffs. 
        In \textbf{F}, all three classes have a cooperation rate of zero.
        In \textbf{G}, non-neighbors and neighbors have identical average reputations.
        In \textbf{H}, cooperation rates between neighbors and non-neighbors are symmetric (solid light green and dashed dark green).)
        Neighborhood size is fixed at $n = 2$. 
        Other parameters: $b = 3$,  $c = 1$, $\assess=\exec=0.1$.
        }
    \label{fig:fig2}
\end{figure}

Indirect reciprocity alone is often insufficient to guarantee stable cooperative behavior. In the most natural setting, where a cooperative player earns a good reputation and a defecting player a bad reputation (i.e., scoring norm), a population of discriminators will be invaded by ALLC and, subsequently, by ALLD. The eventual collapse of cooperation and collective fitness in this simple setting is known as the ``scoring dilemma" \cite{okada_two_2020}, and it has stimulated a body of research into more elaborate methods of assessing reputations in the hopes of ensuring stable cooperation \cite{leimar_evolution_2001,ohtsuki_how_2004,ohtsuki_leading_2006,berger_stability_2016,panchanathan_tale_2003,sasaki_evolution_2017,pacheco_sternjudging_2006,santos_social_2018}.

The first question we study is whether the simple combination of local play with neighbors (whose actions are observable) and global play with non-neighbors (known only by reputation) is sufficient to resolve the scoring dilemma and stabilize cooperation, even without additional complexity in the assessment of reputations. 

The natural analog of DISC in models of direct reciprocity is tit-for-tat (TFT), which discriminates by the most recent action rather than reputation. We therefore consider the strategy pair called tit-for-tat discriminator (TFT--DISC), which uses TFT in local games and DISC in global games. We ask whether TFT--DISC can resolve the scoring dilemma---that is, whether a resident population of TFT--DISC can resist invasion by a mutant ALLC and also by a mutant ALLD. 

In general, it is easier to resist invasion by an ALLD mutant when games are more often global (because DISC is known to resist ALLD in models of indirect reciprocity \cite{sigmund_calculus_2010}), whereas ALLC is easier to resist when games are more often local (because TFT outperforms ALLC in noisy iterated games of infinite length). The key question, then, is whether there is any overlapping regime of local and global games---that is, an intermediate range of $\lambda$---in which TFT--DISC can resist invasion by both ALLD and ALLC.

A resident population of TFT--DISC players can resist local invasion by an ALLC mutant when local games are common (\cref{fig:fig2}A). In local games, both the mutant and the resident neighbors cooperate often (\cref{fig:fig2}B): ALLC cooperates unconditionally with its neighbors, who cooperate in return. In doing so, the resident neighbors earn good reputations. Accordingly, increasing the probability of local play ($\lambda$) improves the average reputation of the resident neighbors (\cref{fig:fig2}C). Once $\lambda$ is sufficiently large and the resident neighbors' reputations sufficiently good, they have an advantage over the mutant in global games: the resident neighbors receive cooperation often from  (\cref{fig:fig2}D, dashed dark green curve), but do not themselves cooperate often with (\cref{fig:fig2}D, light green curve), the resident non-neighbors, who have poor reputations (\cref{fig:fig2}C). As a result, resident neighbors outearn the mutant when local play is sufficiently common (\cref{fig:fig2}A).

By contrast, TFT--DISC players can resist invasion by an ALLD mutant when local games are rare (\cref{fig:fig2}E). In local games, ALLD defects unconditionally against its resident neighbors, who defect in return. The resident neighbors thus earn bad reputations in local games. Accordingly, \textit{decreasing} the probability of local play ($\lambda$) improves the average reputation of the resident neighbors (\cref{fig:fig2}G), which in turn increases their likelihood of receiving cooperation in global games (\cref{fig:fig2}H). The resident neighbors eventually outearn the mutant when local play is sufficiently rare (\cref{fig:fig2}E).

It is useful to consider how the finite neighborhood size supports the robustness of TFT--DISC against mutants: it plays a critical role in robustness against ALLC, but no role in robustness against ALLD. 
In a finite neighborhood, the non-negligible frequency of local play against ALLC allows the TFT--DISC neighbors to sustain cooperation in local games (whereas, in an infinite neighborhood, TFT--DISC could not sustain cooperation due to asymmetric execution errors) (\cref{fig:fig2}B). However, the finite neighborhood size does not help TFT--DISC when the mutant is ALLD. Once ALLD initiates a cascade of defection in its neighborhood, cooperation cannot be restored, since players cannot cooperate by accident.\footnote{The dynamics of local games change substantially when execution errors are symmetric: when both cooperation and defection can occur by mistake, TFT--DISC is able to sustain at least some cooperation in local games against both ALLC and ALLD. We examine symmetric execution errors in \SI\ \cref{sec:symmetric}.} After an initial transient, then, all local games in ALLD's neighborhood converge to mutual defection, regardless of neighborhood size (\cref{fig:fig2}F; \SI\ \cref{sec:neffect}, \cref{fig:SI-neffect1}F). 

The observation that an ALLD mutant eliminates cooperation in its neighborhood provides a key insight: in local games, an ALLD mutant earns just as much as its neighbors (because ALLD and TFT both have zero payoffs), whereas an ALLC mutant earns less than its neighbors (because ALLC cooperates more often than TFT). An ALLC mutant must compensate for this local disadvantage by relying on global games: that is, for ALLC to invade, the frequency of game types must be more strongly weighted toward global games than an ALLD mutant requires the frequency of game types to be weighted toward local games. Therefore, there must be an intermediate range of $\lambda$ for which neither ALLC nor ALLD can invade TFT--DISC.

\begin{figure}[h!]
    \centering
    \includegraphics[width=0.95\linewidth,trim={3.5in 2in 3.5in 2in},clip]{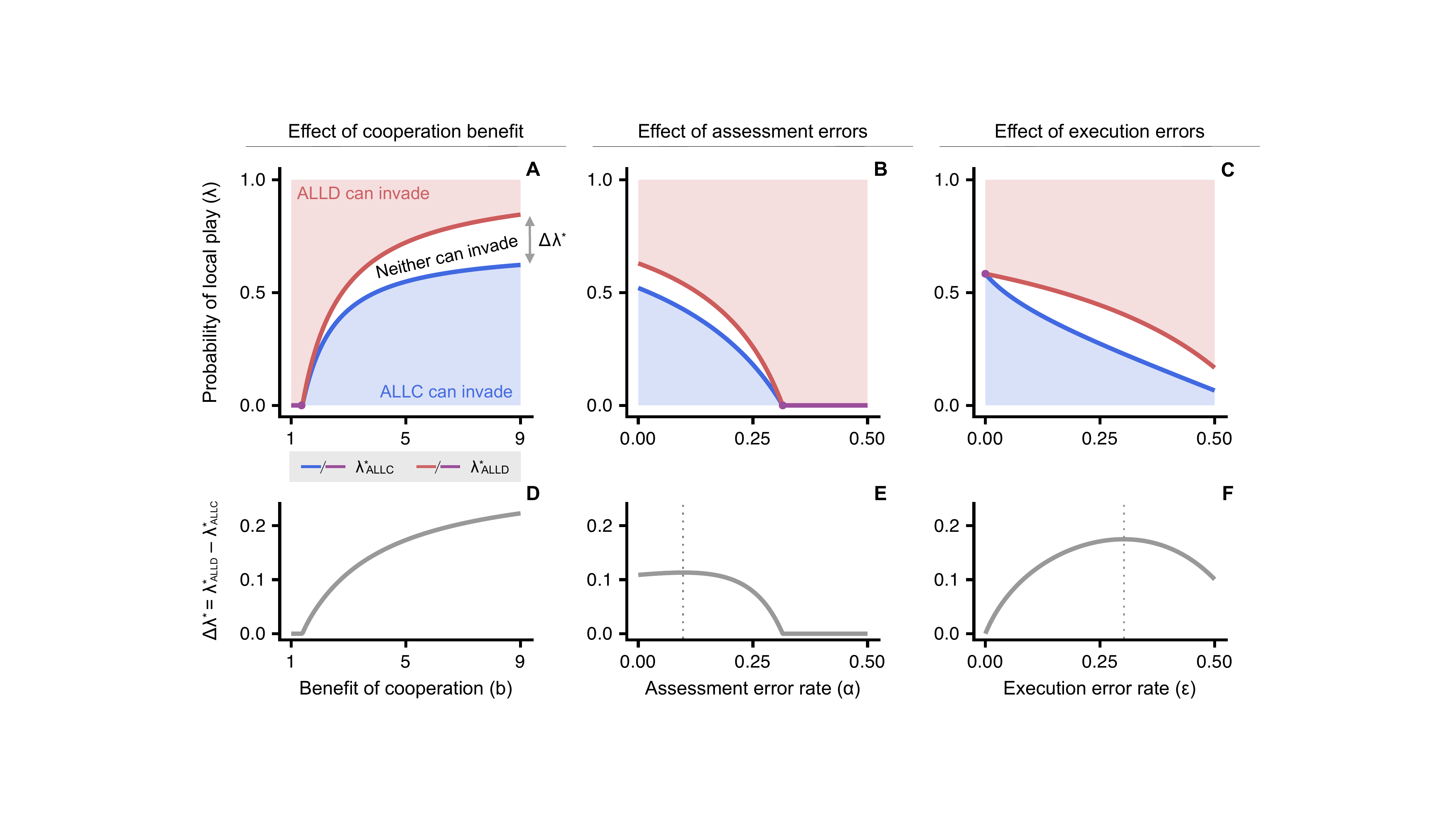}
    \caption{
        \textbf{Intermediate probabilities of local play solve the scoring dilemma.}
        Resident players act as tit-for-tat (TFT) players in local games and discriminators (DISC) in global games. 
        \textbf{A--C}: Critical probability of local play against an ALLC mutant ($\lambda_{\ALLC}^*$, blue and purple curves) and against an ALLD mutant ($\lambda_{\ALLD}^*$, red and purple curves), as a function of the benefit of cooperation $b$ (\textbf{A}), assessment error rate $\assess$ (\textbf{B}), and execution error rate $\exec$ (\textbf{C}). Purple curves indicate regions where $\lambda_{\ALLC}^*=\lambda_{\ALLD}^*$.
        ALLC can invade TFT--DISC locally when $\lambda<\lambda_{\ALLC}^*$ (blue regions); ALLD can invade when $\lambda>\lambda_{\ALLD}^*$ (red regions); and neither can invade when $\lambda_{\ALLC}^*<\lambda<\lambda_{\ALLD}^*$ (white regions). 
        \textbf{D--F}: Range of probabilities of local play that solve the scoring dilemma, defined as $\Delta\lambda^* \coloneqq \lambda_{\ALLD}^* - \lambda_{\ALLC}^*$, as a function of the same parameters as in \textbf{A--C}. 
        In \textbf{E} and \textbf{F}, dashed vertical lines indicate the values of $\assess$ and $\exec$ that maximize $\Delta \lambda^*$ in the respective panels.
        Other parameters: $n = 2$, $c = 1$, $b = 3$ (except in \textbf{A} and \textbf{D}), $\assess=0.1$ (except in \textbf{B} and \textbf{E}), $\exec=0.1$ (except in \textbf{C} and \textbf{F}). 
     }
    \label{fig:fig3}
\end{figure}

The intuition outlined above can be made analytic. When $n$ is sufficiently small, for any set of payoff parameters $b$ and $c$ and error rates $\assess$ and $\exec$, we can analytically determine the probabilities of local gameplay, $\lambda_{\ALLC}^*$ and $\lambda_{\ALLD}^*$, such that TFT--DISC can resist ALLC if $\lambda > \lambda_{\ALLC}^*$ and resist ALLD if $\lambda < \lambda_{\ALLD}^*$ (\textit{\nameref{sec:matmethods}} and \SI\ \cref{sec:analytical}).  
Consistent with our intuition, we find a wide range of parameter values for which $\lambda_{\ALLD}^* > \lambda_{\ALLC}^*$, so that an intermediate probability of local play ($\lambda_{\ALLC}^* < \lambda < \lambda_{\ALLD}^*$) makes TFT--DISC robust against both ALLC and ALLD (\cref{fig:fig3}A--C). Hence, the combination of direct and indirect reciprocity, TFT--DISC, solves the scoring dilemma.

How does the robustness of TFT--DISC depend on game parameters? We quantify robustness using  $\Delta \lambda^* \coloneqq \lambda_{\ALLD}^* - \lambda_{\ALLC}^*$, the range of local-game probabilities that solve the scoring dilemma. Increasing the benefit of cooperation naturally makes it easier for ALLC to invade ($\lambda_{\ALLC}^*$ increases) and harder for ALLD to invade ($\lambda_{\ALLD}^*$ increases)---but the second effect is stronger (\cref{fig:fig3}A)---because ALLC benefits from a larger $b$ in both local and global games, whereas ALLD benefits only in global games---so that, in total, $\Delta \lambda^*$ increases with $b$ (\cref{fig:fig3}D). 
By contrast, the range of $\lambda$ that solves the scoring dilemma can depend on error rates in a complex, non-monotonic manner (\cref{fig:fig3}E and F).

There are some parameter regimes in which TFT--DISC fails to solve the scoring dilemma for any probability of local play ($\lambda$). This occurs when the benefit of cooperation $b$ is too small (\cref{fig:fig3}A), or when either the assessment error rate $\assess$ or the execution error rate $\exec$ is too large (\cref{fig:fig3}B and C). However, these regimes are rare: for most parameter values, some probability of local gameplay solves the scoring dilemma.

In \textit{\nameref{sec:matmethods}}, we derive an analytical condition for TFT--DISC to solve the scoring dilemma for $n=2$. We show that there is a region of parameter space where $\Delta\lambda^* > 0$ if and only if
\begin{equation}
    \left(\dfrac{b}{c}\right) > \left(\dfrac{b}{c}\right)^* = \dfrac{1}{(1-2 \assess) (1-\exec)} \;,
    \label{eq:bccondition}
\end{equation}
consistent with the numerical example in \cref{fig:fig3}A.
We can rewrite this condition in terms of the error rates (\cref{fig:fig3}B and C) as
\begin{equation}
    \assess < \assess^* = \dfrac{1}{2} \left( 1 - \dfrac{1}{\left(b/c\right) \left(1-\exec\right)} \right)
    \qquad \textrm{and} \qquad
    \exec < \exec^* = 1-\dfrac{1}{\left(b/c\right) \left(1-2 \assess\right)} \;.
    \label{eq:errorcondition}
\end{equation}
The scoring dilemma becomes more difficult to solve as $n$ increases, and TFT--DISC can fail to solve the dilemma for sufficiently large $n$ (\cref{fig:SI-neffect1,fig:SI-neffect2}; see also \SI\ \cref{sec:neffect}).
Hence the robust cooperation provided by a combination of direct and indirect reciprocity requires a relatively small number of ``local" friends, whose actions you remember, compared to the vast pool of global interactions governed by public reputations.

\subsection*{Mixed strategies for direct and indirect reciprocity}

So far we have shown how combining indirect reciprocity in global games with direct reciprocity in local games can resolve the scoring dilemma. We assumed that, in local games, players use an $n$-player version of tit-for-tat, conditioning their action towards a neighbor only on that neighbor's most recent action within the neighborhood---and that, in global games, they discriminate by reputation. Behavior toward neighbors therefore depended on observed local actions alone, and behavior toward non-neighbors on public reputations alone.

We now return to the question that motivated this study at the outset: how should we treat those we know well in light of how they are viewed publicly? In practice, behavior toward a neighbor may depend not only on that neighbor's past local actions but also on their public reputation. For example, we may hesitate to cooperate with a neighbor who has a poor public reputation, even if that neighbor recently cooperated in local games; or we may forgive a neighbor's past local defection if that neighbor enjoys good public standing. Such situations are most consequential when the two sources of information conflict---when what we have personally observed about a neighbor disagrees with how the neighbor is regarded at large. How should we integrate direct experience of others with their public reputations?

To address this question, we broaden the space of strategies used in local games. Rather than conditioning on a neighbor's most recent local action alone, as tit-for-tat does, a player may now also condition on that neighbor's public reputation. That is, a strategy for local gameplay may mix direct and indirect reciprocity. We refer to players who combine such a hybrid strategy in local games with the discriminator strategy (DISC) in global games as \textit{cross-scale discriminators}, and denote them by $pq$--DISC.

In local games, a $pq$--DISC player uses a strategy of the form $\mathbf{s}_{(p,q)} = \left(s_{CG}, s_{CB}, s_{DG}, s_{DB}\right) = (1, p, q, 0)$. Here, $p$ denotes the probability of cooperating with a neighbor who cooperated in the most recent local game but has a bad reputation (CB status), so $p$ measures \textit{forgiveness of bad public reputation}. Likewise, $q$ denotes the probability of cooperating with a neighbor who defected in the most recent local game but has a good reputation (DG status), so $q$ measures \textit{forgiveness of local defection}. This strategic space encompasses familiar strategies for local play: $(p,q) = (1,0)$ corresponds to classic TFT, and $(p,q)=(0,1)$ is classical DISC from indirect reciprocity. The space also allows for novel strategies that integrate local and global information. Players with  $(p,q) = (1,1)$ cooperate with any neighbor who either cooperated in the most recent local game or has a good public reputation. Players with $(p,q) = (0,0)$ cooperate only with neighbors who both cooperated in the most recent local game and have good reputations. Strategy $\mathbf{s}_{(1,1)}$ is the most forgiving and $\mathbf{s}_{(0,0)}$ the most strict in this space; more generally, we use $p+q$ as a measure of overall forgiveness, with $p+q>1$ strategies called forgiving and $p+q<1$ called strict. 

\begin{figure}[h!]
    \centering
    \includegraphics[width=0.95\linewidth,trim={4.1in 3.75in 4.1in 3.75in},clip]{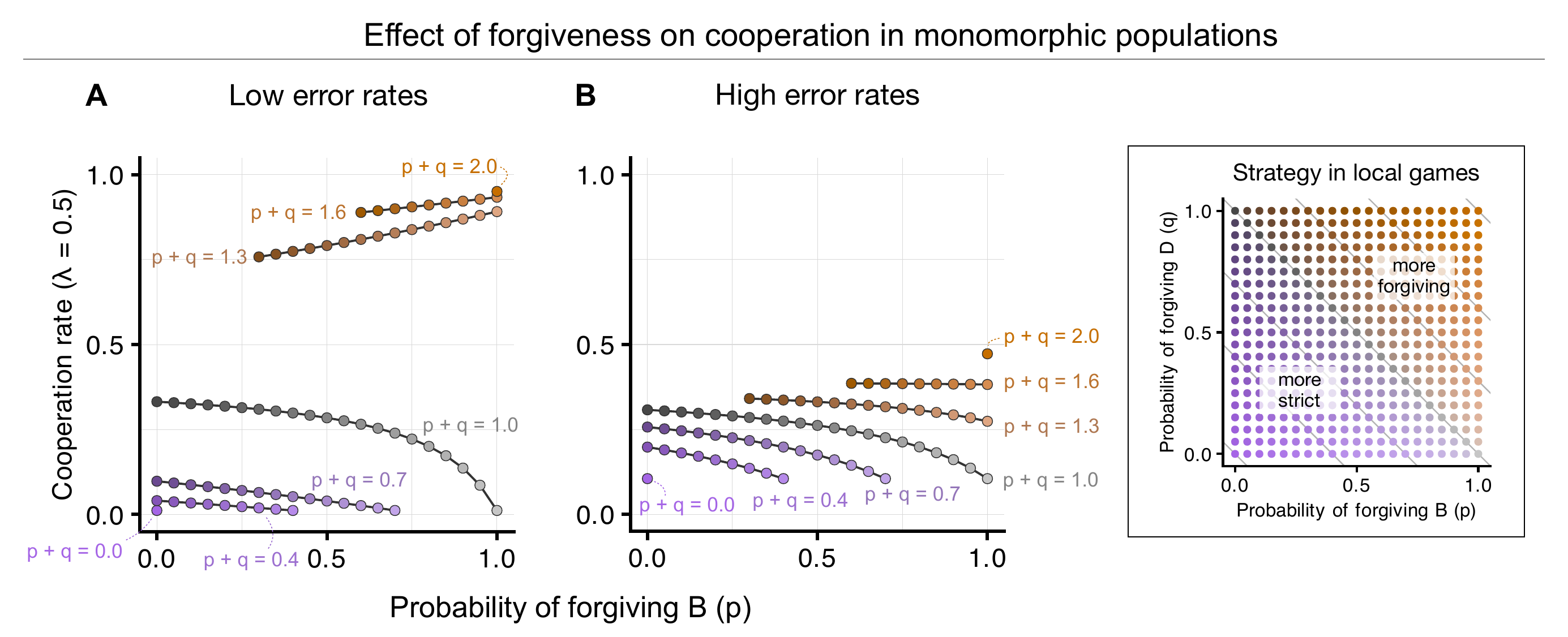}
    \caption{
        \textbf{Forgiveness of neighbors promotes cooperation.}
        Average rates of cooperation in monomorphic populations of cross-scale discriminators ($pq$--DISC) as a function of the probability of forgiving a bad (global) reputation ($p$), for low ($\assess = \exec = 0.0125$; \textbf{A}) and high ($\assess = \exec = 0.2$; \textbf{B}) error rates.
        Colors indicate overall degrees of forgiveness in local games (see 2D legend): shades of orange denote strategies that are overall forgiving ($p+q > 1$), whereas shades of purple denote those that are overall strict ($p+q < 1$).
        Other parameters: $b = 3$, $c = 1$, $n = 2$, $\lambda=0.5$.
    }
    \label{fig:fig4}
\end{figure}

Using this formulation, we first study how forgiveness of neighbors affects fitness. We compute average rates of cooperation in populations of $pq$--DISC for various values of $p$ and $q$ (\cref{fig:fig4}); in a monomorphic population, this rate is proportional to mean fitness.  In the main text, we focus on $n=2$, where each player has one neighbor, and $\lambda = 0.5$, where local and global games occur with equal probability; we report results for $\lambda = 0.1$ and $0.9$ (\cref{fig:SI-dots-grid-l01,fig:SI-dots-grid-l09}) and for $n=3$ (\cref{fig:SI-fig5vn3}) in the \SI.

We find that forgiveness of neighbors tends to promote cooperation (\cref{fig:fig4}). More precisely, cooperation increases with $q$ for fixed $p$; and cooperation increases with $p$ for fixed $q$. Strategies $(p,q) = (1,1)$ and $(0,0)$ produce the maximum and minimum rates of cooperation, respectively. However, for a fixed amount of overall forgiveness $p+q$, which type of forgiveness is better for cooperation---forgiveness of bad global reputation or of local defection---depends on the value of $p+q$ and, to a lesser extent, the error rates. When overall forgiveness is low ($p+q$ small) or errors are common ($\assess$, $\exec$ large), cooperation is maximized when the population favors forgiving  bad global reputations (maximum $q$, minimum $p$; see $p+q\leq 1.0$ in \cref{fig:fig4}A and $p+q\geq 1.6$ in \cref{fig:fig4}B). Only when overall forgiveness is high ($p+q$ large) and errors are rare ($\assess$, $\exec$ small) is cooperation maximized by favoring forgiveness of local defection (maximum $p$, minimum $q$; see $p+q\geq 1.3$ in \cref{fig:fig4}A). These qualitative results hold across a range of error rates (\cref{fig:SI-dots-grid-l05}) and also when interactions are skewed toward local or global games ($\lambda = 0.1$, \cref{fig:SI-dots-grid-l01}; $\lambda = 0.9$, \cref{fig:SI-dots-grid-l09}): in most cases, maximum cooperation is achieved by maximizing $q$. Altogether, forgiving a friend's past local defection tends to yield higher fitness than forgiving their bad public reputation, especially when errors are common or overall forgiveness is limited. In short, cooperation often depends more on overlooking a friend's poor private behavior toward you and your group than on overlooking their poor public reputation.

\begin{figure}[h!]
    \centering
    \includegraphics[width=0.87\linewidth,trim={3.5in 0in 3.5in 0in},clip]{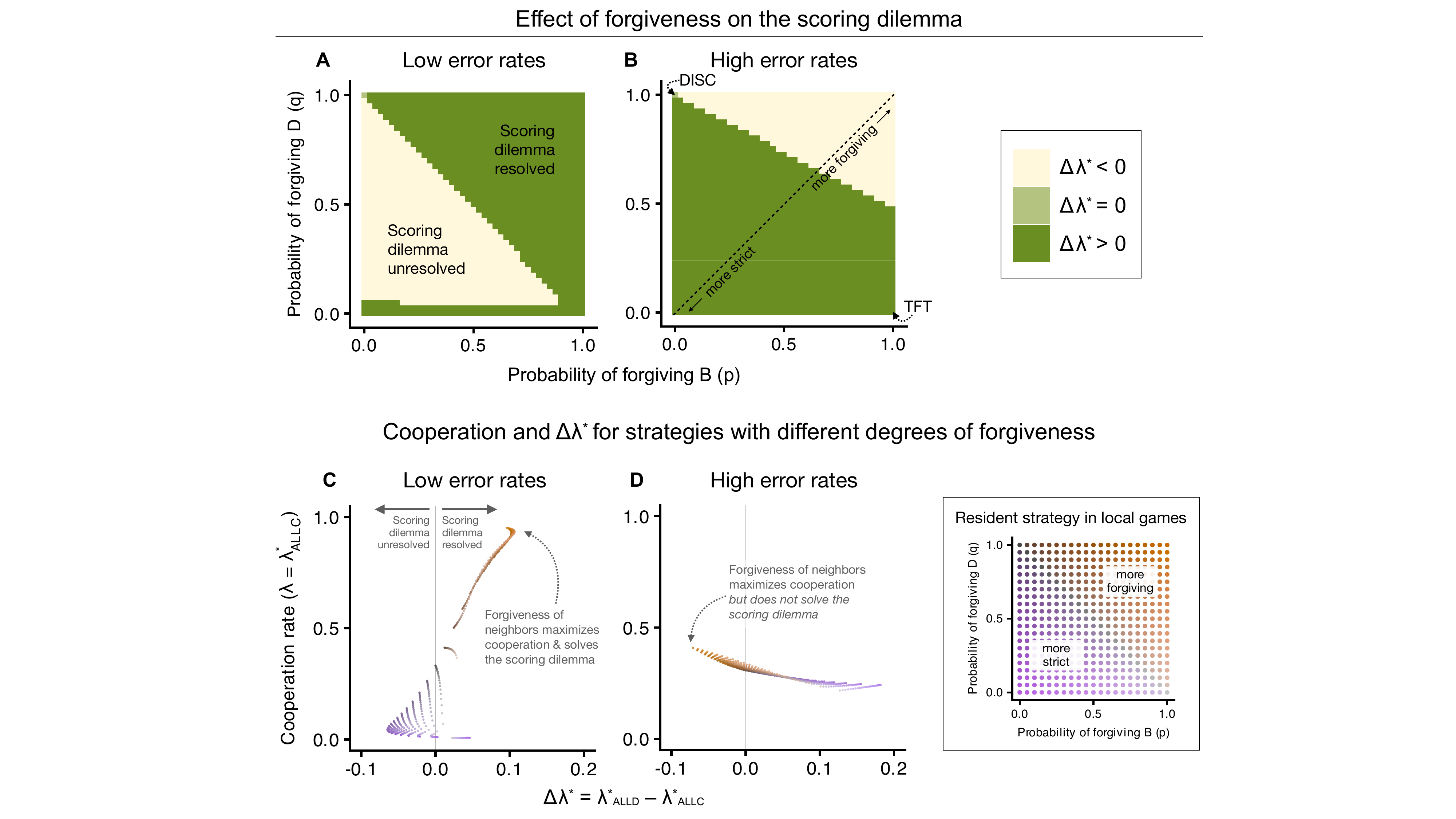}
    \caption{
        \textbf{Effect of forgiveness on the scoring dilemma and fitness.}
        \textbf{A, B}: 
        We classify the two-dimensional space of cross-scale discriminator strategies by whether the scoring dilemma is resolved. Each coordinate $(p,q)$ corresponds to a resident $pq$--DISC strategy. The dilemma is resolved (i.e., there is a range of $\lambda$ for which neither $\ALLC$ nor $\ALLD$ can invade) when $\Delta\lambda^* > 0$ (green); it is unresolved when $\Delta\lambda^* < 0$ (beige) or $\Delta\lambda^* = 0$ (pale green).
        \textbf{C, D}: 
        Cooperation rate and $\Delta\lambda^*$ for strategies with different degrees of forgiveness. Each point corresponds to a resident $pq$--DISC strategy. The coordinates of each point give the value of $\Delta\lambda^*$ (horizontal coordinate) and the cooperation rate at $\lambda = \lambda_{\ALLC}^*$ (vertical coordinate) for the corresponding resident strategy. Colors indicate degrees of forgiveness in local games (see 2D legend): shades of orange denote strategies that are overall forgiving ($p+q > 1$), whereas shades of purple denote those that are overall strict ($p+q < 1$). Vertical lines at $\Delta\lambda^* = 0$ separate regions where the scoring dilemma is resolved ($\Delta\lambda^* > 0$) from those where it is unresolved ($\Delta\lambda^* \leq 0$).
        Error rates are $\assess = \exec = 0.0125$ in \textbf{A} and \textbf{C} (low error rates) and $\assess = \exec = 0.2$ in \textbf{B} and \textbf{D} (high error rates).
        Other parameters: $c = 1$, $b = 3$, $n = 2$. 
        Results across a broader range of error rates are reported in \cref{fig:SI-heatmap-grid,fig:SI-scatter-grid}.
     }
    \label{fig:fig5}
\end{figure}

High levels of cooperation, however, do not guarantee stability against invasion. We therefore turn next to the effects of forgiveness on the scoring dilemma: how does willingness to forgive either a bad global reputation ($p$) or a past local defection ($q$) affect robustness against unconditional cooperators and defectors? Whether forgiving neighbors helps resolve the scoring dilemma depends on both the error rates (\cref{fig:fig5}A vs.\ B) and the neighborhood size (\cref{fig:fig5} vs.\ \cref{fig:SI-fig5vn3}). When assessment and execution errors are rare ($\assess,\exec$ small) or when $n=3$, the strategies that fail to resolve the scoring dilemma are strict (beige regions in \cref{fig:fig5}A, \cref{fig:SI-fig5vn3}A, and \cref{fig:SI-fig5vn3}B). By contrast, and somewhat counterintuitively, when $n=2$ and errors are common ($\assess, \exec$ large) it is the highly forgiving strategies that fail to solve the scoring dilemma (beige region in \cref{fig:fig5}B). Hence, leniency toward neighbors can undermine stability against invasion by unconditional players, but only when $n=2$ and mistakes are frequent. 

These patterns reflect how error rates and neighborhood size affect the fitness of an ALLD mutant and its resident neighbors (\cref{fig:SI-pq11explanatory}). Under low error rates, an ALLD mutant usually has a bad reputation, because assessments are typically accurate. It therefore rarely benefits from the forgiveness of resident neighbors---who cooperate locally only with co-players carrying at least one positive bit of information (CG, CB, or DG)---and so has limited ability to invade (\cref{fig:SI-pq11explanatory}A, C, E, and G). Under high error rates and $n=2$, however, ALLD occasionally earns a good reputation by accident and thus benefits from neighbors' forgiveness, making invasion much easier (\cref{fig:SI-pq11explanatory}B and D). For $n=3$, however, this advantage is offset by the two forgiving resident neighbors, who, unlike TFT--DISC, frequently cooperate with each other. Although ALLD still benefits from their occasional forgiveness, their sustained mutual cooperation raises their payoffs, making it harder for ALLD to invade than for $n=2$ (\cref{fig:SI-pq11explanatory}F and H).

Finally, we examine the interaction between cooperation and robustness: can the strategies that mix local and global information simultaneously maximize fitness and resolve the scoring dilemma? For low error rates, the most forgiving strategy both achieves maximal cooperation and solves the dilemma, for both $n=2$ (\cref{fig:fig5}C) and $n=3$ (\cref{fig:SI-fig5vn3}C). For high error rates, however, the outcome hinges on neighborhood size: for $n=2$, the most forgiving strategy maximizes cooperation but fails to resolve the dilemma (\cref{fig:fig5}D), whereas for $n=3$ it does both---even where TFT--DISC does neither (\cref{fig:SI-fig5vn3}B and D). Thus, maximizing cooperation can conflict with resisting invasion---but only when mistakes are common and neighborhoods are the smallest ($n=2$).

\section{Discussion}

People interact with both friends and strangers throughout their daily lives. Interactions with friends may be shaped by direct observation of their behavior toward one another, whereas strangers may be known only by reputation. Hence direct reciprocity, based on personal experience, and indirect reciprocity, based on reputations, operate simultaneously but in different contexts. Prior work has studied direct and indirect reciprocity in isolation or, more recently, together but in well-mixed populations \cite{schmid_evolution_2021,hubner_stable_2025a,pal_coevolution_2024,yamamoto_tolerant_2025}, effectively assuming that the two mechanisms apply equally to friends and strangers alike. Here we depart from both approaches. Rather than assuming a well-mixed population, we develop a mathematical framework that captures direct reciprocity within small, finite neighborhoods of friends (neighbors) along with indirect reciprocity in a much larger pool of strangers (non-neighbors). This formulation posits that the two types of reciprocity are intercalated in time but separated in space---offering a representation of human social life that is arguably more realistic than prior approaches. 

The combination of direct and indirect reciprocity is not only a natural feature of social life but, as we have shown, strongly beneficial for robust cooperation. Under the image scoring norm, a population of tit-for-tat discriminators (TFT--DISC)---who play tit-for-tat in local games and discriminate by reputation in global games---can resist local invasion by both unconditional cooperators and defectors (ALLC and ALLD), 
thereby solving the scoring dilemma \cite{okada_two_2020}. This robustness requires an intermediate balance of local and global interactions, and the finiteness of the local neighborhood is essential.

Beyond TFT--DISC, we also studied hybrid strategies---cross-scale discriminators---whose behavior toward a neighbor depends on both the neighbor's most recent local action and their public reputation. This allows us to ask how one should treat a neighbor when direct experience and public reputation conflict. We find that forgiveness---overlooking whichever piece of information is unfavorable---tends to promote cooperation, and that forgiving a neighbor's local defection matters more than forgiving a bad reputation. Forgiveness can, however, trade off against robustness: when neighborhoods are the smallest ($n=2$) and errors frequent, the most cooperative strategies become vulnerable to unconditional invaders, but enlarging the neighborhood slightly ($n=3$) restores both maximal cooperation and robustness.

Our results complement recent work on the utility of forgiving strategies under private \cite{berger_stability_2016,schmid_quantitative_2023,michel-mata_evolution_2024,yamamoto_tolerant_2025} and public \cite{park_role_2017,glaubitz_other_2026} reputations. Unlike in prior studies, however, the cross-scale discriminators we study may differentially forgive poor local behavior versus poor global reputation, a distinction typically absent in well-mixed models. The nuanced forgiveness of cross-scale discriminators also differs from the reputation-agnostic forgiveness of generous tit-for-tat ($\svDR=(1,1,Q,Q)$) \cite{sigmund_calculus_2010}. Increasing the generosity ($Q$) in a resident population of GTFT--DISC tends to shrink the parameter region in which the scoring dilemma is resolved (\cref{fig:SI-GTFT}), because GTFT readily extends its generosity to an ALLD mutant. By contrast, cross-scale discriminators can selectively forgive neighbors with at least one positive bit of information (recent local cooperation or good public reputation) while still defecting against neighbors with two negative bits. Selective forgiveness can sometimes allow cooperation and robustness to be achieved simultaneously.

Our framework of local and global reciprocity is notable for its cognitive realism: it asks less of individuals than many other models of reciprocity, in two respects. The first is the demand on memory. Because reputations in our model are public, individuals need not form and store a private opinion of every other player; they recall recent actions only within a small neighborhood and rely on shared reputations for everyone else. This contrasts with recent efforts to unify direct and indirect reciprocity, which assume private assessment \cite{schmid_unified_2021,pal_coevolution_2024,hubner_stable_2025a,yamamoto_tolerant_2025}---requiring each individual to observe, judge, and remember the standing of many or all others in a large population \cite{uchida_effect_2010,uchida_effect_2013,berger_stability_2016,sasaki_evolution_2017,okada_tolerant_2017,okada_solution_2018,fujimoto_evolutionary_2023,kessinger_evolution_2023,michel-mata_evolution_2024,kawakatsu_mechanistic_2024,murase_computational_2024}. Private assessment is not only cognitively demanding but also prone to disagreements that undermine cooperation \cite{uchida_effect_2010,uchida_effect_2013,murase_indirect_2024}. Notably, the large array of proposed remedies to the problem of disagreement---including empathetic perspective taking \cite{radzvilavicius_evolution_2019}, nuanced evaluation \cite{schmid_quantitative_2023}, peer-to-peer gossip \cite{nakamaru_evolution_2004,seki_model_2016,kawakatsu_mechanistic_2024}, institutions that broadcast reputations \cite{kessinger_evolution_2023,kessinger_institutions_2025}, and generous assessment based on a single observation \cite{schmid_evolution_2021} or multiple private observations \cite{berger_stability_2016,michel-mata_evolution_2024}---are all fundamentally mechanisms to synchronize opinions \cite{murase_indirect_2024} that effectively recover the public information our model assumes from the outset.

The second is the simplicity of the social norm. The standard remedy for the scoring dilemma is to invoke higher-order norms, which judge an action using more information---such as the recipient's reputation---to distinguish between justified and unjustified defection \cite{okada_two_2020,okada_review_2020}. Such norms can stabilize cooperation \cite{ohtsuki_how_2004,ohtsuki_leading_2006,pacheco_sternjudging_2006,santos_social_2016,santos_social_2018,fujimoto_evolutionary_2023}, but they are informationally and cognitively demanding, and empirical evidence offers some support for the simpler first-order scoring norm \cite{wedekind_cooperation_2000, milinski_cooperation_2001}. 
Our framework offers a more parsimonious solution: even under image scoring, players who discriminate by reputation can resist invasion provided there is occasional local play among players who directly observe each other's actions. Nonetheless, higher-order norms remain worth studying in our setting, as recent experiments suggest that people use some higher-order information when evaluating others \cite{yamamoto_justified_2020,tateishi_social_2026}. Of particular interest for future work are norms that judge differently in local versus global contexts, for instance holding friends to a higher, or more lenient, moral standard than strangers.

Several assumptions underlying our results merit scrutiny. First, we take execution errors to be asymmetric, so that intended cooperation can fail but defectors never cooperate by accident. This assumption is standard in models of indirect reciprocity, but it has a notable consequence for local games: a single accidental defection triggers a cascade of defection so that TFT cannot sustain cooperation in local games. In reality, however, accidental cooperation may also be possible, and this remains a topic for future study. We speculate that accidental cooperation will reduce the parameter regime in which TFT--DISC solves the scoring dilemma, because it benefits an ALLD mutant more than an ALLC mutant or a TFT--DISC resident.

Another assumption concerns the form of direct reciprocity within a neighborhood: in our model, a focal player's action in a local game may depend on the co-player's most recent action within the neighborhood, regardless of whom it was directed toward. This formulation---which we call \textit{neighborhood direct reciprocity}---is reasonable in small, tight-knit groups whose members care how their friends treat one another as much as how their friends treat them. A finer-grained alternative---\textit{individual direct reciprocity}---would instead condition a focal player's local action on how the co-player has acted toward them specifically, requiring them to track how each neighbor last acted to them. The two formulations are identical for $n=2$; for $n>2$, we speculate that the scoring dilemma would be harder to solve under individual direct reciprocity.

Three additional methodological points are worth noting. First, our invasion criterion is more stringent than the standard condition used in invasion analysis. Rather than asking whether or not a mutant can invade a resident population at large, we ask whether or not it can invade the neighborhood to which it is introduced: we compare the payoff of a mutant to the payoff of the resident neighbors with whom the mutant interacts locally. We adopt this stricter criterion because even a localized invasion by unconditional defectors represents a meaningful breakdown of cooperation within a friend group. A locally robust strategy need not form a Nash equilibrium: even though TFT--DISC is not a best response against itself---because a TFT--DISC non-neighbor could increase their payoff by switching to ALLC (\cref{fig:fig2}A)---TFT--DISC can still be locally robust to an ALLC mutant because the mutant's presence increases the payoffs of the TFT--DISC neighbors.

Second, we have assumed that reputations are updated based on both local and global games. This assumption reflects the fact that people's treatment of those close to them can affect their social standing at large. However, this introduces a strong coupling between the two contexts; future work could relax this coupling by allowing only a subset of local games to affect public reputations or by introducing context-specific reputations. 

Third, we assume games of infinite length, whereas recent models combining direct and indirect reciprocity assume repeated games of finite length \cite{schmid_evolution_2021,hubner_stable_2025a,pal_coevolution_2024}. Long-term iterated play allows us to combine the two modeling approaches: the ODE-based approach from models of indirect reciprocity and the Markov-chain approach from models of direct reciprocity \cite{sigmund_calculus_2010}. In fact, our ODE formulation can be derived as a continuous-time limit of a Markov-chain representation of iterated games generalized to account for local and global games intercalated in time (\SI). Still, extending our framework to finite-length games would be a natural direction for future work. Prior work in well-mixed populations has shown that shorter games tend to favor indirect reciprocity over direct reciprocity, because the former can sustain cooperation even in one-shot interactions, whereas the latter requires sufficiently frequent repeated interactions to be effective \cite{schmid_unified_2021}. How this interplay changes when direct reciprocity is confined to local neighborhoods remains an open question.

Our framework is a step towards a hierarchical account of social behavior, one that distinguishes the rules governing our closest relationships from those governing interactions with the wider world. This account is increasingly needed in modern life. The contrast at the core of our model, between the few whose behavior we witness firsthand and the many we know only by reputation, reflects an increasingly acute distinction in a well-connected world. Modern life has stretched one side of this divide without bound: public reputations, once the product of gossip in small groups, are now aggregated and broadcast at a global scale through ratings, reviews, and social media. How cooperation is sustained when these two scales of information meet, and how we should act when they conflict, is the question our framework was built to address. Much work remains in this direction, including friend groups that change over time, differential treatment of newcomers, and variation or even adaptation of individual rates of local versus global interaction. But the problem of intercalated local and global interactions that shape our social information and behavior is already pressing, and it seems likely only to grow.

\clearpage
\section{Materials and Methods} 
\label{sec:matmethods}

Here we provide additional details of our mathematical model (\textit{\nameref{sec:model}}). We refer the reader to  the \SI\ for detailed derivations. 

\subsection*{Notation}

As a reminder, a strategy in local games has the form $\svDR = \left(\sDR{C}{G}, \sDR{C}{B}, \sDR{D}{G}, \sDR{D}{B} \right)$, where $\sDR{A}{R} \in \left[0,1\right]$ is the probability that a focal player cooperates with a co-player whose most recent local action was $A \in \{C,D\}$ and whose current (public) reputation is $R \in \{G, B\}$. A strategy in global games has the form $\svIR = (\sIR{G}, \sIR{B})$, where $\sIR{R}\in \left[0,1\right]$ is the probability that a focal player cooperates with a co-player whose reputation is $R\in\{G, B\}$.

Regardless of interaction type (local or global), a player intending to cooperate accidentally defects with probability $\exec \in (0, 1/2)$ (\textit{execution error rate}). We denote \textit{effective strategies} that account for these execution errors by $\svDRt = (\sDRt{C}{G}, \sDRt{C}{B}, \sDRt{D}{G}, \sDRt{D}{B}) \in (0,1)^4$ and $\svIRt = (\sIRt{G}, \sIRt{B}) \in (0,1)^2$. Separately, a good reputation is assigned as bad, or vice versa, with probability $\assess \in (0, 1/2)$ (\textit{assessment error rate}). We denote \textit{effective assessment rules} that account for assessment errors by $\avt \coloneqq (\at{C}{}, \at{D}{}) \in (0,1)^2$.

Throughout our analysis, we assume $b > c > 1$ and $n\geq2$.

\subsection*{Status updates after local and global games}

In our analysis, we assume that the statuses of any two players in the population are independent. Under this mean-field assumption, we can directly write down the probabilities governing status updates after each round.

A player's status $AR \in \{CG, CB, DG, DB\}$ is updated each time the player acts as a donor. The new status of a donor after a local game depends on the donor's realized action and the public assessment of that action, but not on the donor's previous status. Hence the conditional probability $\smash{P_{A'R'\mid AR}^{\DR}}$ that a donor who plays a local game against a recipient of status $AR$ is assigned status $A'R'$ is given by
\begin{equation}
    \begin{cases}
    P_{CG\mid AR}^{\DR} &= \sDRt{A}{R} \, \at_{C} \;, \\
    P_{CB\mid AR}^{\DR} &= \sDRt{A}{R} \left(1-\at_{C}\right) \;, \\
    P_{DG\mid AR}^{\DR} &= \left(1-\sDRt{A}{R}\right) \at_{D} \;, \\
    P_{DB\mid AR}^{\DR} &= \left(1-\sDRt{A}{R}\right) \left(1-\at_{D}\right) \;.
    \end{cases}
    \label{eq:probslocal}
\end{equation}

By contrast, after a global game, only the public reputation of the donor is updated, while their most recent local action remains unchanged. This new reputation depends on the donor's realized action and the public assessment of that action. That is, the conditional probability $\smash{P_{R'\mid R}^{\IR}}$ that a donor who plays a global game against a recipient with reputation $R$ is assigned reputation $R'$ is given by
\begin{equation}
    \begin{cases}
    P_{G\mid R}^{\IR} &= \sIRt{R} \at_{C} + (1-\sIRt{R}) \at_{D} \;,\\
    P_{B\mid R}^{\IR} &= \sIRt{R} (1-\at_{C}) + (1-\sIRt{R}) (1-\at_{D}) \;.
    \end{cases}
    \label{eq:probsglobal}
\end{equation}

\subsection*{Monomorphic populations}

We now consider a monomorphic population in which all players adopt the same strategy pair $(\svDR, \svIR)$. Let $h_{AR}$ denote the frequency of players of status $AR$ within the population, with $\sum_{AR} h_{AR}=1$. 

Consider a focal player chosen uniformly at random. Under the mean-field approximation, a randomly selected co-player is of status $AR$ with probability $h_{AR}$. Therefore, the probability that the focal player is assigned status $AR$ after a local game is $\sum_{A'R'} h_{A'R'} P_{AR\mid A'R'}^{\DR}$. Similarly, the probability that the co-player has reputation $R$ is $h_{\bullet R} \coloneqq h_{CR}+h_{DR}$ (the asterisk in $\bullet R$ denotes an arbitrary action), and the probability that the most recent local action of the focal player was $A$ is $h_{A\bullet} \coloneqq h_{AG}+h_{AB}$ (the $\bullet$ in $A\bullet$ denotes an arbitrary reputation). Since the most recent local action is carried over after a global game, the probability that the focal player is assigned status $AR$ after a global game is $h_{A\bullet} \sum_{R'} h_{\bullet R'} P_{R\mid R'}^{\IR}$. The overall probability $P_{AR}$ that a focal player is assigned status $AR$ after a game is therefore a weighted sum of these contributions, with weights given by the probabilities of local and global games ($\lambda$ and $1-\lambda$, respectively):
\begin{equation}
    P_{AR} = \lambda 
        \underbrace{
            \sum_{A'R'} h_{A'R'} P_{AR\mid A'R'}^{\DR} 
        }_{\textrm{changes due to local games}} 
        + \left(1-\lambda\right) 
        \underbrace{
            h_{A\bullet} \sum_{R'} h_{\bullet R'} P_{R\mid R'}^{\IR} 
        }_{\textrm{changes due to global games}} \;.
\end{equation}
The resulting status dynamics can be described by a system of ODEs reported in \cref{eq:hsystem}, analogous to reputation dynamics in models of pure indirect reciprocity \cite{perret_evolution_2021,michel-mata_evolution_2024,kawakatsu_mechanistic_2024}. \Cref{eq:hsystem} can be written in matrix form, which we use throughout the rest of our presentation, as follows:
\begin{equation}
    \dot\vh = - \vh + \lambda \underbrace{ 
        \vh\, \vP^{\DR} 
        }_{ \substack{\textrm{changes due to}\\\textrm{local games}} 
        } + \left(1-\lambda\right) \underbrace{ 
        \vh_{A\bullet} \otimes \left( \vh_{\bullet R}\, \vP^{\IR} \right) 
        }_{ \textrm{changes due to global games} } \;,
    \label{eq:monomorphiceq}
\end{equation}
where $\vh \coloneqq  \left(h_{CG},h_{CB},h_{DG},h_{DB}\right)$ is the status distribution; $\vh_{A\bullet} \coloneqq ( h_{C\bullet}, h_{D\bullet} ) = ( h_{CG}+h_{CB}, h_{DG}+h_{DB} )$ is the marginal distribution over actions; $\vh_{\bullet R} \coloneqq ( h_{\bullet G}, h_{\bullet B} ) = ( h_{CG}+h_{DG}, h_{CB}+h_{DB} )$ is the marginal distribution over reputations; and $\otimes$ denotes the Kronecker product. In addition, we define matrices
\begin{equation} \renewcommand{\arraystretch}{1.5}
    \begin{split}
        \vP^{\DR} 
            & \coloneqq 
            \begin{bmatrix}
                P_{CG \mid CG}^{\DR} & P_{CB \mid CG}^{\DR} & P_{DG \mid CG}^{\DR} & P_{DB \mid CG}^{\DR}
                \\
                P_{CG \mid CB}^{\DR} & P_{CB \mid CB}^{\DR} & P_{DG \mid CB}^{\DR} & P_{DB \mid CB}^{\DR}
                \\
                P_{CG \mid DG}^{\DR} & P_{CB \mid DG}^{\DR} & P_{DG \mid DG}^{\DR} & P_{DB \mid DG}^{\DR}
                \\
                P_{CG \mid DB}^{\DR} & P_{CB \mid DB}^{\DR} & P_{DG \mid DB}^{\DR} & P_{DB \mid DB}^{\DR}
            \end{bmatrix}
            \in [0,1)^{4\times4} 
    \end{split}
    \label{eq:pDRgeneral}
\end{equation}
and
\begin{equation} \renewcommand{\arraystretch}{1.5}
    \begin{split}
        \vP^{\IR}
        & \coloneqq 
        \begin{bmatrix} 
            P_{G\mid G}^{\IR} & P_{B\mid G}^{\IR}
            \\ 
            P_{G\mid B}^{\IR} & P_{B\mid B}^{\IR}
        \end{bmatrix} 
        \in (0,1)^{2\times2} \;,
    \end{split}
    \label{eq:pIRgeneral}
\end{equation}
where $P_{A'R'\mid AR}^{\DR}$ and $P_{R'\mid R}^{\IR}$ are as defined in \cref{eq:probslocal,eq:probsglobal}, respectively.

\subsection*{Invasion analysis}

A key goal is to derive the condition under which tit-for-tat discriminators (TFT--DISC) resist invasion by unconditional cooperators (ALLC) and defectors (ALLD)---and thereby solve the scoring dilemma. To this end, we consider a rare mutant with strategy pair $({\svDR}' ,{\svIR}')$ introduced to a neighborhood $\mathcal{N}$ of a resident population with strategy pair $(\svDR ,\svIR)$. We refer to the resident-type players inside the mutant's neighborhood as \textit{resident neighbors} and those outside as \textit{resident non-neighbors} (\cref{fig:schematic}C). Note that we need not consider a mutant with a different assessment rule: because reputation assessments are public, a rare mutant with a different assessment rule would have a negligible effect on the dynamics of reputations (and therefore statuses). 

We assume the following:
\begin{itemize}
    \item Players can imitate only their neighbors. Invasion therefore succeeds only when the fitness of the mutant exceeds that of the resident neighbors---that is, the mutant has a positive growth rate locally when rare.
    \item Because each neighborhood is much smaller than the population, resident non-neighbors interact with players in $\mathcal{N}$ with vanishing probability. Therefore, the status dynamics (and fitness) of the resident non-neighbors are unaffected by the mutant and the resident neighbors. 
\end{itemize}
Importantly, the resident neighbors and non-neighbors can differ in their status distributions, even in equilibrium, because status dynamics depend on both local and global games. We therefore distinguish among three status distributions: 
$\vh^{\RNN} \coloneqq (h_{CG}^{\RNN},h_{CB}^{\RNN},h_{DG}^{\RNN},h_{DB}^{\RNN})$ for resident non-neighbors, 
$\vh^{\R} \coloneqq(h_{CG}^{\R},h_{CB}^{\R},h_{DG}^{\R},h_{DB}^{\R})$ for resident neighbors, and 
$\vh^{\M} \coloneqq (h_{CG}^{\M},h_{CB}^{\M},h_{DG}^{\M},h_{DB}^{\M})$ for mutants. 

In the large-population limit, the status dynamics of resident non-neighbors, resident neighbors, and mutants can be described by a coupled dynamical system. At equilibrium, the status distributions satisfy the following system of quadratic consistency equations:
\begin{align}
    \vh^{\RNN} & = \lambda\, \vh^{\RNN}\, \vP^{\DR,\,\RNN}  + \left(1-\lambda\right) \vh_{A\bullet}^{\RNN} \otimes \left( \vh_{\bullet R}^{\RNN}\, \vP^{\IR,\,\RNN} \right) \;,
    \label{eq:invasioneqRNN}
    \\
    \vh^{\R} & = \lambda \left(\dfrac{1}{n-1}\,\vh^{\M} + \dfrac{n-2}{n-1}\,\vh^{\R}\right) \vP^{\DR,\,\R}  + \left(1-\lambda\right) \vh_{A\bullet}^{\R} \otimes \left( \vh_{\bullet R}^{\RNN}\, \vP^{\IR,\,\R} \right) \;,
    \label{eq:invasioneqR}
    \\
    \vh^{\M} & = \lambda\, \vh^{\R}\, \vP^{\DR,\,\M}  + \left(1-\lambda\right) \vh_{A\bullet}^{\M} \otimes \left( \vh_{\bullet R}^{\RNN}\, \vP^{\IR,\,\M} \right) \;,
    \label{eq:invasioneqM}
\end{align}
where $\vP^{\DR,\,\RNN}$, $\vP^{\DR,\,\R}$, and $\vP^{\DR,\,\M} \in (0,1)^{4\times 4}$ denote the matrix $\vP^{\DR}$ (\cref{eq:pDRgeneral}) evaluated with the corresponding local strategies ($\svDR$ for $\RNN$ and $\R$, ${\svDR}'$ for $\M$); and $\vP^{\IR,\,\RNN}$, $\vP^{\IR,\,\R}$, and $\vP^{\IR,\,\M} \in (0,1)^{2\times 2}$ denote the matrix $\vP^{\IR}$ (\cref{eq:pIRgeneral}) evaluated with the corresponding global strategies ($\svIR$ for $\RNN$ and $\R$, ${\svIR}'$ for $\M$).
Note that since $\sum_{AR} h_{AR}^{Y} = 1$ for any class $Y$, each of \cref{eq:invasioneqRNN,eq:invasioneqR,eq:invasioneqM} reduces to a 3-dimensional system of quadratic equations. For a resident population of TFT--DISC with either an ALLC or an ALLD mutant (i.e., $(\svDR,\svIR)=((1,1,0,0), (1,0))$, $({\svDR}' ,{\svIR}') = (\mathbf{1}, \mathbf{1})$ or $(\mathbf{0}, \mathbf{0})$), the system \eqref{eq:invasioneqRNN}--\eqref{eq:invasioneqM} has a unique equilibrium (see \SI\ \cref{sec:analytical} for details).

Long-term average payoffs---$\pi_{\RNN}$, $\pi_{\R}$, and $\pi_{\M}$---depend on the average rates of cooperation at equilibrium, as described in the main text (\cref{eq:fitnesseq}).
In local games, the cooperation rate of a class-$X$ player with a class-$Y$ player (with $(X,Y)\!\in\!\{(\RNN,\RNN),\, (\R,\R),\, (\R,\M),\, (\M,\R)\}$) is the strategy-weighted average $\gamma^{\DR}_{X,Y} = \svDRtX{X} \cdot \vh^{\ast,\,Y}$, where $\cdot$ denotes the dot product and $\vh^{\ast,\,Y}$ denotes the equilibrium status distribution in class $Y$.
Similarly, in global games, the cooperation rate of a class-$X$ player with a class-$Y$ player (with $(X,Y)\!\in\!\{(\RNN,\RNN),\, (\RNN,\R),\, (\RNN,\M),\, (\R,\RNN),\, (\M,\RNN)\}$) is the strategy-weighted average 
$\gamma^{\IR}_{X,Y} = \svIRtX{X} \cdot \vh_{\bullet R}^{\ast,\,Y}$, where $\vh_{\bullet R}^{\ast,\,Y} = ( h_{\bullet G}^{\ast,\,Y}, h_{\bullet B}^{\ast,\,Y} )$ denotes the equilibrium distribution of reputations in class $Y$.

\subsection*{Local invasion of TFT--DISC by ALLD}
 
A rare ALLD mutant can locally invade a resident population of TFT--DISC if and only if the mutant fitness $\pi_{\ALLD}$ exceeds the fitness of the resident neighbor(s), $\pi_{\R}$.
Solving $\pi_{\ALLD} > \pi_{\R}$ for $\lambda$, we find that ALLD can invade if $1 > \lambda > \tilde\lambda_{\ALLD}^*$, where 
\begin{equation}
    \tilde\lambda_{\ALLD}^* = 1-\dfrac{1}{\left(b/c\right)(1-2 \assess)(1-\exec)} \;.
    \label{eq:lambdaALLDstar}
\end{equation}
Note that for $\tilde\lambda_{\ALLD}$ to be positive, the following must hold:
\begin{equation}
    \dfrac{b}{c} > \left(\dfrac{b}{c}\right)^* \coloneqq \dfrac{1}{(1-2 \assess)(1-\exec)} \;.
\end{equation}

\subsection*{Local invasion of TFT--DISC by ALLC} \label{sec:ALLCinvasion}

A rare ALLC mutant can locally invade a resident population of TFT--DISC if and only if $\pi_{\R} < \pi_{\ALLC}$.
We consider the cubic polynomial
\begin{equation}
    P(\lambda) \coloneqq a_3 \lambda^3 + a_2 \lambda^2 + a_1 \lambda + a_0 \ \propto \ \left(\pi_{\R} - \pi_{\ALLC} \right) \;,
    \label{eq:polynomial}
\end{equation}
with coefficients
{\footnotesize
\begin{equation}
\begin{split}
    a_3 & = - \left(\dfrac{b}{c}\right) \left(n-1\right) \left(1-2 \assess\right)^2 \left(1-\exec \right)^4 \;,
    \\
    a_2 & = \left(1-2 \assess \right) \left(1-\exec \right)^2 \bigg(
        -\left(n-1\right) \left(1-\exec \right) 
        + \left(\dfrac{b}{c}\right) \Big(
            \left(2-5 \assess \right) \left(n-1\right)
            -\exec  \left(n^2-(1-\assess ) (n-12) (n-1)-9 n+7\right)
            \\ & \quad \quad \quad \quad \quad \quad \quad \quad \quad \quad \quad \quad \quad \quad \quad \quad \quad \quad \quad \quad \quad \quad  
            +\exec^2 \left(n \left(n-2\right)-\left(1-2 \assess \right) \left(n-4\right) \left(n-1\right)\right)
        \Big)
    \bigg) \;,
    \\
    a_1 & = (1-\exec) \bigg(
        \left(n-1\right) \left(\left(1-\exec\right) \left(\left(n-3\right) \exec+1\right)-\assess  \left(\exec \left(\left(3-n\right) 2\exec+n-8\right)+3\right)\right)
    \\ & \quad \quad \quad \quad \quad \quad    
    - \left(\dfrac{b}{c}\right) \Big(
        \left(8 \assess ^2-6 \assess +1\right) \left(n-1\right)
        \\ & \quad \quad \quad \quad \quad \quad \quad \quad \quad \quad \quad    
        + \exec \left(n^2+4 \assess ^2 \left(n-8\right) \left(n-1\right)-2 \assess  \left(\left(n-13\right) n+13\right)-7 n+6\right)
        \\ & \quad \quad \quad \quad \quad \quad \quad \quad \quad  \quad \quad   
        +\exec^2 \left(40 \assess -4 \assess ^2 \left(n-1\right) \left(3 n-11\right)+4 \assess  \left(2 n-11\right) n-2n \left(n-5\right)-9\right)
        % \Big) 
        \\ & \quad \quad \quad \quad \quad \quad \quad \quad \quad \quad \quad + \exec^3 \left(2 \assess -1\right) \left(5 \left(2 \assess -1\right) + n \left(-14 \assess +\left(4 \assess -1\right) n+5\right)\right)
        \Big) 
    \bigg)
    \;,
    \\
    a_0 & = \left(1-\left(\dfrac{b}{c}\right)\left(1-2 \assess \right) \left(1-\exec \right)\right) \left(n-1\right) \left(1+\left(n-2\right)\exec\right)  \left(1-\exec \right) \left(\exec + \assess\left(1-2\exec\right) \right) 
    \;.
\end{split}
\label{eq:coefficients}
\end{equation}
}\!\!
Since $\sgn(P(\lambda)) = \sgn\left(\pi_{\R} - \pi_{\ALLC} \right)$ (see \SI\ \cref{sec:analytical}), ALLC can invade locally if and only if $P(\lambda) < 0$. We therefore analyze the roots and sign structure of $P(\lambda)$.

We first establish a sufficient condition under which there exists a root $\tilde\lambda_{\ALLC}^*\in(0, 1)$ satisfying $P(\tilde\lambda_{\ALLC}^*)=0$. Since we already know that ALLD can invade when for any $\lambda\in(0,1)$ when $b/c<(b/c)^*$ (\cref{eq:lambdaALLDstar}), we focus on the case where $b/c>(b/c)^*$. For $b/c>(b/c)^*$, we have $P(0) = a_0 < 0$; that is, ALLC can invade when all games are global ($\lambda=0$). Moreover, if $b/c>(b/c)^*$ and 
\begin{equation}
    \label{eq:P1positivecondition}
    n < n^* \coloneqq \dfrac{
        2 - \left(b/c\right) \left(3-2 \exec\right) - \sqrt{\left(b/c\right) \left(- 4 \left(1-\exec\right) +  \left(b/c\right) \left(5 - 4 \left(2-\exec\right) \exec\right) \right)}
        }{
        2 \left(1-\left(b/c\right) \left(1-\exec\right)\right)
        } \;,
\end{equation}
then we also have $P(1) = a_3 + a_2 + a_1 + a_0 > 0$; that is, ALLC cannot invade when all games are local ($\lambda=1$). 
By continuity of $P(\lambda)$ and the intermediate value theorem, there is a $\tilde\lambda_{\ALLC}^* \in (0,1)$ satisfying $P(\tilde\lambda_{\ALLC}^*) = 0$, provided that $b/c>(b/c)^*$ and $n<n^*$ \eqref{eq:P1positivecondition}. 

Next, we determine whether $\tilde\lambda_{\ALLC}^*$, when it exists, is unique in the interval $(0,1)$. Since the leading coefficient $a_3$ of the cubic polynomial $P(\lambda)$ is negative for all admissible parameter values, we have $P(\lambda)\to\mp\infty$ as $\lambda\to\pm\infty$. Suppose that $b/c>(b/c)^*$ and $n<n^*$  \eqref{eq:P1positivecondition}. Then $P(1)>0$ and $P(0)<0$, so there must be a root in the interval $(1,+\infty)$ and another in $(-\infty,0)$. Because a cubic polynomial has at most three real roots (counting multiplicity), it follows that there can be at most one root in the interval $(0,1)$ if $b/c>(b/c)^*$ and $n<n^*$ \eqref{eq:P1positivecondition}. 
In particular, this root is the second-largest real root of $P(\lambda)$, which can be expressed analytically using the trigonometric solution to the cubic (see \SI\ \cref{sec:analytical}).

For $n=2$, we can obtain a necessary and sufficient condition for the existence of a unique $\tilde\lambda_{\ALLC}^*$ in the interval $(0, 1)$. First, we have $n^* > 2$ for all admissible error rates ($0<\assess, \exec<1/2$) and game parameters ($b>c>0$), so condition \eqref{eq:P1positivecondition} is satisfied for $n=2$. Therefore, if $b/c>(b/c)^*$ and $n=2$, then $\tilde\lambda_{\ALLC}^*\in(0, 1)$ exists and is unique. Second, if $b/c<(b/c)^*$ and $n=2$, then we have $P(0)>0$ and $P(1)>0$. At the threshold $b/c=(b/c)^*$, we have $P(0) = 0$ and $P'(0) = a_1 > 0$, so the root $\lambda=0$ is simple and $P(\lambda)$ crosses zero from below (and therefore $\lambda=0$ is the second-largest real root). By continuity of roots, and since $P(1)>0$, reducing $b/c$ shifts this root into the region $\lambda < 0$. In this case, the cubic polynomial $P(\lambda)$ must have two roots in $(-\infty,0)$ and one in $(1,\infty)$, since its leading coefficient $a_3$ is negative. It follows that there cannot be a root in the interval $(0,1)$ for $b/c<(b/c)^*$. Therefore, for $n=2$, $\tilde\lambda_{\ALLC}^*$ exists and is unique in the interval $(0, 1)$ if and only if $b/c>(b/c)^*$.

\subsection*{Effective critical probabilities with a TFT--DISC resident population}

In our formulation thus far, the critical probabilities $\tilde \lambda_{\ALLD}^*$ and $\tilde \lambda_{\ALLC}^*$ may lie outside the feasible domain $[0,1]$. For example, we know from \cref{eq:lambdaALLDstar} that, for $b/c < (b/c)^*$, we have $\tilde\lambda_{\ALLD}^* < 0$, which means that ALLD can invade for all admissible values of $\lambda$. To obtain thresholds that are meaningful for game dynamics, we define \textit{effective critical probabilities} on the feasible domain $[0,1]$ as follows. For ALLD, we know that there is a single threshold $\tilde\lambda_{\ALLD}^*$ that can lie in $(0,1)$  (\cref{eq:lambdaALLDstar}), so we define $\lambda_{\ALLD}^* = \tilde\lambda_{\ALLD}^*$ if $\tilde\lambda_{\ALLD}^* >0$ and $0$ otherwise. For ALLC, we define $\lambda_{\ALLC}^*$ to be the smallest root of $P(\lambda)=0$ in the interval $(0,1)$, if such a root exists, and $0$ otherwise. Accordingly, ALLD can invade for all $\lambda > \lambda_{\ALLD}^*$ and ALLC can invade for all $\lambda < \lambda_{\ALLC}^*$; if there are multiple roots of $P(\lambda)=0$, ALLC may also invade for some $\lambda > \lambda_{\ALLC}^*$.

For $n=2$, we have already shown that a unique root $\tilde\lambda_{\ALLC}^*\in(0,1)$ exists if and only if $b/c>(b/c)^*$. In this case, we can write the effective critical probabilities as 
\begin{equation}
    \begin{split}
        \lambda_{\ALLD}^* 
        =
        \begin{cases}
            \tilde\lambda_{\ALLD}^* & \text{if } b/c > (b/c)^* \;, \\
            0 & \text{if } b/c \leq (b/c)^* \;,
        \end{cases}
        \qquad\text{and}\qquad
        \lambda_{\ALLC}^* 
        = \begin{cases}
            \tilde\lambda_{\ALLC}^* & \text{if } b/c > (b/c)^* \;, \\
            0 & \text{if } b/c \leq (b/c)^* \;,
        \end{cases}
    \end{split}
    \label{eq:lambdaeff}
\end{equation}
which are shown in \cref{fig:fig3}A--C.

\subsection*{Condition under which TFT--DISC solves the scoring dilemma for $n=2$}

We have established that, for $n=2$ and sufficiently large benefit-to-cost ratio ($b/c>(b/c)^*$), the polynomial $P(\lambda)$ satisfies $P(0) < 0$ and $P(1) > 0$, and it has exactly one root $\lambda_{\ALLC}^*(=\tilde\lambda_{\ALLC}^*)$ in the interval $(0,1)$. In this case, for any $\lambda\in(0,1)$, $P(\lambda) > 0$ implies $\lambda_{\ALLC}^*<\lambda$. 
In particular, for $b/c>(b/c)^*$, evaluating $P(\lambda)$ at $\lambda=\lambda_{\ALLD}^*$  yields $P(\lambda_{\ALLD}^*) > 0$ and, consequently, $\lambda_{\ALLC}^* < \lambda_{\ALLD}^*$. 
By contrast, for $b/c<(b/c)^*$, we have $\lambda_{\ALLC}^*=\lambda_{\ALLD}^*=0$ \eqref{eq:lambdaeff}.
Hence, there is an nonempty interval $(\lambda_{\ALLC}^*, \lambda_{\ALLD}^*)$ in which TFT--DISC is robust against both ALLC and ALLD---that is, TFT--DISC solves the scoring dilemma---if and only if $b/c>(b/c)^*$, as reported in \cref{eq:bccondition}.

\subsection*{Stochastic simulations in finite populations}

To verify that our mean-field analysis provides a good approximation to finite populations, we performed Monte Carlo simulations in Julia \cite{bezanson_julia_2017}. Each population consists of $N=120$ players (one mutant and $N-1$ residents) partitioned into neighborhoods of size $n$. The $n-1$ residents in the mutant's neighborhood are resident neighbors, while all residents outside the mutant's neighborhood are resident non-neighbors. Each player $i$ is assigned fixed local and global strategies, given by $\smash{(\sDR{i,\,C}{G}, \sDR{i,\,C}{B}, \sDR{i,\,D}{G}, \sDR{i,\,D}{B} )}$ and $\smash{(\sIR{i,\,C}, \sIR{i,\,D} )}$, respectively. 
Public reputations and most recent local actions are initialized at random. Each round proceeds as follows.

\subsubsection*{Player selection} First, a focal player $i$ is selected uniformly at random from the $N$ players. Then an interaction type is selected: if a randomly generated number is less than $\lambda$, then the interaction is local, and a co-player $j$ is selected uniformly at random from among $i$'s neighbors; otherwise, the interaction is global, and a co-player $j$ is selected uniformly at random from among $i$'s non-neighbors. 

\subsubsection*{Gameplay} Focal player $i$ and co-player $j$ each act once as a donor. In a local game, $i$ chooses an action toward $j$ according to $i$'s local strategy: $i$ accesses $j$'s most recent local action $A_j$ and public reputation $R_j$, cooperating with probability $\smash{s_{i,\,A_jR_j}^{\DR}}$ and defecting otherwise. In a global game, $i$ chooses an action toward $j$ according to $i$'s global strategy: $i$ accesses $j$'s public reputation $R_j$, cooperating with probability $\smash{s_{i,\,R_j}^{\IR}}$ and defecting otherwise. Similarly, $j$ acts toward $i$ according to $j$'s local or global strategy, depending on the interaction type. Each action is independently subject to execution error: if a randomly generated number is less than $\exec$, then cooperation is flipped to defection, but not vice versa. 
Players accrue payoffs: player $i$ ($j$) earns payoff $b$ if $j$ ($i$) cooperates and pays a cost $c$ if $i$ ($j$) cooperates.

\subsubsection*{Status updates} Regardless of interaction type, the reputations of focal player $i$ and co-player $j$ are updated after gameplay. Actions are assessed according to the scoring norm: player $i$ ($j$) is assigned a good reputation if they cooperated and a bad reputation if they defected. Each reputation update is independently subject to assessment error: if a randomly generated number is less than $\assess$, then the reputation of $i$ ($j$) is flipped from good to bad, and vice versa. The newly assigned reputations are then stored in a public reputation vector. 
When the interaction type is local (but not when it is global), the most recent local actions of $i$ and $j$ are also updated.

Each simulation was run for $2\times 10^7$ rounds, with the initial $2\times 10^6$ rounds discarded as burn-in (\cref{fig:SI-comparison}). Average payoffs computed from stochastic simulations show good agreement with theoretical predictions from our mean-field analysis (\cref{eq:hsystem,eq:fitnesseq}).

\section*{Data Availability}
Simulation code is available in a public repository on GitHub (\text{https://github.com/marikawakatsu/HybridDRIR}).

\section*{Acknowledgments}
M.K.\ acknowledges support from the James S.\ McDonnell Foundation (doi:10.37717/2021-3209). Y.M.\ acknowledges support by JSPS KAKENHI Grant Number JP25K07145 and from RIKEN Pioneering Project ``Planetary Resilience Science for Safeguarding the Global Commons.'' GitHub Copilot was used to assist code development; the authors assume responsibility for the content of the manuscript.

%%%%%%%%%%%%%%%%%%%%%%%%%%%
% References
%%%%%%%%%%%%%%%%%%%%%%%%%%%
\newpage \small
\bibliographystyle{unsrtnat}
\bibliography{bib_arxiv}
\normalfont

%%%%%%%%%%%%%%%%%%%%%%%%%%%
% SI header
%%%%%%%%%%%%%%%%%%%%%%%%%%%
\clearpage
\appendix
\setcounter{section}{0}
\renewcommand{\thesection}{S\arabic{section}}

\numberwithin{figure}{section}
\renewcommand{\thefigure}{S\arabic{figure}}
\numberwithin{equation}{section}
\renewcommand{\theequation}{S\arabic{equation}}

\newgeometry{margin=0.65in}

\section*{\huge Supplementary Information} \vspace{2ex}

% show toc
\addtocontents{toc}{\protect\setcounter{tocdepth}{2}} 
{\small\tableofcontents} 
\vspace{3ex}

% refresh footer 
\setlength{\headwidth}{\textwidth}
\fancyhf{} 
\fancyfoot[R]{\footnotesize\hfill \textit{Page \thepage\ of \pageref{LastPage}}}

\small
%%%%%%%%%%%%%%%%%%%%%%%%%%%
% SI TEXT
%%%%%%%%%%%%%%%%%%%%%%%%%%%

\renewcommand{\R}[0]{\res{\textrm{R-N}}}
\renewcommand{\RNN}[0]{\resnn{\textrm{R-NN}}}
\renewcommand{\M}[0]{\mut{\textrm{M}}}
\renewcommand{\ALLD}[0]{{\color[HTML]{CD5C5C}\textrm{ALLD}}} % indianred
\renewcommand{\ALLC}[0]{{\color[HTML]{4169e1}\textrm{ALLC}}} % royalblue

\section{Full model description}
\label{sec:fullmodel}

Here we provide a complete description of our model. Some material from the main text is repeated to ensure that the description in this section is self-contained.

\subsection{Setup} 

We consider a population of $N$ players partitioned into neighborhoods of $n$ players each, with $N\gg n \geq 2$. Throughout our analysis, we assume a large population and consider the limit $N\to\infty$. For a given focal player, the $n-1$ players in the focal player's neighborhood are called \textit{neighbors}, while all others in the population are called \textit{non-neighbors}. 

Players engage in infinitely many rounds of pairwise interactions. In each round, a focal player is chosen uniformly at random from the population. An interaction type is then selected: with probability $\lambda$, the focal player engages in a \textit{local game} and is paired with a randomly selected neighbor (\cref{fig:schematic}A); with probability $1-\lambda$, the focal player engages in a \textit{global game} and is paired with a randomly selected non-neighbor (\cref{fig:schematic}B). 

Regardless of interaction type (local or global), interactions take the form of donation games. Each interacting pair plays the game twice, with each player acting once as a \textit{donor} and once as a \textit{recipient}. In each game, the donor chooses one of two actions: cooperate ($C$), i.e., pay a cost $c>0$ to provide a benefit $b>c$ to the recipient, or defect ($D$), i.e., incur no cost and provide no benefit. 

\subsection{Strategy in local games}  

In a local game, the action of a focal player toward a neighbor may depend on the neighbor's past local action or the neighbor's current public reputation. Specifically, a strategy in local games takes the form
\begin{equation}
    \svDR = \left(\sDR{C}{G}, \sDR{C}{B}, \sDR{D}{G}, \sDR{D}{B}\right) \;,
    \label{eq:stratlocalgeneral}
\end{equation}
where $\sDR{A}{R} \in \left[0,1\right]$ is the probability that a focal player cooperates with a co-player whose most recent local action was $A \in \{C,D\}$ and has reputation $R \in \{G,B\}$.

For example, $\svDR=(1,1,1,1)$ corresponds to unconditional cooperation and $(0,0,0,0)$ to unconditional defection. Strategies of the form $\svDR=(\rho,\rho,\sigma,\sigma)$ ignore reputations altogether and correspond to memory-1 reactive strategies in models of repeated games: for instance, $(1,1,0,0)$ is tit-for-tat (TFT), and $(1,1,Q,Q)$ with $Q>0$ to generous tit-for-tat (GTFT). 

\subsection{Strategy in global games}

In a global game, the action of a focal player toward a non-neighbor may depend on the non-neighbor's reputation. A strategy in global games takes the form
\begin{equation*}
    \svIR = (\sIR{G}, \sIR{B}) \;,
\end{equation*} 
where $\sIR{R}\in \left[0,1\right]$ is the probability that a focal player cooperates with a co-player with reputation $R\in\{G, B\}$. 
For example, $\svIR=(1,1)$ corresponds to unconditional cooperation, $(0,0)$ to unconditional defection, and $(1,0)$ to the discriminator (DISC) strategy in models of indirect reciprocity.

\subsection{Reputation assessment}
After a round of either local or global games, a third-party observer assesses the action of every player according to an \textit{assessment rule}. An assessment rule (also called a \textit{social norm}) governs how an observer judges a donor based on the donor's action toward a recipient \cite{sigmund_calculus_2010}. We consider first-order assessment rules of the form
\begin{equation*}
    \av = \left(\ant{C}{}, \ant{D}{}\right)
\end{equation*}
where $\ant{A}{} \in \left[0,1\right]$ is the probability that a donor who takes action $A \in \{C,D\}$ earns a good reputation. Our analysis focuses on the \textit{image scoring} norm, $\av = \left(1, 0\right)$, which judges cooperation as good and defection as bad regardless of the recipient's reputation. 

Future extensions of the model could consider higher-order assessment rules \cite{ohtsuki_how_2004,santos_social_2018}. For instance, second-order norms take the form $\av = \left(\ant{C}{G}, \ant{C}{B}, \ant{D}{G}, \ant{D}{B}\right)$, where $\ant{A}{R} \in \left[0,1\right]$ is the probability that a donor who takes action $A \in \{C,D\}$ against a recipient with reputation $R \in \{G,B\}$ earns a good reputation.

\subsection{Errors}
Our model allows for two types of errors. With probability $\assess$ (\textit{assessment error rate}), a good reputation is assigned as bad, or vice versa. Separately, with probability $\exec$ (\textit{execution error rate}), a player who intends to cooperate accidentally defects. The introduction of these errors effectively rescales strategies and assessment rules. We denote \textit{effective strategies} that account for execution errors by 
\begin{equation}
    \begin{split}
        \svDRt & = \left(1-\exec\right)\svDR = \big(\sDRt{C}{G}, \sDRt{C}{B}, \sDRt{D}{G}, \sDRt{D}{B} \big) \;,
        \\
        \svIRt & = \left(1-\exec\right)\svIR = \big(\sIRt{G}, \sIRt{B}\big) \;,
    \end{split}
    \label{eq:errorsasymmetricSI}
\end{equation}    
where $\sDRt{A}{R},\sIRt{R}\in[0,1)$. Similarly, we denote an \textit{effective assessment rule} that accounts for assessment errors by
\begin{align}
    \avt & = \left(1-\assess\right)\av + \assess \left(\mathbf{1} - \av\right) = \left(\at{C}{}, \at{D}{}\right) \;,
    \label{eq:errorsasymmetricSIa}
\end{align}
where $\at{A}{} \in (0,1)$.

\subsection{Markov-chain representation of status dynamics in a monomorphic population}
\label{sec:markov}

As in the main text, we first consider a monomorphic population in which all players adopt the same strategy pair $(\svDR,\svIR)$. In a given round, each player can be in one of four statuses: $CG, CB, DG$, or $DB$. The first letter denotes the player's most recent local action ($C$ or $D$); the second denotes the player's current public reputation ($G$ or $B$). We denote by $h_{AR}$ the frequency of status $AR$ in the population.

In the main text, we approximate the mean-field dynamics of $\vh = (h_{AR})$ using a system of ODEs (\cref{eq:hsystem}). Here, we show that \cref{eq:hsystem} can also be derived as a continuous-time limit of a discrete-time Markov chain describing the status dynamics.

Under a mean-field approximation in a monomorphic population, symmetry across neighborhoods means that the probability that a player is in a particular status is independent of neighborhood identity. Moreover, because interaction partners are paired at random---either within neighborhoods for local games, or between neighborhoods for global games---the joint distribution of statuses for any interacting pair can be approximated at the population level. We can therefore represent iterated game dynamics as a discrete-time Markov chain defined on the set of all 16 possible combinations of statuses for an interacting pair. That is, the state space of this Markov chain is $\mathcal{M} = \{(A_1, A_2; R_1, R_2)\} = \{C, D\}^2 \times \{G, B\}^2$, where subscripts indicate players 1 and 2.

In each round, an interacting pair plays the donation game twice (with each player acting once as a donor and once as a recipient), and each donation game induces an update to the status $A_iR_i$ of its donor $i$ without affecting the status of its recipient. We first consider a round of local games between player 1 of status $A_1R_1$ and player 2 of status $A_2R_2$. The conditional probability $P_{A'R'\mid AR}^{\DR}$ that a player who plays a local game against $AR$ is assigned status $A'R'$ has the form
\begin{equation*}
    P_{A'R'\mid AR}^{\DR} = \mathbb{P}(A' \mid \textrm{local game vs }AR) \ \mathbb{P}\left(R' \mid A'\right) \;,
\end{equation*}
where the first factor is the probability that the donor takes action $A'$ and the second factor is the probability that action $A'$ is assessed as $R'$. More explicitly (see also \cref{eq:probslocal} in \nameref{sec:matmethods}),
\begingroup\small
\begin{equation*}
    \begin{split}
        \begin{aligned}
            P_{CG \mid CG}^{\DR} &= \sDRt{C}{G} \at{C}{}\;,
            &\quad 
            P_{CB \mid CG}^{\DR} &= \sDRt{C}{G} (1-\at{C}{})\;,
            &\quad
            P_{DG \mid CG}^{\DR} &= (1-\sDRt{C}{G}) \at{D}{}\;,
            &\quad 
            P_{DB \mid CG}^{\DR} &= (1-\sDRt{C}{G}) (1-\at{D}{})\;,
            \\
            P_{CG \mid CB}^{\DR} &= \sDRt{C}{B} \at{C}{}\;, 
            &\quad 
            P_{CB \mid CB}^{\DR} &= \sDRt{C}{B} (1-\at{C}{})\;, 
            &\quad
            P_{DG \mid CB}^{\DR} &= (1-\sDRt{C}{B}) \at{D}{}\;,
            &\quad 
            P_{DB \mid CB}^{\DR} &= (1-\sDRt{C}{B}) (1-\at{D}{})\;,
            \\
            P_{CG \mid DG}^{\DR} &= \sDRt{D}{G} \at{C}{}\;,
            &\quad 
            P_{CB \mid DG}^{\DR} &= \sDRt{D}{G} (1-\at{C}{})\;, 
            &\quad
            P_{DG \mid DG}^{\DR} &= (1-\sDRt{D}{G}) \at{D}{}\;, 
            &\quad 
            P_{DB \mid DG}^{\DR} &= (1-\sDRt{D}{G}) (1-\at{D}{})\;,
            \\
            P_{CG \mid DB}^{\DR} &= \sDRt{D}{B} \at{C}{}\;,
            &\quad 
            P_{CB \mid DB}^{\DR} &= \sDRt{D}{B} (1-\at{C}{})\;, 
            &\quad
            P_{DG \mid DB}^{\DR} &= (1-\sDRt{D}{B}) \at{D}{}\;,
            &\quad 
            P_{DB \mid DB}^{\DR} &= (1-\sDRt{D}{B}) (1-\at{D}{})\;,
        \end{aligned}
    \end{split}
\end{equation*}
\endgroup
where, as a reminder, $\sDRt{C}{G}$, $\sDRt{C}{B}$, $\sDRt{D}{G}$, and $\sDRt{D}{B}$ are entries of the error-modified local strategy vector $\svDRt$ (\cref{eq:errorsasymmetricSI}); and $\at{C}{}$ and $\at{D}{}$ are entries of the error-modified assessment rule vector $\av$ (\cref{eq:errorsasymmetricSIa}).

After a round of local games, player 1 is assigned status $A_1'R_1'$ with probability $P_{A_1'R_1'\mid A_2R_2}^{\DR}$, and, independently, player 2 is assigned status $A_2'R_2'$ with probability $P_{A_2'R_2'\mid A_1R_1}^{\DR}$. Hence, the component of the transition probability matrix associated with local games is given by
\begin{equation} \renewcommand*{\arraystretch}{1.3}
    \raisetag{-0.6ex}
    \tcboxmath[noborderbox]{
    \mathbf{\tilde P}^{\DR}
    \coloneqq 
    \Big[
        P_{(A_1, A_2; R_1, R_2)\to (A_1', A_2'; R_1', R_2')}^{\DR}
    \Big]
    = 
    \Big[
        P_{A_1'R_1'\mid A_2R_2}^{\DR}
        \cdot 
        P_{A_2'R_2'\mid A_1R_1}^{\DR}
    \Big] 
    \in [0, 1)^{16\times 16}
    \;,
    }
    \label{eq:pdrtilde}
\end{equation}

Next, we consider a round of global games between player 1 of status $A_1R_1$ and player 2 of status $A_2R_2$. By definition, a player's most recent \textit{local} action does not change following a \textit{global} game. The conditional probability $P_{R'\mid R}^{\IR}$ that a player who plays a global game against a recipient with reputation $R$ is assigned reputation $R'$ has the form: 
\begin{equation*}
    P_{R' \mid R}^{\IR} = \sum_{A''\in\{C,D\}} \mathbb{P}\left(A'' \mid \textrm{global game vs }R\right) \mathbb{P}\left(R' \mid A''\right) \;,
\end{equation*}
where the first factor in each summand is the probability that the donor takes action $A''$, and the second factor in each summand is the probability that action $A''$ is assessed as $R'$.
More explicitly (see also \cref{eq:probsglobal} in \nameref{sec:matmethods}),
\begin{equation*} 
    \begin{split}
        \begin{aligned}
            P_{G \mid G}^{\IR} & = \sIRt{G} \cdot \at{C}{} + \big(1-\sIRt{G}\big) \at{D}{}\;, 
            &\quad
            P_{B \mid G}^{\IR} & = \sIRt{G} \big(1-\at{C}{}\big) + \big(1-\sIRt{G}\big) \big(1-\at{D}{}\big)\;,
            \\ 
            P_{G \mid B}^{\IR} & = \sIRt{B} \cdot \at{C}{} + \big(1-\sIRt{B}\big) \at{D}{}\;,
            &\quad
            P_{B \mid B}^{\IR} & = \sIRt{B} \big(1-\at{C}{}\big) + \big(1-\sIRt{B}\big) \big(1-\at{D}{}\big)\;,
        \end{aligned}
    \end{split}
\end{equation*}
where, as a reminder, $\sIRt{G}$ and $\sIRt{B}$ are entries of the error-modified global strategy vector $\svIRt$ (\cref{eq:errorsasymmetricSI}); and $\at{C}{}$ and $\at{D}{}$ are entries of the error-modified assessment rule vector $\av$ (\cref{eq:errorsasymmetricSIa}).

After a round of global games, player 1 is assigned reputation $R_1'$ with probability $P_{R_1'\mid R_2}^{\IR}$, and, independently, player 2 is assigned reputation $R_2'$ with probability $P_{R_2'\mid R_1}^{\IR}$. Hence, the 4-by-4 submatrix governing reputation dynamics in global games is given by
\begin{equation}
\renewcommand*{\arraystretch}{1.3}
    \raisetag{-0.6ex}
    \tcboxmath[noborderbox]{
    \mathbf{\tilde P}^{\IR,\,\text{sub}}
    \coloneqq
    \Big[
        P^{\IR}_{R_1'\mid R_2}
        \cdot
        P^{\IR}_{R_2'\mid R_1}
    \Big]
    \in (0,1)^{4\times4}
    \;.
    }
    \label{eq:pIRtildesub}
\end{equation}
Since global games do not alter players' most recent local actions, transitions out of state $(A_1,A_2;R_1,R_2)$ after a round of global games must preserve entries $A_1$ and $A_2$. Therefore, the component of the transition matrix associated with global games has a block-diagonal structure: 
\begin{equation} \renewcommand*{\arraystretch}{1.3}
    \raisetag{-0.6ex}
    \tcboxmath[noborderbox]{
    \mathbf{\tilde P}^{\IR}
    \coloneqq \begin{bmatrix}
        \mathbf{\tilde P}^{\IR,\, \textrm{sub}} & \mathbf{0} & \mathbf{0} & \mathbf{0}
        \\
        \mathbf{0} & \mathbf{\tilde P}^{\IR,\, \textrm{sub}} & \mathbf{0} & \mathbf{0}
        \\
        \mathbf{0} & \mathbf{0} & \mathbf{\tilde P}^{\IR,\, \textrm{sub}} & \mathbf{0}
        \\
        \mathbf{0} & \mathbf{0} & \mathbf{0} & \mathbf{\tilde P}^{\IR,\, \textrm{sub}}
    \end{bmatrix} \in [0, 1)^{16\times 16}
    \;,
    }\label{eq:pirtilde}
\end{equation}

The full transition probability matrix associated with the Markov chain defined on $\mathcal{M}$ is a weighted sum of \cref{eq:pdrtilde,eq:pirtilde}, with weights determined by the probabilities of local and global games:
\begin{equation} \renewcommand*{\arraystretch}{1.3}
    \raisetag{-0.6ex}
    \tcboxmath[noborderbox]{
    \vP \coloneqq 
    \lambda\, \mathbf{\tilde P}^{\DR} + (1-\lambda)\, \mathbf{\tilde P}^{\IR} 
    \in [0, 1)^{16\times 16}
    \;.
    }
    \label{eq:fullchainSI}
\end{equation}
We can also write the transition matrix in entry-wise form as
\begin{equation} 
    \Big[ P_{(A_1, A_2; R_1, R_2)\to (A_1', A_2'; R_1', R_2')} \Big] = 
    \lambda\ \Big[ 
    \overbrace{
        \underbrace{
            P_{A_1'R_1' \mid A_2R_2}^{\DR}
        }_{\substack{\text{player 1 acts}\\\text{as a donor and}\\\text{is assigned }A_1'R_1'}}
        \cdot 
        \underbrace{
            P_{A_2'R_2' \mid A_1R_1}^{\DR}
        }_{\substack{\text{player 2 acts}\\\text{as a donor and}\\\text{is assigned }A_2'R_2'}}  
    }^{\textrm{changes due to local games}}
    \Big] 
    + 
    \left(1-\lambda\right) 
    \Big[ 
    \overbrace{
        \delta_{A_2,A_2'} \!\!\!\!
        \underbrace{
            P_{R_1' \mid R_2}^{\IR}
        }_{\substack{\text{player 1 acts}\\\text{as a donor and}\\\text{is assigned }R_1'}} 
        \!\!\!\! \cdot \
        \delta_{A_1,A_1'} \!\!\!\!
        \underbrace{
            P_{R_2' \mid R_1}^{\IR}
        }_{\substack{\text{player 2 acts}\\\text{as a donor and}\\\text{is assigned }R_2'}} 
        \!\!\!\!
    }^{\textrm{changes due to global games}}    
    \Big]
    \;,
    \label{eq:fullchain}
\end{equation}
where $\delta$ denotes the Kronecker delta function ($\delta_{A,A'}=1$ if $A=A'$, $0$ otherwise).

Analysis of long-term game dynamics requires solving for the stationary distribution of the Markov chain associated with the transition probability matrix $\vP$ (\cref{eq:fullchainSI}). By considering the sign structure of $\vP$, we can deduce that, for positive probabilities of local play ($\lambda>0$) and the scoring norm ($\av=(1,0)$), the Markov chain converges to a unique stationary distribution, corresponding to the left dominant eigenvector of $\vP$:
\begin{itemize}
    \item If $\sDR{D}{G}>0$ and $\sDR{D}{B}=0$, then all entries in row 13 and column 13 of $\vP$ are positive. Thus state $(D,D;G,G)$ communicates with every other state, meaning that the chain is irreducible. Since entry (13, 13) is positive, the chain is also aperiodic. 
    
    If $\sDR{D}{G}=0$ and $\sDR{D}{B}>0$, then all entries in row 16 and column 16 of $\vP$ are positive. Thus $(D,D;B,B)$ communicates with every other state, meaning that the chain is irreducible. Since entry (16, 16) is positive, the chain is also aperiodic. 
    
    In either case, the Markov chain is irreducible and aperiodic, and so it converges to a unique stationary distribution.
    
    \item If $\sDR{D}{G}=\sDR{D}{B}=0$, then the matrix $\vP$ has the following block structure:
    \begingroup\renewcommand\arraystretch{1.5}
        \begin{equation*}
            \vP = \begin{bmatrix}
                \vP_{1,1}^{+} & \vP_{1,2} & \vP_{1,3} & \vP_{1,4}^{+}
                \\
                \mathbf{0} & \vP_{2,2}^{+} & \vP_{2,3} & \vP_{2,4}^{+}
                \\
                \mathbf{0} & \vP_{3,2} & \vP_{3,3}^{+} & \vP_{3,4}^{+}
                \\
                \mathbf{0} & \mathbf{0} & \mathbf{0} & \vP_{4,4}^{+}
            \end{bmatrix} \;,
        \end{equation*}
        where $\vP_{i,j}^{+}$ denotes a 4-by-4 matrix whose entries are strictly positive for all admissible parameter values, and $\vP_{i,j}$ (without a superscript) denotes a non-negative matrix that contains zeros for some parameter values. 
        
        In this case, the Markov chain associated with $\vP$ is reducible. However, the states corresponding to the fourth block---$(D,D;G,G)$, $(D,D;G,B)$, $(D,D;B,G)$, and $(D,D;B,B)$---form a closed communicating class: transitions out of the fourth block are impossible, whereas transitions from the first three blocks to the fourth block occur with positive probabilities (through $\vP_{1,4}^{+}$, $\vP_{2,4}^{+}$, and $\vP_{3,4}^{+}$). All states outside of the fourth block are therefore transient. Since $\vP_{4,4}^{+}$ is strictly positive, the Markov chain restricted to this class is irreducible and aperiodic. Therefore, the full chain associated with $\vP$ converges to a unique stationary distribution supported on the fourth block.
    \endgroup
\end{itemize}

\subsection{ODE representation of status dynamics in a monomorphic population}
\label{sec:markovtoode}

For ease of computation and interpretation, we seek to describe the dynamics of the marginal distribution over single-player statuses $\mathcal{S} = \{CG, CB, DG,$ $DB\}$. We let $\vh(k) = \left(h_{CG}(k),h_{CB}(k),h_{DG}(k),h_{DB}(k)\right)$ denote the distribution of statuses in round $k$, where $h_{AR}(k)$ is the probability that a randomly selected player is of status $AR\in\mathcal{S}$. Below, we show that the ODE representation reported in the main text (\cref{eq:hsystem}) can be derived as a continuous-time limit of the Markov-chain representation described in the previous section.

Under the mean-field approximation, the statuses of any two players are assumed to be independent. Hence the probability that a focal player interacts with a co-player of status $AR$ in round $k$ is $h_{AR}(k)$. To obtain the expected distribution of statuses in round $k+1$, we average the pairwise transition probabilities over the distribution of co-player statuses (color added for emphasis only):
\begin{equation}
    \mathbb{E}\left[h_{{\color{OliveGreen}AR}}(k+1) \mid \vh(k)\right] = \sum_{A_1R_1\in\mathcal{S}} \sum_{A_2R_2\in\mathcal{S}} h_{A_1R_1}(k)\ h_{A_2R_2}(k) \sum_{A_2'R_2'\in\mathcal{S}} P_{(A_1,A_2;R_1,R_2) \to ({\color{OliveGreen}A},A_2';{\color{OliveGreen}R},R_2')} \;.
    \label{eq:stationaritySI}
\end{equation}
Substituting \cref{eq:fullchain} into \cref{eq:stationaritySI} yields
\begin{equation*}
    \begin{split}
        \mathbb{E}\left[h_{{\color{OliveGreen}AR}}(k+1) \mid \vh(k)\right] & = \lambda\ \overbrace{ \sum_{A_1R_1} \sum_{A_2R_2} h_{A_1R_1}(k)\ h_{A_2R_2}(k) \sum_{A_2'R_2'} P_{{\color{OliveGreen}AR} \mid A_2R_2}^{\DR} \cdot P_{A_2'R_2'\mid A_1R_1}^{\DR} }^{(1)}
        \\
        & \qquad\qquad\qquad\qquad + \left(1-\lambda\right) \underbrace{ \sum_{A_1R_1} \sum_{A_2R_2} h_{A_1R_1}(k)\ h_{A_2R_2}(k) \sum_{A_2'R_2'} \delta_{A_2,A_2'} P_{{\color{OliveGreen}R} \mid R_2}^{\IR} \cdot \delta_{A_1,{\color{OliveGreen}A}}\,P_{R_2' \mid R_1}^{\IR} }_{(2)} \;.
    \end{split}
\end{equation*}
Since $\sum_{A_2'R_2'} P_{A_2'R_2' \mid A_1R_1}^{\DR} = 1$ and $\sum_{A_1R_1} h_{A_1R_1} = 1$, expression (1) simplifies to
\begin{equation*}
    (1) = \sum_{A_2R_2} h_{A_2R_2}(k) P_{{\color{OliveGreen}AR} \mid A_2R_2}^{\DR} \;. 
\end{equation*}
Since $\sum_{A_2'R_2'} \delta_{A_2,A_2'} P_{R_2' \mid R_1}^{\IR} = 1$ and $\sum_{A_1R_1} \delta_{A_1,{\color{OliveGreen}A}} h_{A_1R_1} = \sum_{R_1} h_{{\color{OliveGreen}A}R_1}$, expression (2) simplifies to
\begin{equation*}
    \begin{split}
        (2) & = \sum_{A_2R_2} h_{A_2R_2}(k) P_{{\color{OliveGreen}R} \mid R_2}^{\IR} \sum_{A_1R_1} \delta_{A_1,{\color{OliveGreen}A}} h_{A_1R_1}(k) \sum_{A_2'R_2'} \delta_{A_2,A_2'} P_{R_2' \mid R_1}^{\IR}
        = \sum_{R_1\in\{G,B\}}  h_{{\color{OliveGreen}A}R_1}(k) \sum_{A_2R_2\in\mathcal{S}} h_{A_2R_2}(k) P_{{\color{OliveGreen}R} \mid R_2}^{\IR}  \;.
    \end{split}
\end{equation*}
Hence, we obtain
\begin{equation} 
        \mathbb{E}\left[h_{{\color{OliveGreen}AR}}(k+1) \mid \vh(k)\right] = 
            \lambda \underbrace{ 
                \sum_{A'R'} h_{A'R'}(k) P_{{\color{OliveGreen}AR} \mid A'R'}^{\DR}
            }_{\textrm{changes due to local games}}
            + \left(1-\lambda\right) \underbrace{ 
                h_{{\color{OliveGreen}A}\bullet}(k) \sum_{R'} h_{\bullet R'}(k) P_{{\color{OliveGreen}R} \mid R'}^{\IR}
                }_{\textrm{changes due to global games}} \;,
\end{equation}
where we have used the notations $h_{A\bullet} \coloneqq \sum_{R\in\{G,B\}} h_{AR}$ and $h_{\bullet R} \coloneqq \sum_{A\in\{C,D\}} h_{AR}$. Note that this update rule is nonlinear in $\vh$ due to the dependence of the second term on $h_{A\bullet}$ and $h_{\bullet R}$. In matrix form, the corresponding expected increment is
\begin{equation}
    \mathbb{E}\left[\vh(k+1) - \vh(k) \mid \vh(k)\right] = - \vh(k) + \lambda \underbrace{ \vh(k)\, \vP^{\DR} }_{ \substack{\textrm{changes due to}\\\textrm{local games}} } + \left(1-\lambda\right) \underbrace{ \vh_{A\bullet}(k) \otimes \left( \vh_{\bullet R}(k)\, \vP^{\IR} \right) }_{ \textrm{changes due to global games} } \;,
    \label{eq:stationaritymatrixSI}
\end{equation} 
where, as in \textit{\nameref{sec:matmethods}}, $\otimes$ denotes the Kronecker product; $\vh_{A\bullet} \coloneqq ( h_{C\bullet}, h_{D\bullet} )$ and $\vh_{\bullet R} \coloneqq ( h_{\bullet G}, h_{\bullet B} )$ are the marginal distributions over actions and reputations, respectively; and
\begin{equation*} \renewcommand{\arraystretch}{1.5}
    \vP^{\DR} 
    \coloneqq 
    \begin{bmatrix}
        P_{CG \mid CG}^{\DR} & P_{CB \mid CG}^{\DR} & P_{DG \mid CG}^{\DR} & P_{DB \mid CG}^{\DR}
        \\
        P_{CG \mid CB}^{\DR} & P_{CB \mid CB}^{\DR} & P_{DG \mid CB}^{\DR} & P_{DB \mid CB}^{\DR}
        \\
        P_{CG \mid DG}^{\DR} & P_{CB \mid DG}^{\DR} & P_{DG \mid DG}^{\DR} & P_{DB \mid DG}^{\DR}
        \\
        P_{CG \mid DB}^{\DR} & P_{CB \mid DB}^{\DR} & P_{DG \mid DB}^{\DR} & P_{DB \mid DB}^{\DR}
    \end{bmatrix} 
    \qquad\text{and}\qquad
    \vP^{\IR} 
    \coloneqq \begin{bmatrix} 
        P_{G \mid G}^{\IR} & P_{B \mid G}^{\IR}
        \\ 
        P_{G \mid B}^{\IR} & P_{B \mid B}^{\IR}
    \end{bmatrix} \;,
\end{equation*}
as defined in \cref{eq:pDRgeneral,eq:pIRgeneral} in \textit{\nameref{sec:matmethods}}. Taking the continuous-time limit of the difference equation in \cref{eq:stationaritymatrixSI} yields the ODE system reported in the main text (\cref{eq:hsystem}). At equilibrium, the status distribution $\vh$ satisfies
\begin{equation}
    \vh = \lambda\, \vh \vP^{\DR}  + \left(1-\lambda\right) \vh_{A\bullet} \otimes \left( \vh_{\bullet R}\, \vP^{\IR} \right) \;,
    \label{eq:consistency-dr-repupdate}
\end{equation}
which we refer to as a consistency equation.

\subsection{Consistency equations for invasion analysis}
\label{sec:invasioneqSI}

Next, we consider a single mutant with strategy pair $({\svDR}', {\svIR}')$ introduced to a neighborhood $\mathcal{N}$ within an otherwise monomorphic population with strategy pair $(\svDR, \svIR)$. We refer to the resident-type players inside the mutant's neighborhood as \textit{resident neighbors} and those outside as \textit{resident non-neighbors} (\cref{fig:schematic}C).

Our model assumes a global population of size $N$ that is structured into neighborhoods of size $n$, with $n\ll N$. In the limit $N\to\infty$, the mutant is vanishingly rare relative to the global population, and each neighborhood is negligibly small. Consequently, resident non-neighbors interact with members of neighborhood $\mathcal{N}$ (the mutant and the resident neighbors) with negligible probability, and so their status dynamics are unaffected by the mutant. However, because local games occur in finite neighborhoods, the mutant affects the status dynamics (and fitness) of resident neighbors. 

Based on these assumptions, we can write down a system of consistency equations for invasion analysis. For ease of interpretation, we denote the three classes using both shorthand notations and colors: $\RNN$ for resident non-neighbors, $\R$ for resident neighbors, and $\M$ for mutants. We define separate status distributions for the three classes: $\vh^{\RNN} \coloneqq (h_{CG}^{\RNN},h_{CB}^{\RNN},h_{DG}^{\RNN},h_{DB}^{\RNN})$ for resident non-neighbors, $\vh^{\R} \coloneqq(h_{CG}^{\R},h_{CB}^{\R},h_{DG}^{\R},h_{DB}^{\R})$ for resident neighbors, and $\vh^{\M} \coloneqq (h_{CG}^{\M},h_{CB}^{\M},h_{DG}^{\M},h_{DB}^{\M})$ for mutants.

Similarly to before, we denote the marginal distributions over actions and reputations by $\vh^{X}_{\bullet R} \coloneqq \left( h^{X}_{\bullet G},\, h^{X}_{\bullet B} \right) = ( h^{X}_{CG} + h^{X}_{DG},$ $h^{X}_{CB} + h^{X}_{DB} ) \;$ and $\vh^{X}_{A\bullet} \coloneqq \left( h^{X}_{C\bullet},\, h^{X}_{D\bullet} \right) = \left( h^{X}_{CG} + h^{X}_{CB},\, h^{X}_{DG} + h^{X}_{DB} \right)$, respectively, for $X\in\{\RNN,\R,\M\}$.
We also define
\begingroup\renewcommand*{\arraystretch}{1.7} 
\begin{equation}
    \begin{split}
    \vP^{\DR,\, X} 
    \coloneqq \begin{bmatrix}
        P^{\DR,\, X}_{CG \mid CG} & P^{\DR,\, X}_{CB \mid CG} & P^{\DR,\, X}_{DG \mid CG} & P^{\DR,\, X}_{DB \mid CG}
        \\
        P^{\DR,\, X}_{CG \mid CB} & P^{\DR,\, X}_{CB \mid CB} & P^{\DR,\, X}_{DG \mid CB} & P^{\DR,\, X}_{DB \mid CB}
        \\
        P^{\DR,\, X}_{CG \mid DG} & P^{\DR,\, X}_{CB \mid DG} & P^{\DR,\, X}_{DG \mid DG} & P^{\DR,\, X}_{DB \mid DG}
        \\
        P^{\DR,\, X}_{CG \mid DB} & P^{\DR,\, X}_{CB \mid DB} & P^{\DR,\, X}_{DG \mid DB} & P^{\DR,\, X}_{DB \mid DB}
    \end{bmatrix}
    \qquad\text{and}\qquad
    \vP^{\IR,\, X}
    & \coloneqq \begin{bmatrix} 
        P^{\IR,\, X}_{G \mid G} & P^{\IR,\, X}_{B \mid G}
        \\ 
        P^{\IR,\, X}_{G \mid B} & P^{\IR,\, X}_{B \mid B}
    \end{bmatrix} 
    \;,
\end{split}
\label{eq:pdririnvasion}
\end{equation}
\endgroup
where, for each class $X\in\{\RNN,\R,\M\}$, $\vP^{\DR,X}$ and $\vP^{\IR,X}$ denote the matrices $\vP^{\DR}$ (\cref{eq:pDRgeneral}) and $\vP^{\IR}$ (\cref{eq:pIRgeneral}) evaluated with the corresponding local and global strategies (i.e., $(\svDR, \svIR)$ for $\RNN$ and $\R$, $({\svDR}', {\svIR}')$ for $\M$), respectively. Here $P^{\DR,\, X}_{A'R' \mid AR}$ denotes the probability that a focal player in class $X\in\{\RNN,\R,\M\}$ is assigned status $A'R'$ after a local game against a co-player of status $AR$; and $P^{\IR,\, X}_{R' \mid R}$ denotes the probability that a focal player in class $X$ is assigned reputation $R'$ after a global game against a co-player with reputation $R$.

Then, the status distributions $\vh^{\RNN}$, $\vh^{\R}$, and $\vh^{\M}$ satisfy the following system of consistency equations at equilibrium:
\begin{equation} 
    \label{eq:invasioneq}
    \raisetag{-0.6ex}
    \tcboxmath[noborderbox]{
    \begin{aligned}
        \vh^{\RNN} & = \lambda\, \vh^{\RNN}\, \vP^{\DR,\,\RNN}  + \left(1-\lambda\right) \vh_{A\bullet}^{\RNN} \otimes \left( \vh_{\bullet R}^{\RNN}\, \vP^{\IR,\,\RNN} \right) \;,
        \\
        \vh^{\R} & = \lambda \left(\dfrac{1}{n-1}\,\vh^{\M} + \dfrac{n-2}{n-1}\,\vh^{\R}\right) \vP^{\DR,\,\R}  + \left(1-\lambda\right) \vh_{A\bullet}^{\R} \otimes \left( \vh_{\bullet R}^{\RNN}\, \vP^{\IR,\,\R} \right) \;,
        \\
        \vh^{\M} & = \lambda\, \vh^{\R}\, \vP^{\DR,\,\M}  + \left(1-\lambda\right) \vh_{A\bullet}^{\M} \otimes \left( \vh_{\bullet R}^{\RNN}\, \vP^{\IR,\,\M} \right) \;,
    \end{aligned}
    }
\end{equation}
as reported in \cref{eq:invasioneqRNN,eq:invasioneqR,eq:invasioneqM} in \textit{\nameref{sec:matmethods}}.

In component form, the consistency equations for the resident non-neighbors are
\begin{equation}
    h^{\RNN}_{AR}
    =
    \lambda
    \sum_{A'R'} h^{\RNN}_{A'R'}\,P^{\DR,\,\RNN}_{AR\mid A'R'}
    +
    \left(1-\lambda\right)
    h^{\RNN}_{A\bullet} \sum_{R'} h^{\RNN}_{\bullet R'} \,P^{\IR,\,\RNN}_{R\mid R'}
    \;.
    \label{eq:consistencyRNNSI}
\end{equation}
The consistency equations for the resident neighbors are
\begin{equation}
    h^{\R}_{AR}
    =
    \lambda
    \sum_{A'R'}
    h^{\R/\M}_{A'R'}(n)
    \,P^{\DR,\,\R}_{AR\mid A'R'}
    +
    \left(1-\lambda\right)
    h^{\R}_{A\bullet}
    \sum_{R'}
    h^{\RNN}_{\bullet R'}
    \,P^{\IR,\,\R}_{R\mid R'}
    \;,
    \label{eq:consistencyRSI}
\end{equation}
where
\begin{equation}
    h_{AR}^{\R/\M}\left(n\right) \coloneqq \dfrac{1}{n-1}\, h^{\M}_{AR} +  \dfrac{n-2}{n-1} \, h^{\R}_{AR}
\end{equation}
is the probability that, in a local game, a focal resident neighbor in a neighborhood of size $n$---containing one mutant and $n-1$ resident neighbors, including the focal player---interacts with a co-player in $AR$ status.

Finally, the consistency equations for the mutant are:
\begin{equation}
    h^{\M}_{AR}
    =
    \lambda
    \sum_{A'R'}
    h^{\R}_{A'R'}
    \,P^{\DR,\,\M}_{AR\mid A'R'}
    +
    \left(1-\lambda\right)
    h^{\M}_{A\bullet}
    \sum_{R'}
    h^{\RNN}_{\bullet R'}
    \,P^{\IR,\,\M}_{R\mid R'}
    \;.
    \label{eq:consistencyMSI}
\end{equation}

\subsection{Long-term average payoffs}

In both local and global games, the long-term average cooperation rate of a focal player depends on both their own strategy and the equilibrium status distribution among potential co-players. We denote by $\gamma_{X\to Y}^{\DR}$ the average cooperation rate of a class $X$ player with a class $Y$ player in a local game:
\begin{align*}
    \gamma_{\RNN\,\to\,\RNN}^{\DR} & 
    \coloneqq \svDRtX{\RNN} \cdot \vh^{\ast,\,\RNN} 
    \;,
    \\
    \gamma_{\R\,\to\,\R}^{\DR} & 
    \coloneqq \svDRtX{\R} \cdot \vh^{\ast,\,\R} 
    \;,
    \\
    \gamma_{\R\,\to\,\M}^{\DR} & \coloneqq 
    \svDRtX{\R} \cdot \vh^{\ast,\,\M} 
    \;,
    \\
    \gamma_{\M\,\to\,\R}^{\DR} & \coloneqq 
    \svDRtX{\M} \cdot \vh^{\ast,\,\R} 
    \;,
\end{align*}
where $\cdot$ denotes the dot product and $\vh^{\ast,\,X} \coloneqq \big( h^{\ast,\,X}_{CG},\,h^{\ast,\,X}_{CB},\,h^{\ast,\,X}_{DG},\, h^{\ast,\,X}_{DB} \big)$ denotes the equilibrium status distribution in class $X$. Similarly, we denote by $\gamma_{X\to Y}^{\IR}$ the long-term average cooperation rate of a class-$X$ player with a class-$Y$ player in a global game:
\begin{align*}
    \gamma_{\RNN\,\to\,\RNN}^{\IR} & 
    \coloneqq \svIRtX{\RNN} \cdot \vh^{\ast,\,\RNN}_{\bullet R}
    \;,
    \\
    \gamma_{\R\,\to\,\RNN}^{\IR} & 
    \coloneqq \svIRtX{\R} \cdot \vh^{\ast,\,\RNN}_{\bullet R}
    \;,
    \\
    \gamma_{\RNN\,\to\,\R}^{\IR} & 
    \coloneqq \svIRtX{\RNN} \cdot \vh^{\ast,\,\R}_{\bullet R}
    \;,
    \\
    \gamma_{\M\,\to\,\RNN}^{\IR} & 
    \coloneqq \svIRtX{\M} \cdot \vh^{\ast,\,\RNN}_{\bullet R}
    \;,
    \\
    \gamma_{\RNN\,\to\,\M}^{\IR} & 
    \coloneqq \svIRtX{\RNN} \cdot \vh^{\ast,\,\M}_{\bullet R}
    \;,
\end{align*}
where $\vh^{\ast,\,X}_{\bullet R} \coloneqq \big( h^{\ast,\,X}_{\bullet G},\, h^{\ast,\,X}_{\bullet B} \big) = \big( h^{\ast,\,X}_{CG} + h^{\ast,\,X}_{DG},\, h^{\ast,\,X}_{CB} + h^{\ast,\,X}_{DB} \big)$ denotes the equilibrium distribution of reputations in class $X$.

The long-term average payoff of class $X$, denoted $\pi_X$, is given by the expected benefit received minus expected cost paid per round, weighted by the probabilities of local and global games:
\begin{equation}
    \begin{split}
        \pi_{\RNN} & = \lambda \left(b-c\right) \, \gamma_{\RNN\,\to\,\RNN}^{\DR} + \left(1-\lambda\right) \left(b-c\right) \gamma_{\RNN\,\to\,\RNN}^{\IR} \;,
        \\
        \pi_{\R} 
        & = \lambda 
        \left(
        \dfrac{1}{n-1} \left(b\, \gamma_{\M\,\to\,\R}^{\DR}
        -c \, \gamma_{\R\,\to\,\M}^{\DR} \right)
        + \dfrac{n-2}{n-1}\, \left(b-c\right) \, \gamma_{\R\,\to\,\R}^{\DR}
        \right) 
        + \left(1-\lambda\right) \left(b\,\gamma_{\RNN\,\to\,\R}^{\IR} -c\,\gamma_{\R\,\to\,\RNN}^{\IR} \right)  \;,
        \\
        \pi_{\M} & = \lambda \left(b\, \gamma_{\R\,\to\,\M}^{\DR} -c\, \gamma_{\M\,\to\,\R}^{\DR} \right)  + \left(1-\lambda\right) \left(b\, \gamma_{\RNN\,\to\,\M}^{\IR} - c\,\gamma_{\M\,\to\,\RNN}^{\IR} \right)  \;,
    \end{split}
    \label{eq:fitnesseqSI}
\end{equation}
as reported in \cref{eq:fitnesseq} in the main text.

\section{Invasibility analysis for TFT--DISC}
\label{sec:analytical}

Our goal is to derive conditions under which tit-for-tat discriminators (TFT--DISC) resist invasion by unconditional cooperators and defectors---and thereby solve the scoring dilemma. To this end, we now apply the consistency equations derived in \cref{sec:invasioneqSI} to cases where the resident strategy is TFT--DISC and the mutant strategy is ALLC or ALLD. As a reminder, TFT--DISC players act as TFT players in local games and as DISC players in global games: that is, $\svDRX{\RNN}=\svDRX{\R}=(1,1,0,0)$ and $\svIRX{\RNN}=\svIRX{\R}=(1,0)$.

\subsection{Analytical expression for $\tilde\lambda_{\ALLD}^*$}

We first consider an ALLD mutant, i.e., with $\svDRX{\M}=\svDRX{\ALLD}=(0,0,0,0)$ and $\svIRX{\M}=\svIRX{\ALLD}=(0,0)$. Solving the system of consistency equations (\cref{eq:invasioneq}) with this mutant yields a unique solution given by
\begingroup
\renewcommand\arraystretch{1.5}
\begin{equation*}
    \begin{split}
        \vh^{*,\,\RNN} = \vh^{*,\,\R} & = 
        \begin{bmatrix} 
            0 & 0 & \dfrac{\assess}{\exec+\left(1-\exec\right) \left(\lambda + \left(1-\lambda\right) 2\assess\right)} & 1-\dfrac{\assess}{\exec+\left(1-\exec\right) \left(\lambda + \left(1-\lambda\right) 2\assess\right)} 
        \end{bmatrix}
        \\
        \vh^{*,\,\ALLD} & = 
        \begin{bmatrix}
            0 & 0 & \assess & 1-\assess
        \end{bmatrix} \;.
    \end{split}
\end{equation*}
\endgroup
Substituting these into \cref{eq:fitnesseqSI} yields the following long-term average payoffs:
\begin{equation*}
    \begin{split}
        \pi_{\RNN} = \pi_{\R} & = \left(1-\exec\right) \dfrac{\left(b-c\right) \assess \left(1-\lambda\right)}{\exec + \left(1-\exec\right) \left(\lambda + \left(1-\lambda\right) 2\assess\right)} \;,
        \\
        \pi_{\ALLD} & = \left(1-\exec\right) b \assess \left(1-\lambda\right) \;.
    \end{split}
\end{equation*}

An ALLD mutant can invade a TFT--DISC resident population whenever $\pi_{\R} < \pi_{\ALLD}$. For $\lambda\in[0,1]$, this condition simplifies to $1 > \lambda > \tilde \lambda_{\ALLD}^*$, where 
\begin{equation}
    \raisetag{-0.6ex}
    \tcboxmath[noborderbox]{
        \tilde \lambda_{\ALLD}^* = 1-\dfrac{1}{\left(b/c\right)(1-2 \assess)(1-\exec)} \;,
    }
    \label{eq:lambdaALLDstarSI}
\end{equation}
as reported in \cref{eq:lambdaALLDstar} in \textit{\nameref{sec:matmethods}}.

\subsection{Analytical expression for $\tilde\lambda_{\ALLC}^*$}

Next, we consider an ALLC mutant, i.e., with $\svDRX{\M}=\svDRX{\ALLC}=(1,1,1,1)$ and $\svIRX{\M}=\svIRX{\ALLC}=(1,1)$. Solving the consistency equations (\cref{eq:invasioneq}) with this mutant yields a unique solution given by
\begingroup\footnotesize
\renewcommand\arraystretch{1.5}
\begin{equation*}
    \begin{split}
        \vh^{*,\,\RNN} & = 
        \begin{bmatrix} 
            0 & 0 & \dfrac{\assess}{\exec+\left(1-\exec\right) \left(\lambda + \left(1-\lambda\right) 2\assess\right)} & 1-\dfrac{\assess}{\exec+\left(1-\exec\right) \left(\lambda + \left(1-\lambda\right) 2\assess\right)} 
        \end{bmatrix}
        \\
        \vh^{*,\,\R} & = 
            \dfrac{1}{\left(1 + \left(n-2\right)\exec\right) \left(\exec + \left(1-\exec\right) \left(\lambda + \left(1-\lambda\right)2\assess\right)\right)}
        \\
        & \qquad\qquad\qquad\qquad 
        \times 
        \begin{bmatrix}
            \left(1 - \exec\right)^2 \left(\assess + \left(1 - 2 \assess\right) \left(\exec + \left(1 - \exec\right) 2 \assess\right) \lambda + \left(1 - 2 \assess\right)^2 \left(1 - \exec\right) \lambda^2 \right) 
            \\[1pt]
            \left(1 - \exec\right)^2 \left(\assess + \left(1 - 2 \assess\right) \exec + \left(1 - 2 \left(1 - \assess\right)\left(\exec + \left(1 - \exec\right) 2 \assess\right)\right) \lambda - \left(1 - 2 \assess\right)^2 \left(1 - \exec\right) \lambda^2 \right) 
            \\[1pt]
            \assess \left(n - \exec\right) \exec 
            \\[1pt]
            \left(n - \exec\right) \exec \left(\assess + \left(1 - 2 \assess\right) \exec + \left(1 - 2 \assess\right)\left(1 - \exec\right) \lambda \right)
        \end{bmatrix}^{T}
        \\
        \vh^{*,\,\ALLC} & = \begin{bmatrix}
            \left(1 - \exec\right)\bigl(1 - \assess - \exec \left(1 - 2 \assess\right)\left(1 - \lambda\right)\bigr) 
            \\[1pt]
            \left(1 - \exec\right)\bigl(\assess + \exec \left(1 - 2 \assess\right)\left(1 - \lambda\right)\bigr) 
            \\[1pt]
            \exec \bigl(\assess + \left(1 - \exec\right)\left(1 - 2 \assess\right)\left(1 - \lambda\right)\bigr) 
            \\[1pt]
            \exec \bigl(1 - \assess - \left(1 - \exec\right)\left(1 - 2 \assess\right)\left(1 - \lambda\right)\bigr)
        \end{bmatrix}^{T}
    \end{split}
\end{equation*}
\endgroup

Substituting these into \cref{eq:fitnesseqSI} yields the following long-term average payoffs:
\begingroup\footnotesize
\begin{equation*}
    \begin{split}
        \pi_{\RNN} & = \left(1-\exec\right) 
        \dfrac{\left(b - c\right)\assess\left(1 - \lambda\right)}{\exec + \left(1 - \exec\right)\left(\lambda + \left(1 - \lambda\right) 2 \assess\right)} 
        \;,
        \\
        \pi_{\R} & = \left(1-\exec\right) \Biggl(
        \dfrac{\left(b - c\right)\assess\left(1 - \lambda\right)}{\exec + \left(1 - \exec\right)\left(\lambda + \left(1 - \lambda\right) 2 \assess\right)}
        \\
        & \qquad\qquad\qquad\qquad + \dfrac{\left(n - 1\right)\left(b\left(1 + \left(1 - 2 \assess\right)\left(1 - \exec\right)^2\right) - c\left(1 - \exec\right)\right) - b\left(n - 2\right)\left(1 - \exec\right)\exec}{\left(n - 1\right)\left(1 + \left(n - 2\right)\exec\right)} \lambda
        - \dfrac{b\left(1 - 2 \assess\right)\left(1 - \exec\right)^2}{1 + \left(n - 2\right)\exec} \lambda^2
        \Biggr) \;,
        \\
        \pi_{\ALLC} & = 
        \left(1 - \exec\right)
        \bigl(b\left(1 - \exec - \assess\left(1 - 2 \exec\right)\left(1 - \lambda\right)\right) - c\bigr) \;.
    \end{split}
\end{equation*}
\endgroup

An ALLC mutant can invade a TFT--DISC resident population whenever $\pi_{\R} < \pi_{\ALLC}$. We consider the difference
\begin{equation}
    \begin{split}
        \pi_{\R} - \pi_{\ALLC}
        & = \dfrac{
            c
            }{
            \left(n - 1\right)\left(1 + \left(n - 2\right)\exec\right)\left(\exec + \left(1 - \exec\right)\left(\lambda + \left(1 - \lambda\right) 2 \assess \right)\right)
            } 
            \cdot P\left(\lambda\right)
    \end{split} \;,
    \label{eq:payoffdiffALLC}
\end{equation}
where $P\left(\lambda\right) \coloneqq a_3 \lambda^3 + a_2 \lambda^2 + a_1 \lambda + a_0$ is a cubic polynomial in $\lambda$ with coefficients
\begin{equation*}
\begin{split}
    a_3 & = - \left(\dfrac{b}{c}\right) \left(n-1\right) \left(1-2 \assess\right)^2 \left(1-\exec \right)^4 \;,
    \\
    a_2 & = \left(1-2 \assess \right) \left(1-\exec \right)^2 \bigg(
        -\left(n-1\right) \left(1-\exec \right) 
        + \left(\dfrac{b}{c}\right) \Big(
            \left(2-5 \assess \right) \left(n-1\right)
            -\exec  \left(n^2-(1-\assess ) (n-12) (n-1)-9 n+7\right)
            \\ & \quad \quad \quad \quad \quad \quad \quad \quad \quad \quad \quad \quad \quad \quad \quad \quad \quad \quad \quad \quad \quad \quad  
            +\exec^2 \left(n \left(n-2\right)-\left(1-2 \assess \right) \left(n-4\right) \left(n-1\right)\right)
        \Big)
    \bigg) \;,
    \\
    a_1 & = (1-\exec) \bigg(
        \left(n-1\right) \left(\left(1-\exec\right) \left(\left(n-3\right) \exec+1\right)-\assess  \left(\exec \left(\left(3-n\right) 2\exec+n-8\right)+3\right)\right)
    \\ & \quad \quad \quad \quad \quad \quad    
    - \left(\dfrac{b}{c}\right) \Big(
        \left(8 \assess ^2-6 \assess +1\right) \left(n-1\right)
        \\ & \quad \quad \quad \quad \quad \quad \quad \quad \quad \quad \quad    
        + \exec \left(n^2+4 \assess ^2 \left(n-8\right) \left(n-1\right)-2 \assess  \left(\left(n-13\right) n+13\right)-7 n+6\right)
        \\ & \quad \quad \quad \quad \quad \quad \quad \quad \quad  \quad \quad   
        +\exec^2 \left(40 \assess -4 \assess ^2 \left(n-1\right) \left(3 n-11\right)+4 \assess  \left(2 n-11\right) n-2n \left(n-5\right)-9\right)
        \\ & \quad \quad \quad \quad \quad \quad \quad \quad \quad \quad \quad + \exec^3 \left(2 \assess -1\right) \left(5 \left(2 \assess -1\right) + n \left(-14 \assess +\left(4 \assess -1\right) n+5\right)\right)
        \Big) 
    \bigg)
    \;,
    \\
    a_0 & = \left(1-\left(\dfrac{b}{c}\right)\left(1-2 \assess \right) \left(1-\exec \right)\right) \left(n-1\right) \left(1+\left(n-2\right)\exec\right)  \left(1-\exec \right) \left(\exec + \assess\left(1-2\exec\right) \right) 
    \;.
\end{split}
\end{equation*}
as reported in \cref{eq:coefficients} in \textit{\nameref{sec:matmethods}}.
Since the prefactor in \cref{eq:payoffdiffALLC} is strictly positive for all admissible parameter values, we have $\sgn\left(\pi_{\R} - \pi_{\ALLC}\right) = \sgn\left(P(\lambda)\right)$. In other words, ALLC can invade whenever $P(\lambda) < 0$.

In \textit{\nameref{sec:matmethods}}, we have established that if  
\begin{equation}
    \dfrac{b}{c}>\left(\dfrac{b}{c}\right)^* 
    \quad \textrm{and}\quad
    n < n^* \coloneqq \dfrac{
        2 - \left(b/c\right) \left(3-2 \exec\right) - \sqrt{\left(b/c\right) \left(- 4 \left(1-\exec\right) +  \left(b/c\right) \left(5 - 4 \left(2-\exec\right) \exec\right) \right)}
        }{
        2 \left(1-\left(b/c\right) \left(1-\exec\right)\right)
        } \;,
\end{equation}
as reported in \cref{eq:P1positivecondition}, then the polynomial $P(\lambda)$ has three real roots and the second-largest real root root corresponds to $\tilde\lambda_{\ALLC}^*$, the critical value of $\lambda$ \textit{below} which ALLC can invade.
In this case, we can use the trigonometric solution to the cubic to obtain an analytical expression for $\lambda_{\ALLC}^*$.
We let
\begin{equation*}
    \begin{split}
        T \coloneqq \dfrac{3 a_3 a_1 - a_2^2}{3 a_3^2} \;, \qquad\qquad
        U \coloneqq \dfrac{2 a_2^3 - 9 a_3 a_2 a_1 + 27 a_3^2 a_0}{27 a_3^3} \;,
    \end{split}
\end{equation*}
where $a_0, a_1, a_2, a_3$ are the coefficients of $P(\lambda)$, and we define the angle
\begin{equation*}
    \phi \coloneqq \arccos \left(\dfrac{3U}{2T}\sqrt{-\dfrac{3}{T}}\right) \;, \quad \phi\in[0,\pi] \;.
\end{equation*}
Then the roots of the polynomial $P(\lambda)$ are given by
\begin{equation*}
    \lambda_k^* = 2\sqrt{-\dfrac{T}{3}} \cdot \cos \left(\dfrac{\phi + 2\pi k}{3}\right) - \dfrac{a_2}{3 a_3}, \quad k=0,1,2 \;.
\end{equation*}

Finally, we consider the relative ordering of $\lambda_k^*$. Since $\phi\in[0,\pi]$ by convention, the ordering of the cosine factor is $\cos (\frac{\phi + 2\pi \cdot 0}{3}) > \cos (\frac{\phi + 2\pi \cdot 2}{3}) > \cos (\frac{\phi + 2\pi \cdot 1}{3})$. Our desired root (second-largest real root) therefore corresponds to $k=2$:
\begin{equation}
    \raisetag{-0.6ex}
    \tcboxmath[noborderbox]{
        \tilde\lambda_{\ALLC}^* \coloneqq \lambda_2^* = 2\sqrt{-\dfrac{T}{3}} \cdot \cos \left(\dfrac{\phi + 4\pi}{3}\right) - \dfrac{a_2}{3 a_3} \;.
    }
    \label{eq:lambdaALLCstarSI}
\end{equation}

\subsection{Condition under which TFT--DISC solves the scoring dilemma for $n=2$: an alternative approach}

For $n=2$, we have shown in the main text that $\lambda_{\ALLD}^*>\lambda_{\ALLC}^*$ if and only if $b/c > \left(b/c\right)^* = 1/((1-2\assess)(1-\exec))$ (\cref{eq:bccondition}), using a combination of the intermediate value theorem and a sign analysis of the polynomial $P(\lambda)$ (see \textit{\nameref{sec:matmethods}}). 
Here we present an alternative approach based on local gradient comparison that proves the \textit{existence} of a parameter region that solves the scoring dilemma (i.e., region with $\Delta \lambda^* > 0$) in the neighborhood of the critical threshold $\left(b/c\right)^*$. While this result is weaker than the one reported in the main text, we include it here because it provides intuition for how the two curves in \cref{fig:fig3}A ($\lambda_{\ALLD}^*$ and $\lambda_{\ALLC}^*$) behave near the threshold.

At the critical benefit-to-cost ratio $\left(b/c\right)^*$, the critical probabilities $\tilde\lambda_{\ALLD}^*$ and $\tilde\lambda_{\ALLC}^*$ both cross zero (for any $n\geq 2$), since $\tilde\lambda_{\ALLD}^*|_{\left(b/c\right)^*} = 0$ and $P(0)|_{\left(b/c\right)^*} = 0$. The partial derivative of $\tilde\lambda_{\ALLD}^*$ with respect to $b/c$, evaluated at $b/c = (b/c)^*$, is
\begin{align*}
    \dfrac{\partial \tilde\lambda_{\ALLD}^*}{\partial \left(b/c\right)} \biggr\rvert_{\left(b/c\right)^*} 
    & = \dfrac{1}{\left(b/c\right)^*}
\end{align*}
by direct computation. The partial derivative of $\tilde\lambda_{\ALLC}^*$ with respect to $\left(b/c\right)$ can be obtained through implicit differentiation: since $\tilde\lambda_{\ALLC}^*(b/c)$ is defined implicitly by $P(b/c,\tilde\lambda_{\ALLC}^*\left(b/c\right)) = 0$ (\cref{eq:polynomial} in the main text),
\begin{align*}
    \dfrac{\partial \tilde\lambda_{\ALLC}^*}{\partial \left(b/c\right)} 
    & 
    = -\dfrac{\partial P/\partial (b/c)}{\partial P/\partial \lambda}
    \;.
\end{align*}
Evaluating this at $((b/c)^*, 0)$ for $n=2$ yields
\begin{align*}
    \dfrac{\partial \tilde\lambda_{\ALLC}^*}{\partial \left(b/c\right)} \biggr\rvert_{\left(b/c\right)^*} 
    & 
    = \dfrac{1}{\left(b/c\right)^*} \cdot \dfrac{\exec + \assess\left(1-2\exec\right)}{\left(b/c\right)^* \exec + \assess\left(1-2\exec\right)}
    \;,
\end{align*}
for $n=2$.
Since $(b/c)^*>1$ for $\assess\in(0,1/2)$ and $\exec\in(0,1/2)$, we have
\begin{equation*}
    \dfrac{\partial \tilde\lambda_{\ALLD}^*}{\partial \left(b/c\right)} \biggr\rvert_{\left(b/c\right)^*} > \dfrac{\partial \tilde\lambda_{\ALLC}^*}{\partial \left(b/c\right)} \biggr\rvert_{\left(b/c\right)^*} \;.
\end{equation*}
Therefore, for $n=2$, there exists $\delta>0$ such that $\tilde\lambda_{\ALLD}^* - \tilde\lambda_{\ALLC}^* > 0$ for all $\left(b/c\right) \in \left(\left(b/c\right)^*, \left(b/c\right)^* + \delta \right)$. Since $\lambda_{\ALLD}^* = \tilde\lambda_{\ALLD}^*$ and $\lambda_{\ALLC}^* = \tilde\lambda_{\ALLC}^*$ for $\left(b/c\right) > \left(b/c\right)^*$, it follows that $\Delta \lambda^* = \lambda_{\ALLD}^* - \lambda_{\ALLC}^*>0$ in the same interval.

\subsection{Effect of neighborhood size}
\label{sec:neffect}

In the main text, we have considered how the probability of local games ($\lambda$) affects the robustness of TFT--DISC against ALLC or ALLD. Here we consider the effect of neighborhood size ($n$), starting with an ALLC mutant (\cref{fig:SI-neffect1}A--D). A key observation is that when neighborhoods have finite size, the non-negligible frequency of play against the ALLC mutant allows the TFT--DISC resident neighbors to sustain cooperation in local games (\cref{fig:SI-neffect1}B; see also \cref{fig:fig2}B). In fact, if the neighborhood size is small enough, then the TFT--DISC resident neighbors can receive higher total payoffs than the ALLC mutant (\cref{fig:SI-neffect1}A), due to a payoff advantage in global games (\cref{fig:SI-neffect1}D). As the neighborhood size increases, however, resident neighbors cooperate less often with other resident neighbors, even as they continue to cooperate frequently with the ALLC mutant (\cref{fig:SI-neffect1}B). This undermines the reputation of the resident neighbors (\cref{fig:SI-neffect1}C), which in turn decreases their likelihood of receiving cooperation in global games (\cref{fig:SI-neffect1}D). As a result, the mutant eventually outearns the resident neighbors when neighborhood size is sufficiently large (\cref{fig:SI-neffect1}A). 

By contrast, when the mutant is ALLD, neighborhood size $n$ has no effect on the payoffs of any player (\cref{fig:SI-neffect1}E--H). This is because ALLD eliminates cooperation in local games: once an ALLD mutant initiates a cascade of defection in its neighborhood, cooperation cannot be restored locally, since players cannot cooperate by accident. Therefore, all local games in the neighborhood of an ALLD mutant converge to mutual defection, regardless of $n$ (\cref{fig:SI-neffect1}F). Local payoffs (\cref{fig:SI-neffect1}F), average reputations (\cref{fig:SI-neffect1}H), and consequently global payoffs (\cref{fig:SI-neffect1}G) are therefore all independent of $n$. 

\subsection{Effect of symmetric execution errors}
\label{sec:symmetric}

So far we have explored asymmetric execution errors, whereby a player intending to cooperate can accidentally defect, but the reverse is not possible. This assumption is standard in models of indirect reciprocity \cite{ohtsuki_how_2004}. An alternative assumption, commonly used in models of direct reciprocity \cite{sigmund_calculus_2010}, is that execution errors are symmetric, so that intended cooperation can be accidentally implemented as defection, and vice versa, with equal probability. Although accidental cooperation may be less realistic than accidental defection, this assumption provides a useful benchmark. 

We implement symmetric execution errors by modifying the strategy vectors (\cref{eq:errorsasymmetricSI}) as follows:
\begin{equation*}
    \begin{split}
        \svDRt & = \left(1-\exec\right)\svDR + \exec \left(\mathbf{1} - \svDR \right) \;,
        \\
        \svIRt & = \left(1-\exec\right)\svIR + \exec \left(\mathbf{1} - \svIR \right) \;.
    \end{split}
\end{equation*}
As before, we substitute these expressions into \cref{eq:pdririnvasion} to obtain a system of consistency equations (\cref{eq:consistencyRNNSI,eq:consistencyRSI,eq:consistencyMSI}), and we solve that system to obtain long-term average payoffs for resident non-neighbors, resident neighbors, and mutants. 

We find that the critical probabilities of local play, $\lambda_{\ALLD}^*$ and $\lambda_{\ALLC}^*$, coincide under symmetric execution errors: that is, $\Delta\lambda^* = 0$ regardless of parameters (\cref{fig:SI-symmetric}). Consequently, for any $\lambda\in[0,1]$, either ALLC or ALLD can invade TFT--DISC. Hence, TFT--DISC cannot solve the scoring dilemma when execution errors are symmetric.

\newpage
\section*{\Large Supplementary Figures}
\addcontentsline{toc}{section}{Supplementary Figures}

%%%%%%%%%%%%%%%%%%%%%%%%%%%
% SI FIGURES
%%%%%%%%%%%%%%%%%%%%%%%%%%%

\begin{figure}[h!]
    \centering
    \includegraphics[width=0.99\linewidth,trim={6.5in 0in 6.5in 0in},clip]{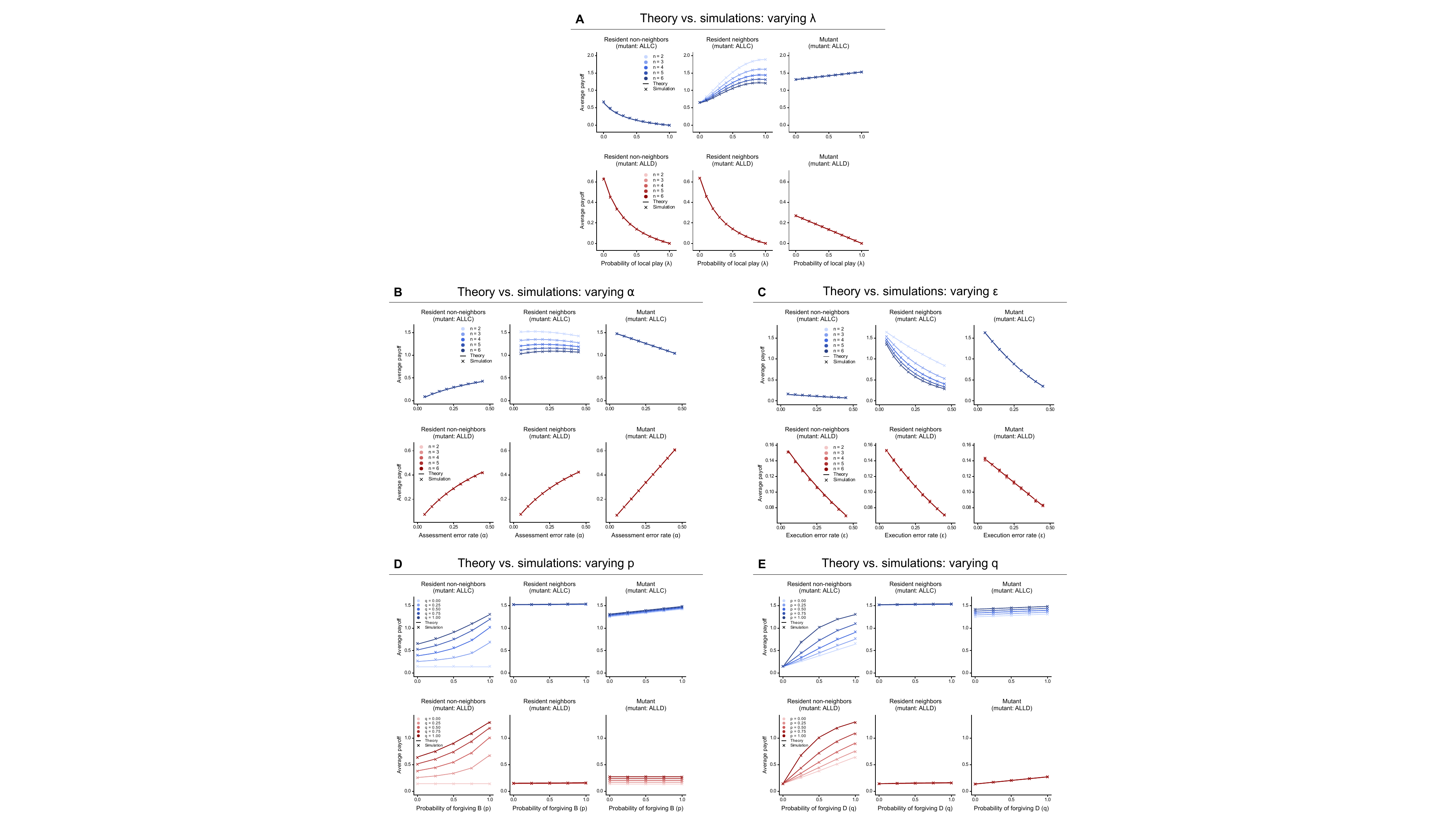}
    \caption{
        \textbf{Comparison between theoretical predictions and stochastic simulations.}
        Average payoffs by player class across a range of parameters: varying $\lambda$ (\textbf{A}), $\assess$ (\textbf{B}), $\exec$ (\textbf{C}), $p$ (\textbf{D}), and $q$ (\textbf{E}).
        Solid lines show theoretical predictions based on the consistency equations (\cref{eq:consistencyRNNSI,eq:consistencyRSI,eq:consistencyMSI});
        ``x'' marks show the long-term outcomes of stochastic simulations in finite populations.
        In each panel, columns correspond to player class: resident non-neighbors (left column), resident neighbors (middle column), and mutant (right column); rows correspond to mutant type: ALLC (top row) and ALLD (bottom row).
        Colors denote values of $n$ in \textbf{A--C}, values of $q$ in \textbf{D}, and values of $p$ in \textbf{E}.
        Theory and simulations are in close agreement across all parameter conditions.
        Simulation parameters: population size $N=120$, total $2\times 10^7$ rounds with $2\times 10^6$ of burn-in rounds.
        Other parameters: $b=3$, $c=1$, $\lambda=0.5$ (except in A), $\assess=\exec=0.1$ (except in \textbf{B} and \textbf{C}), $n=2$ (except in \textbf{A--C}). 
    }
    \label{fig:SI-comparison}
\end{figure}

\clearpage
\begin{figure}[p]
    \centering
    \includegraphics[width=0.91\linewidth,trim={1in -1in 1in -1in},clip]{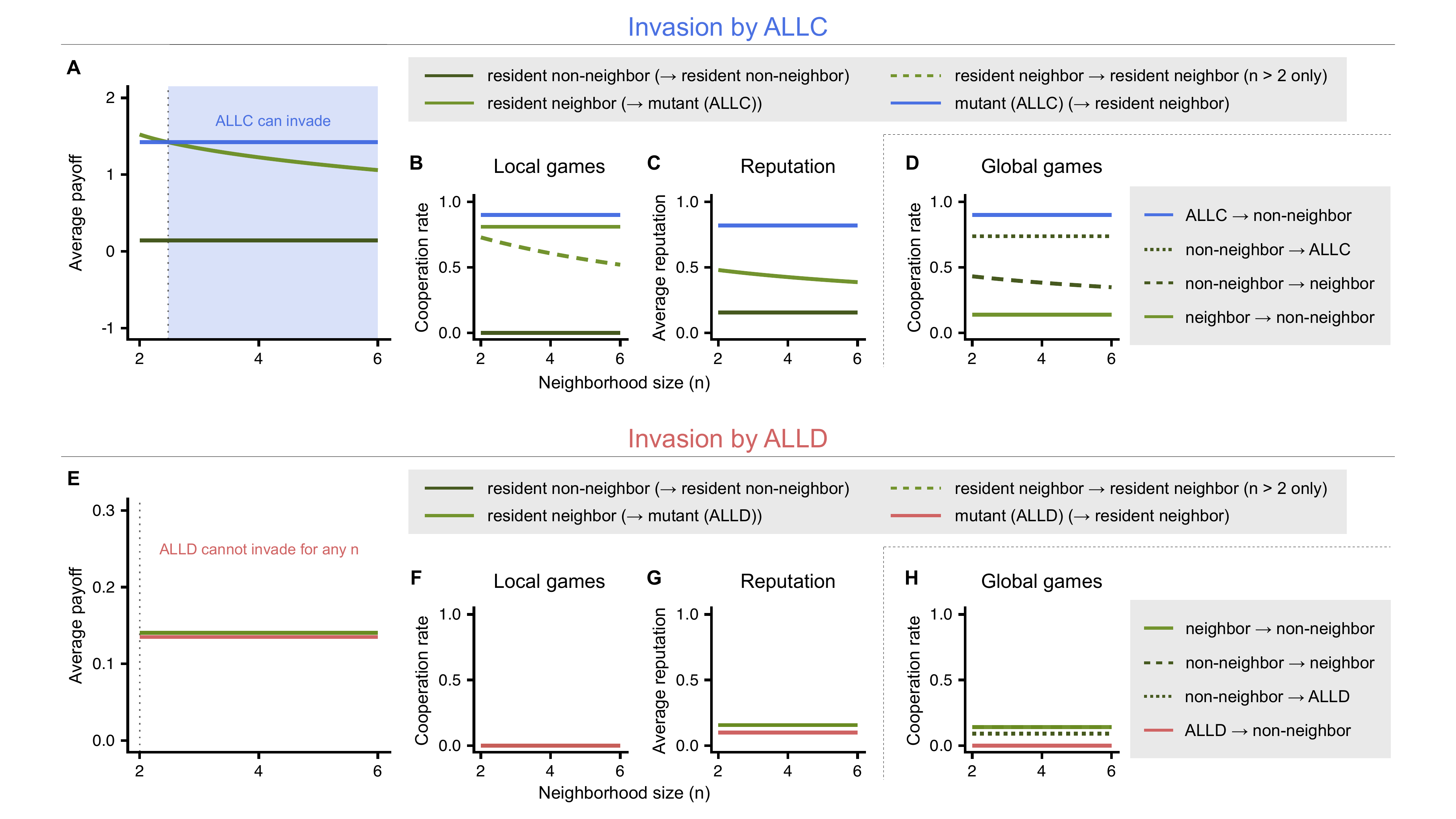}
    \caption{
        \textbf{Effect of neighborhood size on the invasibility of tit-for-tat discriminators under hybrid reciprocity.}
        As in \cref{fig:fig2}, but with varying neighborhood size $n$. 
        Resident players (both non-neighbors and neighbors) act as tit-for-tat (TFT) players in local games and discriminators (DISC) in global games (TFT--DISC).
        Top row (\textbf{A--D}) corresponds to invasion by a mutant cooperator (ALLC), and bottom row (\textbf{E--H}) to invasion by a mutant defector (ALLD).
        Panels show average payoff (\textbf{A}, \textbf{E}), cooperation rate in local games (\textbf{B}, \textbf{F}), average reputation (\textbf{C}, \textbf{G}), and cooperation rate in global games (\textbf{D}, \textbf{H}), as a function of the probability of local play ($\lambda$).
        Colors and line types indicate player classes (or pairs of classes) as shown; in all cases, colors denote the player class of the donor, and line types distinguish between the player classes of the recipient where necessary. 
        Note that some curves overlap in panels \textbf{E--H}.
        In \textbf{E} and \textbf{G}, resident non-neighbors (dark green) and resident neighbors (light green) have identical curves. 
        In \textbf{F}, all four curves overlap (payoff per local game is zero).
        In \textbf{H}, resident non-neighbors cooperate with resident neighbors (light green) at the same rate as resident neighbors do with resident non-neighbors (dashed dark green).
        Other parameters: $b = 3$,  $c = 1$, $\assess=\exec=0.1$, $\lambda = 0.5$.
        }
    \label{fig:SI-neffect1}
\end{figure}

\clearpage
\begin{figure}[p]
    \centering
    \includegraphics[width=0.6\linewidth,trim={5in 2in 5in 2in},clip]{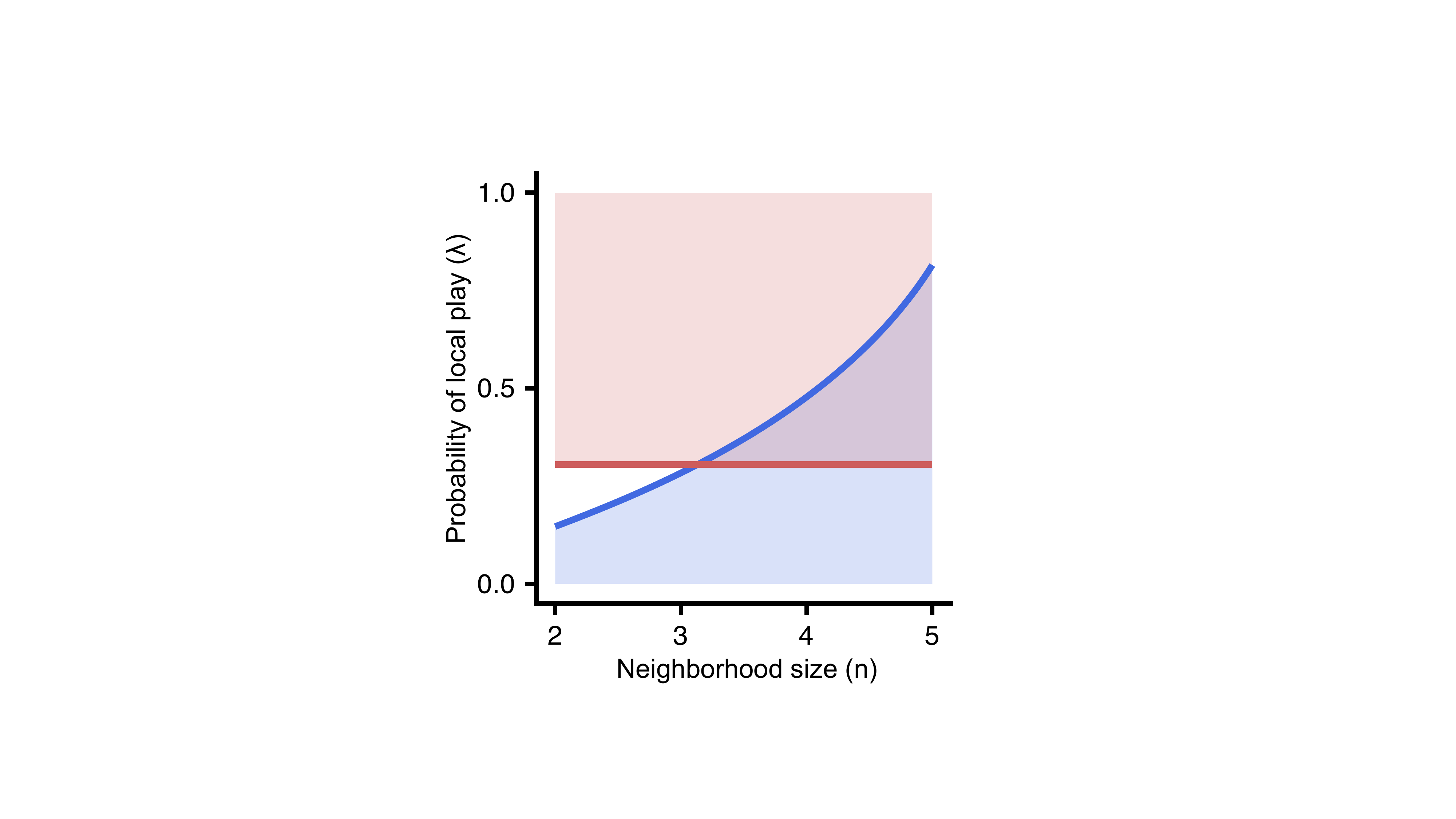}
    \caption{
        \textbf{Effect of neighborhood size $n$ on the scoring dilemma.}
        Resident players act as tit-for-tat (TFT) players in local games and discriminators (DISC) in global games. 
        Figure shows the effective critical probabilities of local play against an ALLC mutant ($\lambda_{\ALLC}^*$, blue curve) and against an ALLD mutant ($\lambda_{\ALLD}^*$, red curve), as a function of neighborhood size $n$. 
        ALLC can invade TFT--DISC locally when $\lambda<\lambda_{\ALLC}^*$ (blue region); ALLD can invade when $\lambda>\lambda_{\ALLD}^*$ (red region); and neither can invade when $\lambda_{\ALLC}^*<\lambda<\lambda_{\ALLD}^*$ (white region). 
        The scoring dilemma is resolved for sufficiently small neighborhood size $n$.
        Other parameters: $n = 2$, $c = 1$, $b = 3$, $\assess=0.1$, $\exec=0.4$.
     }
    \label{fig:SI-neffect2}
\end{figure}

\clearpage
\begin{figure}
    \centering
    \includegraphics[width=0.91\linewidth,trim={3in 0.85in 3in 0.85in},clip]{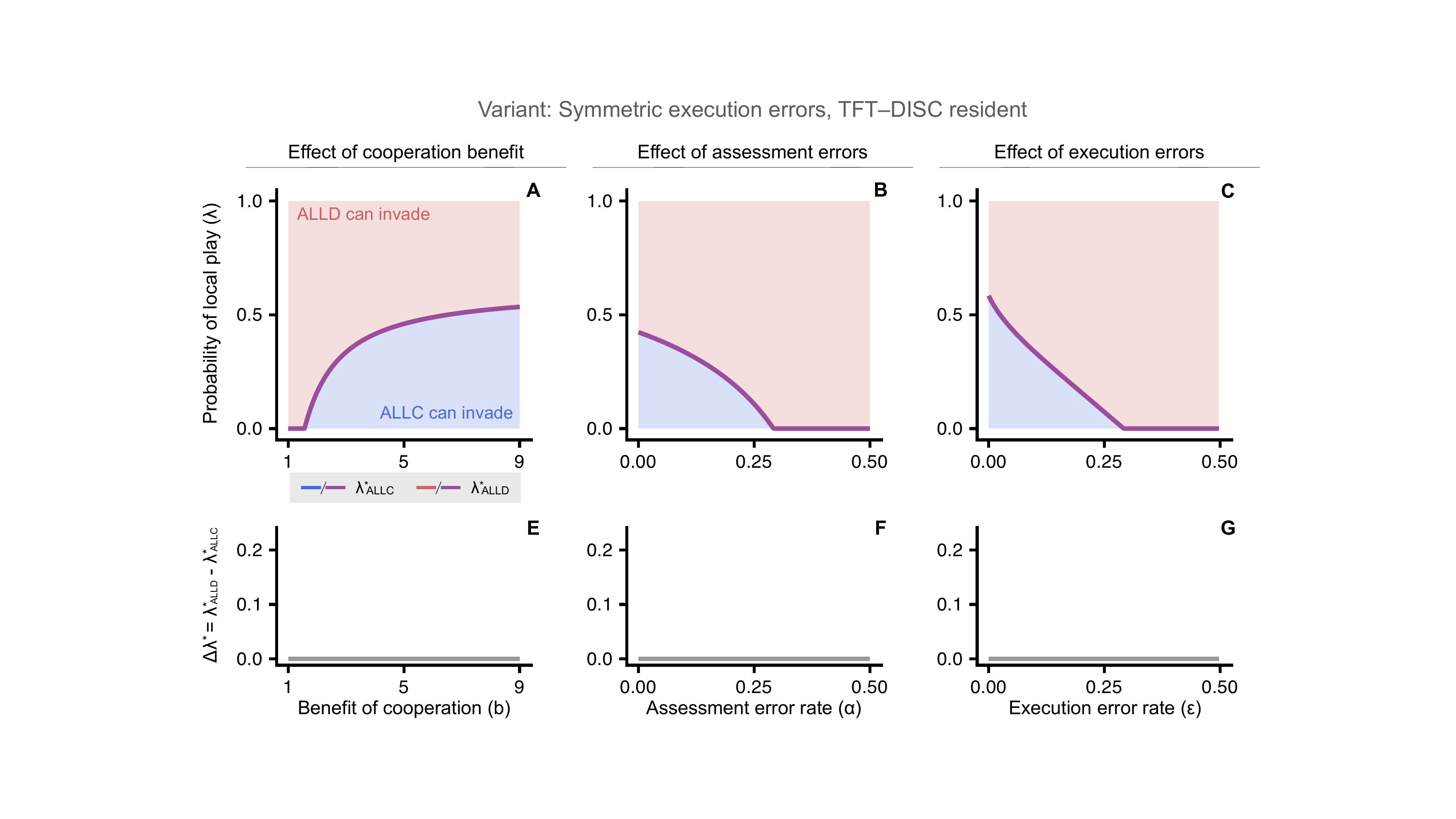}
    \caption{
        \textbf{The scoring dilemma persists when execution errors are symmetric and resident plays TFT--DISC.}
        As in \cref{fig:fig3}, but with symmetric execution errors.
        Resident players act as tit-for-tat (TFT) players in local games and discriminators (DISC) in global games. 
        Since $\lambda_{\ALLC}^*$ and $\lambda_{\ALLD}^*$ coincide (purple curves in \textbf{A--D}; $\Delta\lambda^*=0$, \textbf{E--H}), for any $\lambda\in[0,1]$ either ALLC or ALLD can invade TFT--DISC. Hence, TFT--DISC cannot solve the scoring dilemma under symmetric execution errors. 
        }
    \label{fig:SI-symmetric}
\end{figure}

\clearpage

\begin{figure}
    \centering
    \includegraphics[width=0.91\linewidth,trim={2.8in 0in 2.8in 0in},clip]{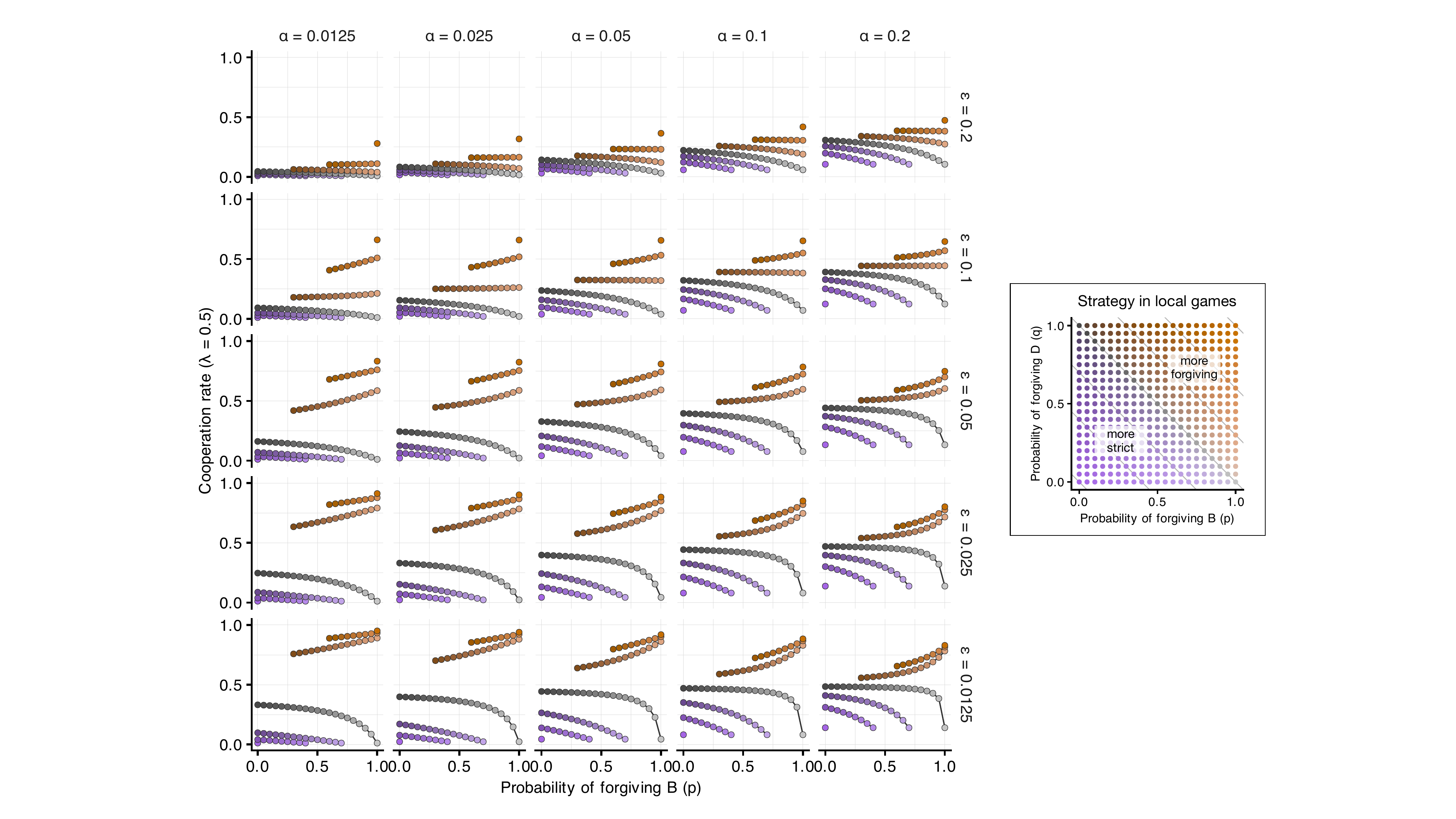}
    \caption{
        \textbf{Effect of forgiveness on cooperation in monomorphic populations.}
        As in \cref{fig:fig4}, but for various assessment error rates ($\assess$, columns) and execution error rates ($\exec$, rows). 
        Panels show average rates of cooperation in monomorphic populations of cross-scale discriminators ($pq$--DISC) as a function of the probability of forgiving a bad (global) reputation ($p$).
        Colors indicate degrees of forgiveness in local games (see 2D legend): shades of orange denote strategies that are overall forgiving ($p+q > 1$), whereas shades of purple denote those that are overall strict ($p+q < 1$).
        Other parameters: $b = 3$, $c = 1$, $n = 2$, $\lambda=0.5$.
    }
    \label{fig:SI-dots-grid-l05}
\end{figure}

\begin{figure}
    \centering
    \includegraphics[width=0.91\linewidth,trim={2.8in 0in 2.8in 0in},clip]{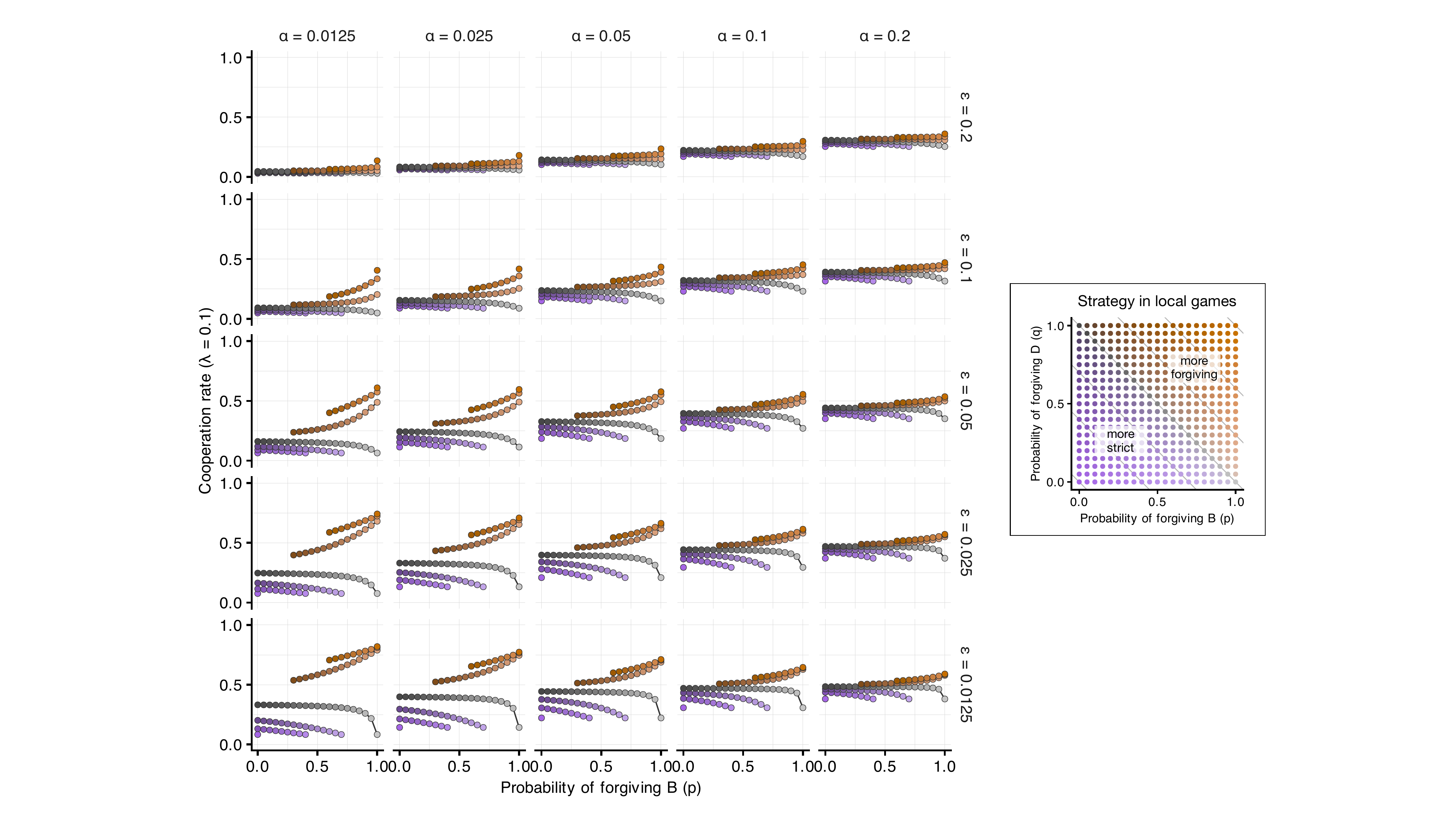}
    \caption{As in \cref{fig:SI-dots-grid-l05}, but with $\lambda = 0.1$.}
    \label{fig:SI-dots-grid-l01}
\end{figure}

\begin{figure}
    \centering
    \includegraphics[width=0.91\linewidth,trim={2.8in 0in 2.8in 0in},clip]{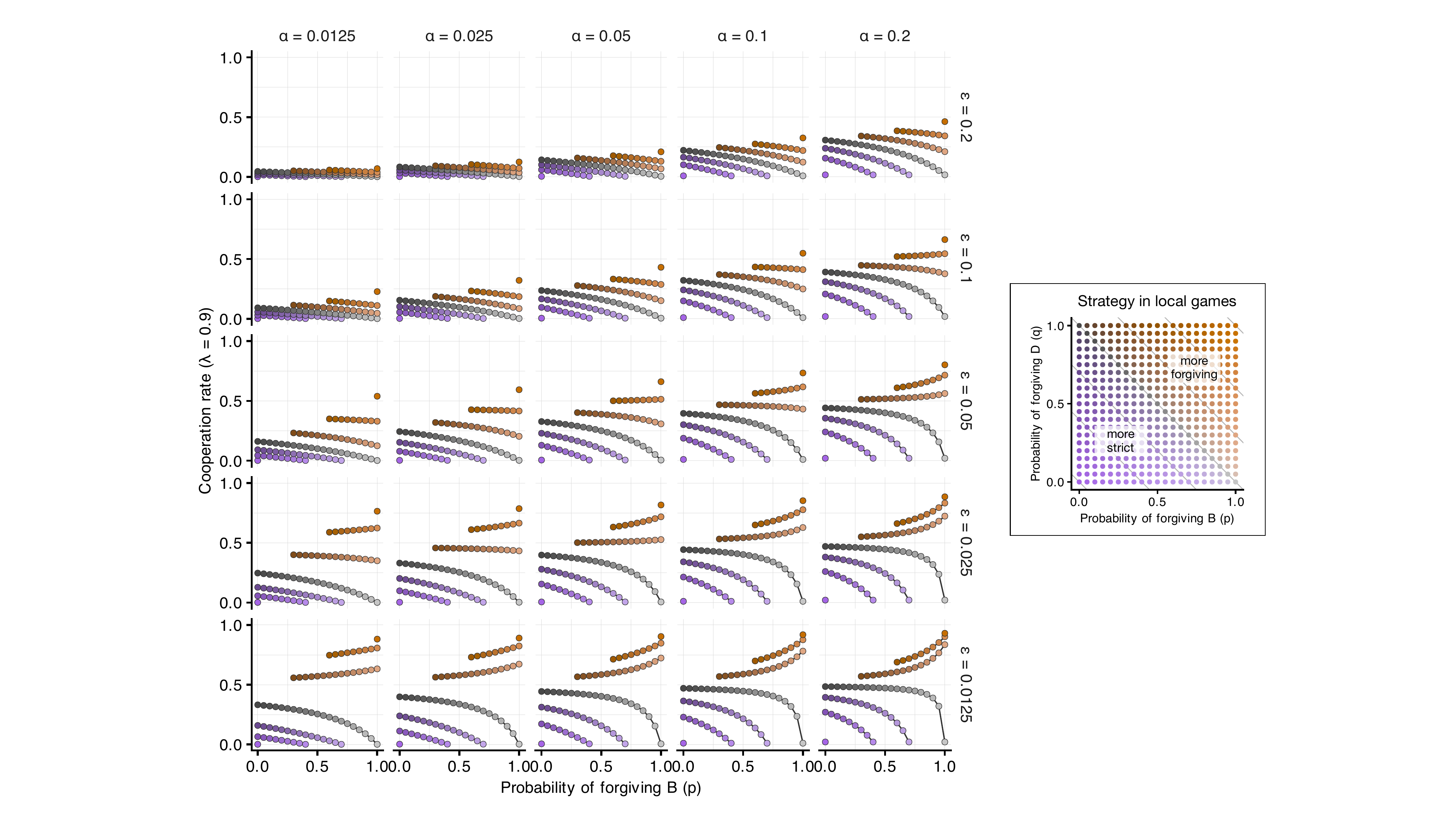}
    \caption{As in \cref{fig:SI-dots-grid-l05}, but with $\lambda = 0.9$.}
    \label{fig:SI-dots-grid-l09}
\end{figure}

\begin{figure}
    \centering
    \includegraphics[width=0.91\linewidth,trim={3.5in 0in 3.5in 0in},clip]{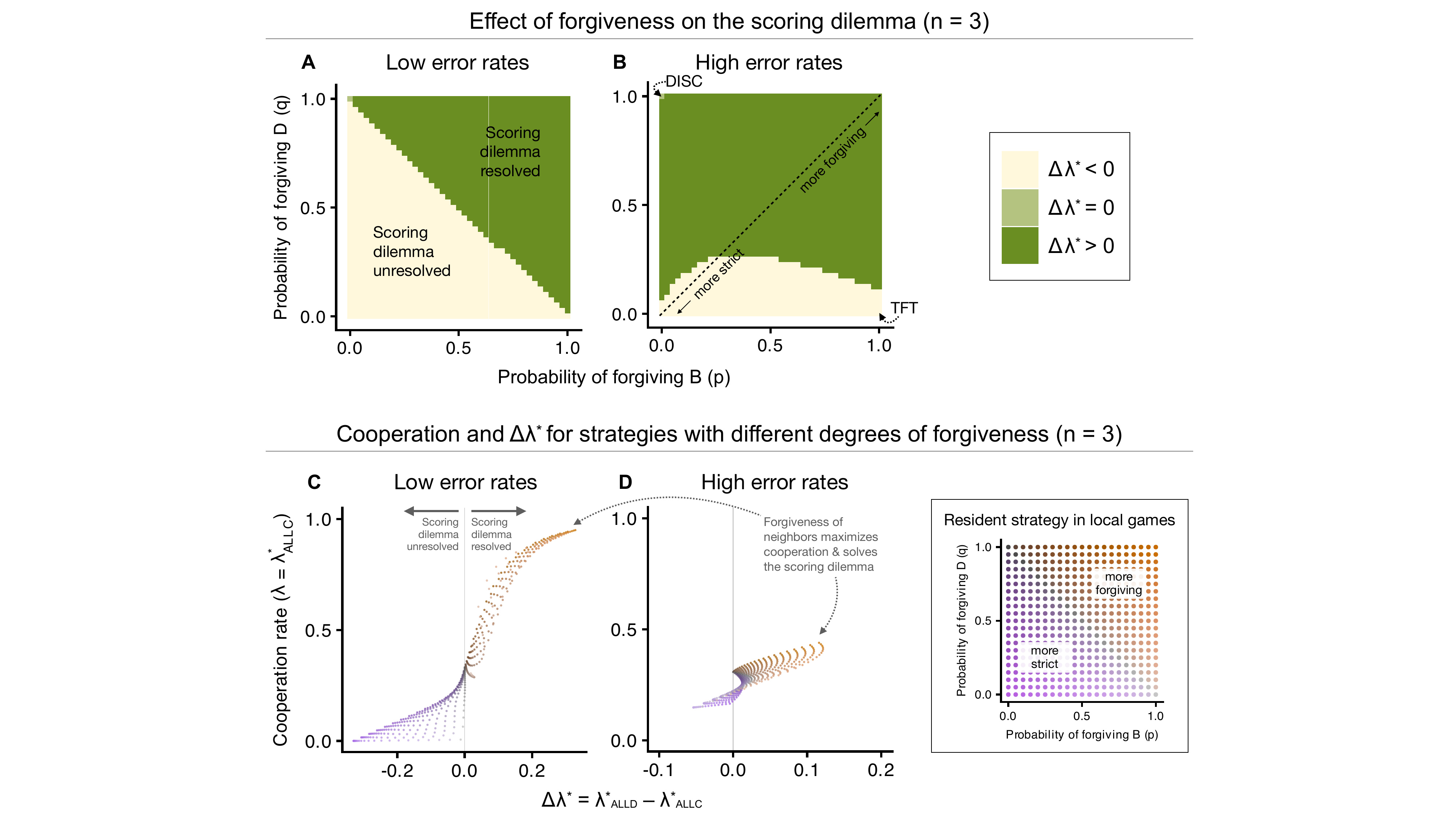}
    \caption{
        \textbf{Effect of forgiveness on the scoring dilemma and fitness ($n=3$).}
        As in \cref{fig:fig5}, but for neighborhood size $n = 3$. 
        \textbf{A, B}: When $n=3$, forgiving strategies ($p+q$ large) tend to solve the scoring dilemma under both low (\textbf{A}) and high (\textbf{B}) error rates. In particular, whereas highly forgiving strategies fail to solve the scoring dilemma when $n=2$ and errors are common (\cref{fig:fig5}B, beige region), they do so when $n=3$ (\textbf{B}).
        \textbf{C, D}: When $n=3$, the most forgiving strategy both achieves maximal cooperation and solves the dilemma under both low (\textbf{C}) and high (\textbf{D}) error rates. In particular, whereas maximizing cooperation is at odds with solving the scoring dilemma when $n=2$ and errors are common (\cref{fig:fig5}D, beige region), this tradeoff is absent when $n=3$ (\textbf{D}).
        Error rates are $\assess = \exec = 0.0125$ in \textbf{A, C} (low error rates) and $\assess = \exec = 0.2$ in \textbf{B, D} (high error rates). Other parameters: $c = 1$, $b = 3$, $n = 3$. 
    }
    \label{fig:SI-fig5vn3}
\end{figure}

\begin{figure}
    \centering
    \includegraphics[width=0.91\linewidth,trim={1.5in 0.25in 1.5in 0.25in},clip]{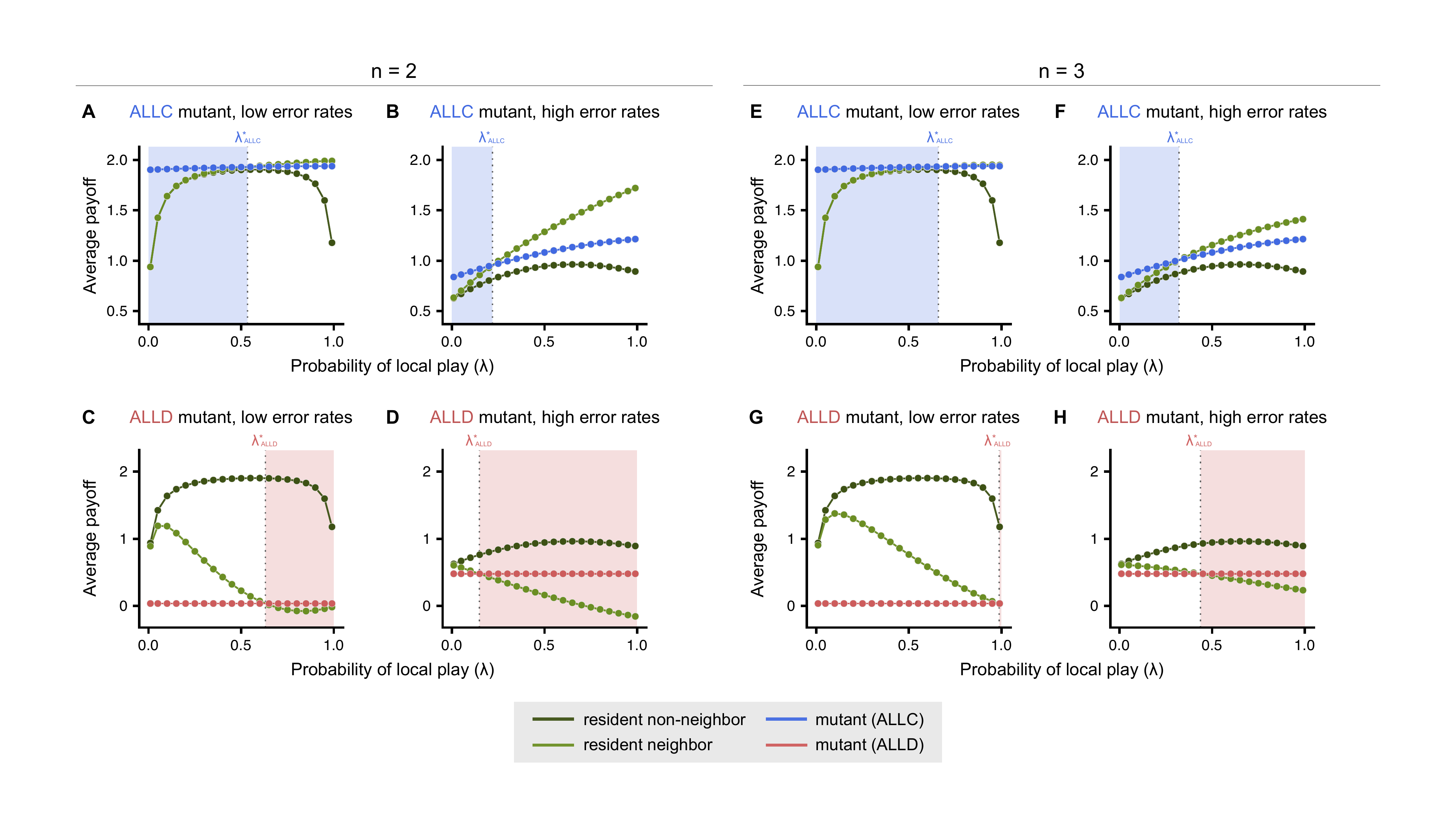}
    \caption{
        \textbf{Effect of neighborhood size and error rates on the scoring dilemma for the most forgiving $pq$--DISC strategy ($(p,q)=(1,1)$).}
        Average payoff as a function of the probability of local play ($\lambda$) when the resident population adopts the most forgiving strategy $(p,q)=(1,1)$ in local games and DISC in global games. Neighborhood size is $n=2$ in \textbf{A--D} and $n=3$ in \textbf{E--H}.
        A rare ALLC mutant (\textbf{A, B, E, F}) or ALLD mutant (\textbf{C, D, G, H}) is introduced into a neighborhood of the resident population, under low error rates (\textbf{A, C, E, G}; $\assess=\exec=0.0125$) or high error rates (\textbf{B, D, F, H}; $\assess=\exec=0.2$). Colors denote player classes as indicated in the legend.
        Under low error rates, an ALLD mutant is almost always assigned a bad reputation, because ALLD defects deterministically (under asymmetric execution errors) and assessments are typically accurate. As a result, ALLD rarely benefits from the forgiveness of resident neighbors (red curves in \textbf{C} and \textbf{G}), who cooperate locally only with those who have at least one positive bit of information---i.e., those who either cooperated in the most recent local game or have good reputations. This makes it particularly difficult for ALLD to invade. As a result, under low error rates, the scoring dilemma is resolved---i.e., $\lambda_{\ALLD}^*>\lambda_{\ALLC}^*$ such that there is an intermediate region of $\lambda$ for which $(p,q)=(1,1)$ can resist both ALLC and ALLD---for both $n=2$ (\textbf{A} and \textbf{C}) and $n=3$ (\textbf{E} and \textbf{G}).
        Under high error rates and $n=2$, however, the ALLD mutant is occasionally assigned a good reputation by accident, and therefore occasionally benefits from the forgiveness of resident neighbors (red curve in \textbf{D}). This makes it substantially easier for ALLD to invade. Consequently, the scoring dilemma is no longer resolved for $n=2$ under high error rates (\textbf{B} and \textbf{D}).
        However, when the neighborhood size increases to $n=3$, the scoring dilemma is once again resolved under high error rates (\textbf{F} and \textbf{H}). In this case, the ALLD mutant has two resident neighbors, who frequently cooperate with one another because they are maximally forgiving. Although ALLD also benefits from their occasional forgiveness, the sustained mutual cooperation between the two resident neighbors increases their payoffs (light green curve in \textbf{H}), thus making it more difficult for ALLD to invade relative to the $n=2$ case.
        Other parameters: $c = 1$, $b = 3$. 
    }
    \label{fig:SI-pq11explanatory}
\end{figure}

\begin{figure}
    \centering
    \includegraphics[width=0.91\linewidth,trim={3.5in 0in 3.5in 0in},clip]{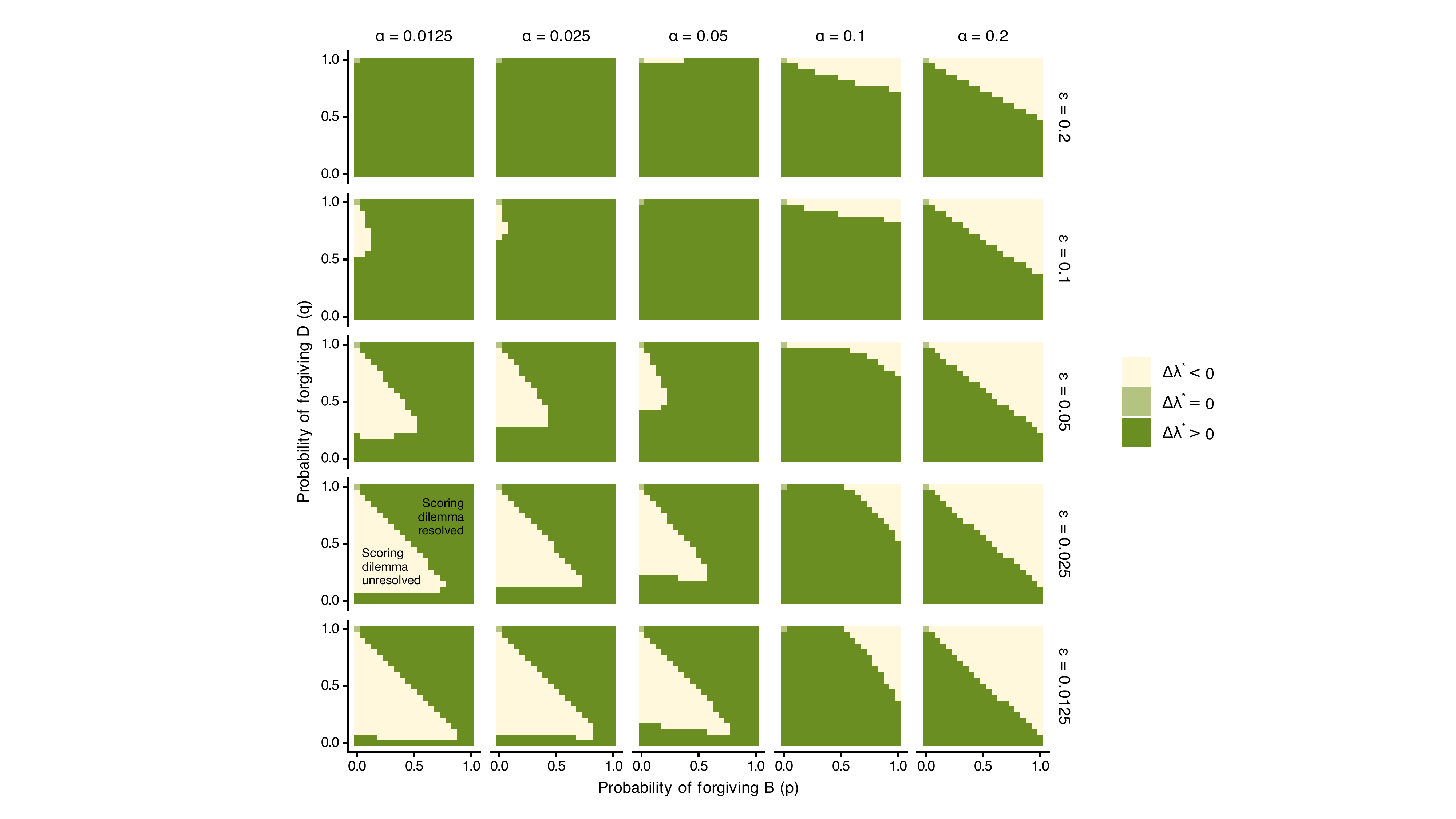}
    \caption{
        \textbf{Effect of forgiveness on the scoring dilemma ($n=2$).}
        As in \cref{fig:fig5}A and B, but for various assessment error rates ($\assess$, columns) and execution error rates ($\exec$, rows). We classify the two-dimensional space of cross-scale discriminator strategies by whether the scoring dilemma is resolved.
        Each coordinate $(p,q)$ corresponds to a resident $pq$--DISC strategy.
        The dilemma is resolved when $\Delta\lambda^* > 0$ (green); it is unresolved when $\Delta\lambda^* < 0$ (beige) or $\Delta\lambda^* = 0$ (pale green).
        Other parameters: $c = 1$, $b = 3$, $n = 2$. 
    }
    \label{fig:SI-heatmap-grid}
\end{figure}

\begin{figure}
    \centering
    \includegraphics[width=0.91\linewidth,trim={2.75in 0in 2.75in 0in},clip]{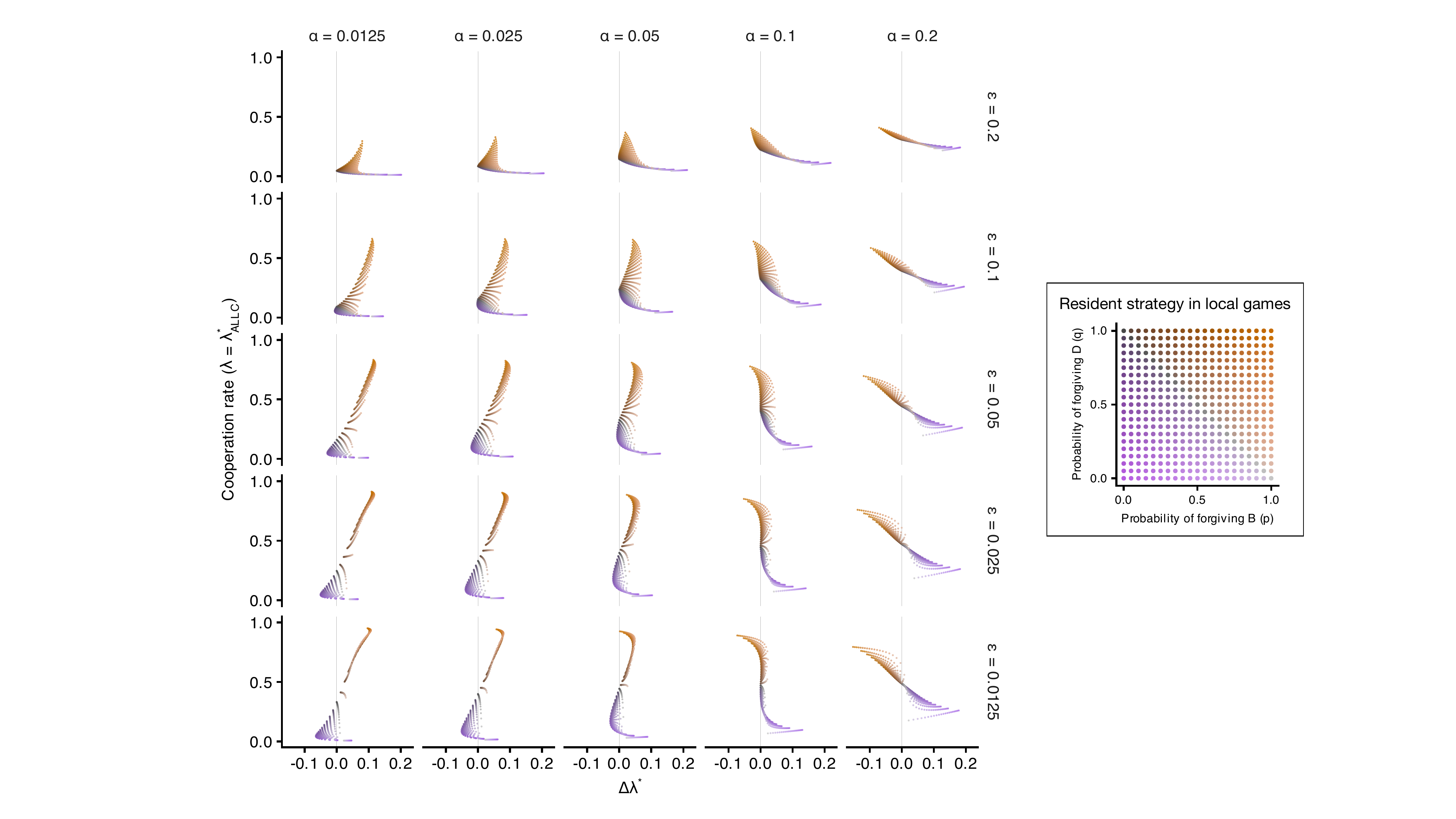}
    \caption{
        \textbf{Effect of forgiveness on the scoring dilemma and fitness ($n=2$).}
        As in \cref{fig:fig5}C and D, but for various assessment error rates ($\assess$, columns) and execution error rates ($\exec$, rows). 
        The resident population plays $pq$--DISC, with varying $p$ and $q$. Each point corresponds to a resident strategy in local games. 
        The coordinates of each point give the value of $\Delta\lambda^*$ (horizontal coordinate) and the cooperation rate at $\lambda = \lambda_{\ALLC}^*$ (vertical coordinate) for the corresponding resident strategy.
        Colors indicate degrees of forgiveness in local games (see 2D legend): shades of orange denote strategies that are overall forgiving ($p+q > 1$), whereas shades of purple denote those that are overall strict ($p+q < 1$).
        Vertical lines at $\Delta\lambda^* = 0$ separate regions where the scoring dilemma is resolved ($\Delta\lambda^* > 0$) from those where it is unresolved ($\Delta\lambda^* \leq 0$).
        Other parameters: $c = 1$, $b = 3$, $n = 2$. 
    }
    \label{fig:SI-scatter-grid}
\end{figure}

\begin{figure}[h!]
    \centering
    \includegraphics[width=0.6\linewidth,trim={5in 2in 5in 2in},clip]{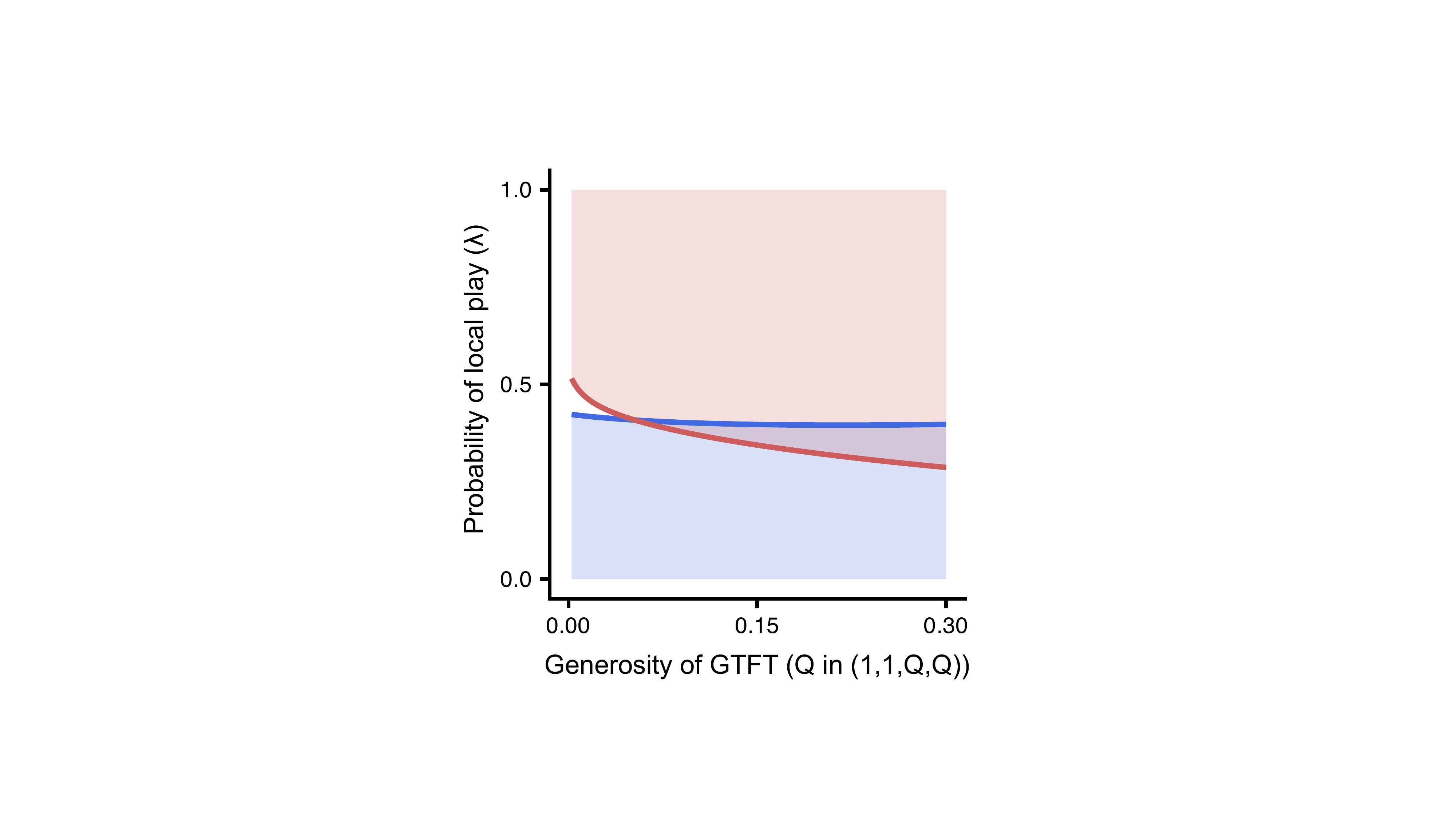}
    \caption{
        \textbf{Effect of reputation-agnostic generosity on the scoring dilemma.}
        Resident players act as generous tit-for-tat (GTFT) players in local games and discriminators (DISC) in global games. 
        Figure shows the effective critical probabilities of local play against an ALLC mutant ($\lambda_{\ALLC}^*$, blue curve) and against an ALLD mutant ($\lambda_{\ALLD}^*$, red curve), as a function of the generosity of GTFT (i.e., $Q$ in $\svDR=(1,1,Q,Q)$). 
        ALLC can invade TFT--DISC locally when $\lambda<\lambda_{\ALLC}^*$ (blue region); ALLD can invade when $\lambda>\lambda_{\ALLD}^*$ (red region); and neither can invade when $\lambda_{\ALLC}^*<\lambda<\lambda_{\ALLD}^*$ (white region). 
        The scoring dilemma is resolved for sufficiently small $Q$.
        Other parameters: $n = 2$, $c = 1$, $b = 3$, $\assess=\exec=0.1$.
     }
    \label{fig:SI-GTFT}
\end{figure}

\end{document}